\documentclass[aps,pra,reprint,amsmath,amssymb,superscriptaddress,onecolumn,longbibliography, notitlepage,nofootinbib]{revtex4-2}
\usepackage{mathtools}
\usepackage{siunitx}
\usepackage[hidelinks]{hyperref}
\usepackage{float}

\usepackage{physics}

\usepackage{bbm}

\usepackage[labelfont=bf, justification=RaggedRight, figurename={Figure }, tablename={Table}, labelsep=period]{caption}

\usepackage{xcolor}

\begin{document}

\title{Quantum-limited stochastic optical neural networks operating at a few quanta per activation}

\author{Shi-Yuan~Ma}
\email{sm2725@cornell.edu}
\affiliation{School of Applied and Engineering Physics, Cornell University, Ithaca, NY 14853, USA}

\author{Tianyu~Wang}
\affiliation{School of Applied and Engineering Physics, Cornell University, Ithaca, NY 14853, USA}

\author{Jérémie~Laydevant}
\affiliation{School of Applied and Engineering Physics, Cornell University, Ithaca, NY 14853, USA}
\affiliation{USRA Research Institute for Advanced Computer Science, Mountain View, CA 94035, USA}

\author{Logan~G.~Wright} 
\thanks{Present address: Department of Applied Physics, Yale University, New Haven, CT 06511, USA}
\affiliation{School of Applied and Engineering Physics, Cornell University, Ithaca, NY 14853, USA}
\affiliation{NTT Physics and Informatics Laboratories, NTT Research, Inc., Sunnyvale, CA 94085, USA}

\author{Peter~L.~McMahon}
\email{pmcmahon@cornell.edu}
\affiliation{School of Applied and Engineering Physics, Cornell University, Ithaca, NY 14853, USA}
\affiliation{Kavli Institute at Cornell for Nanoscale Science, Cornell University, Ithaca, NY 14853, USA}

\begin{abstract}

Energy efficiency in computation is ultimately limited by noise, with quantum limits setting the fundamental noise floor. Analog physical neural networks hold promise for improved energy efficiency compared to digital electronic neural networks. However, they are typically operated in a relatively high-power regime so that the signal-to-noise ratio (SNR) is large (\textgreater10), and the noise can be treated as a perturbation. We study optical neural networks where all layers except the last are operated in the limit that each neuron can be activated by just a single photon, and as a result the noise on neuron activations is no longer merely perturbative. We show that by using a physics-based probabilistic model of the neuron activations in training, it is possible to perform accurate machine-learning inference in spite of the extremely high shot noise (SNR $\sim$ 1). We experimentally demonstrated MNIST handwritten-digit classification with a test accuracy of 98\% using an optical neural network with a hidden layer operating in the single-photon regime; the optical energy used to perform the classification corresponds to just 0.038 photons per multiply–accumulate (MAC) operation. Our physics-aware stochastic training approach might also prove useful with non-optical ultra-low-power hardware.

\end{abstract}

\maketitle

\section{Introduction}
\label{sec:intro}

The development and widespread use of very large neural networks for artificial intelligence \cite{lecun2015deep,canziani2016analysis} has motivated the exploration of alternative computing paradigms---including analog processing---in the hope of improving both energy efficiency and speed \cite{markovi2020physics,christensen20222022}. 
Photonic implementations of neural networks using analog optical systems have experienced a resurgence of interest over the past several years \cite{shen2017deep,lin2018all,rios2019memory,wetzstein2020inference,xu202111,feldmann2021parallel,zhou2021large,wang2022optical,davis2022frequency,ashtiani2022chip,sludds2022delocalized}. However, analog processors---including those constructed using optics---inevitably have noise and typically also suffer from imperfect calibration and drift. These imperfections can result in degraded accuracy for neural-network inference performed using them \cite{shen2017deep,moon2019enhancing,joshi2020accurate,semenova2022understanding}. To mitigate the impact of noise, noise-aware training schemes have been developed \cite{klachko2019improving,zhou2020noisy,yang2022tolerating,wright2022deep,semenova2022noise,wu2022harnessing,anderson2024optical,jiang2023physical}. These schemes treat the noise as a relatively small perturbation to an otherwise deterministic computation, either by explicitly modeling the noise as the addition of random variables to the processor's output or by modeling the processor as having finite bit precision. 
Recent demonstrations of ultra-low optical energy usage in optical neural networks (ONNs) \cite{wang2022optical,sludds2022delocalized} have used merely hundreds to thousands of photons (SNR $\lesssim10^2$) to represent the neuron pre-activation signal prior to photodetection (or equivalently, $<1$ photon per MAC). However, they were still in the regime where the noise is a small perturbation. More typically, millions of photons per activation are used to achieve reliable, accurate results \cite{feldmann2021parallel,zhou2021large,sludds2022delocalized,bernstein2023single}.
In this paper we address the following question: 
what happens if we use such weak optical signals in a ONN that each photodetector in a neural-network layer receives at most just one, or perhaps two or three, photons? As we will explain, in this scenario, the training methods that treat the photodetector outputs as deterministic values with a small amount of noise added to them, such as used in Refs.~\cite{wang2022optical,sludds2022delocalized}, would fail to achieve high machine-learning accuracy, so a new approach is needed.

Physical systems are subject to various sources of noise. While some noise can be reduced through improvements to the hardware, some noise is fundamentally unavoidable, especially when the system is operated with very little power---which is an engineering goal for neural-network processors. Shot noise is a fundamental noise that arises from the quantized, i.e., discrete, nature of information carriers: the discreteness of energy in the case of photons in optics, and of discreteness of charge in the case of electrons in electronics \cite{beenakker2003quantum}. A shot-noise-limited measurement of a signal encoded with an average of $N_\textrm{p}$ photons (quanta) will have an SNR that scales as $\sqrt{N_\textrm{p}}$ \cite{agarwal2012quantum}.\footnote{The \textit{shot-noise limit}, which is sometimes also referred to as the \textit{standard quantum limit} \cite{machida1987observation}, can be evaded if, instead of encoding the signal in a thermal or coherent state of light, a quantum state---such as an intensity-squeezed state or a Fock state---is used. In this paper we consider only the case of \textit{classical} states of light for which shot noise is present and the shot-noise limit applies.} To achieve a suitably high SNR, ONNs typically use a large number of quanta for each detected signal. In situations where the optical signal is limited to just a few photons, photodetectors measure and can count individual quanta. Single-photon detectors (SPDs) are highly sensitive detectors that---in the typical \textit{click detector} setting---report, with high fidelity, the absence of a photon (\textit{no click}) or presence of one or more photons (\textit{click}) during a given measurement period \cite{hadfield2009single}. In the quantum-noise-dominated regime of an optical signal with an average photon number of about 1 impinging on an SPD, the measurement outcome will be highly stochastic, resulting in a very low SNR (of about 1).\footnote{Again, this is under the assumption that the optical signal is encoded in an optical state that is subject to the shot-noise limit---which is the case for classical states of light.} Conventional noise-aware-training algorithms are not able to achieve high accuracy with this level of noise. \textbf{Is it possible to operate ONNs in this very stochastic regime and still achieve high accuracy in deterministic classification tasks?} The answer is \textit{yes}, and in this work we will show how.

\begin{figure}
\includegraphics [width=.68\textwidth]{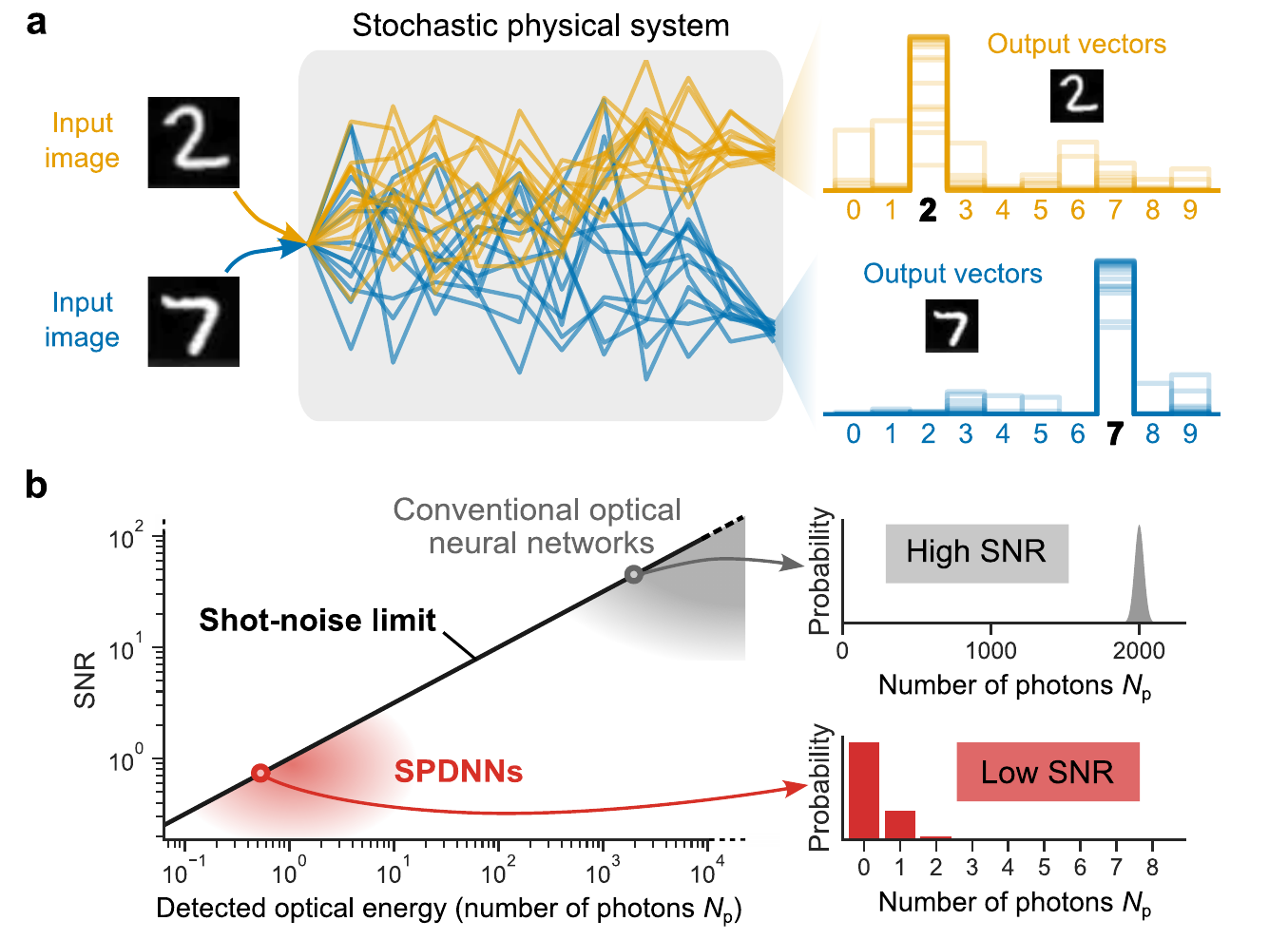}
\caption{\textbf{Deterministic inference using noisy neural-network hardware.}
\textbf{a}, The concept of a stochastic physical neural network performing a classification task. Given a particular input image to classify, repetitions exhibits variation (represented by different traces of the same color), but the class is predicted nearly deterministically.
\textbf{b}, The single-to-noise ratio (SNR) of single-photon-detection neural networks (SPDNNs) compared to conventional optical neural networks (ONNs). Conventional ONNs operate with high photon budgets (SNR $\gg 1$) to obtain reliable results, whereas SPDNNs operate with low photon budgets---of up to just a few detected photons per shot (SNR $\sim 1$). The relation between the detected optical energy (in number of photons $N_\textrm{p}$) and SNR is SNR $=\sqrt{N_\textrm{p}}$, which is known as the shot-noise limit.}
\label{fig:intro}
\end{figure}

The stochastic operation of neural networks has been extensively studied in computer science as part of the broader field of stochastic computing \cite{alaghi2013survey}. In the field of machine learning, binary stochastic neurons (BSNs) have been used to construct stochastic neural networks \cite{ackley1985learning,neal1990learning,neal1992connectionist,bengio2013estimating,tang2013learning,raiko2014techniques,hubara2016binarized}, with training being a major focus of study. Investigations of hardware implementations of stochastic computing neural networks, such as those in Refs.~\cite{ji2015hardware,lee2017energy} (with many more surveyed in Ref.~\cite{liu2020survey}), have typically been for deterministic complementary metal--oxide--semiconductor (CMOS) electronics, with the stochasticity introduced by random-number generators. While many studies of binary stochastic neural networks have been conducted with standard digital CMOS processors, there have also been proposals to construct them from beyond-CMOS hardware, motivated by the desire to minimize power consumption: direct implementation of binary stochastic neurons using bistable systems that are noisy by design---such as low-barrier magnetic tunnel junctions (MTJs)---has been explored \cite{vodenicarevic2017low,hassan2019low,chowdhury2023full}, and there have also been proposals to realize hardware stochastic elements for neural networks that could be constructed with noisy CMOS electronics or other physical substrates \cite{hylton2021vision,coles2023thermodynamic}. ONNs in which noise has been intentionally added \cite{wu2022harnessing,wu2022photonic,ma2023stochastic} have also been studied. Our work with low-photon-count optics is related but distinct from many of the studies cited here in its motivating assumption: instead of desiring noise and stochastic behavior---and purposefully designing devices to have them, we are concerned with situations in which physical devices have large and unavoidable noise but where we would like to nevertheless construct deterministic classifiers using these devices because of their potential for low-energy computing (Figure \ref{fig:intro}).

The \textbf{key idea} in our work is that incorporating a \textit{physics-based, probabilistic model} of the highly stochastic photodetector outputs into the training algorithm can result in high-accuracy, deterministic inference with the ONN hardware. When ONNs are operated in the approximately-1-photon-per-neuron-activation regime and the detectors are SPDs, it is natural to consider the neurons as binary stochastic neurons: the output of an SPD is binary (\textit{click} or \textit{no click}) and fundamentally stochastic. Instead of trying to train the ONN as a deterministic neural network that has very poor numerical precision, one can instead train it as a binary stochastic neural network, adapting some of the methods from the last decade of machine-learning research on stochastic neural networks \cite{bengio2013estimating,tang2013learning,gu2015muprop,hubara2016binarized,liu2020survey} and using a physics-based model of the stochastic single-photon detection (SPD) process during training. We call this \textit{physics-aware stochastic training}. While a high SNR for the output of the final layer is likely essential for deterministic inference, \textbf{our approach allows all the previous layers to be operated in the highly stochastic regime with SNR near 1} (Figure~\ref{fig:intro}a). This is in contrast with an approach like quantization-aware training \cite{gu2021o2nn,wang2022optical,anderson2024optical}, which is only able to deal with quasi-deterministic systems where \textbf{the SNR in every layer is high}.

We experimentally implemented a stochastic ONN using as a building block an optical matrix-vector multiplier \cite{wang2022optical} modified to have SPDs at its output: we call this a \textit{single-photon-detection neural network} (SPDNN). We present results showing that high classification accuracy can be achieved even when the number of photons per neuron activation is approximately 1, and even without averaging over multiple shots. We also studied in simulation how larger, more sophisticated stochastic ONNs could be constructed and what their performance on CIFAR-10 image classification would be.

\begin{figure}[htp]
\includegraphics [width=.99\textwidth]{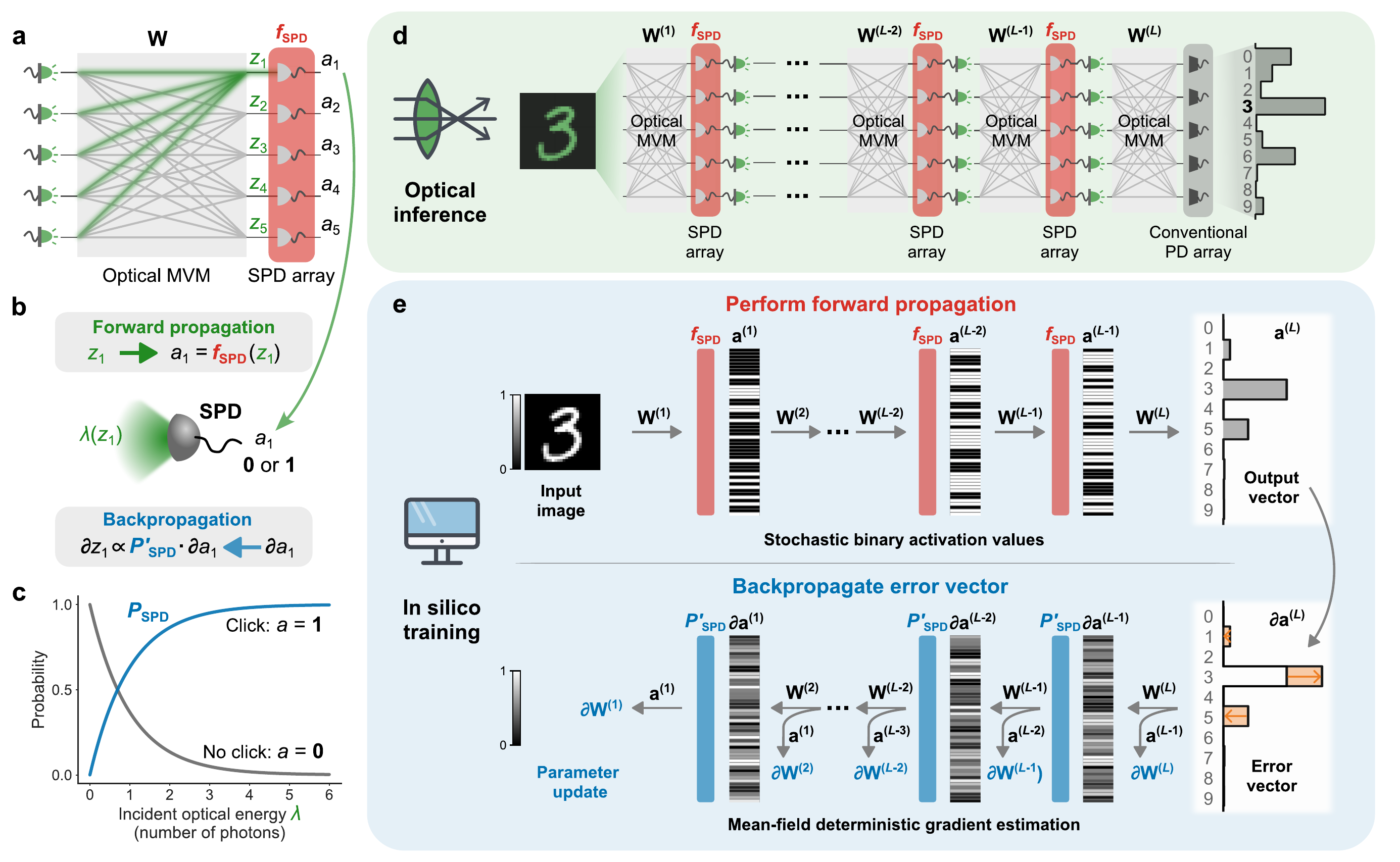}
\caption{\textbf{Single-photon-detection neural networks (SPDNNs): \textit{physics-aware stochastic training} and \textit{inference}.}
\textbf{a}, A single layer of an SPDNN, comprising an optical matrix-vector multiplier (optical MVM, in grey) and single-photon detectors (SPDs; in red), which perform stochastic nonlinear activations. Each output neuron's value is computed by the physical system as $a_i=f_{\text{SPD}}(z_i)$), where $z_i$ is the weighted sum (shown in green) of the input neurons to the $i$th output neuron computed as part of the optical MVM, and $a_i$ is the stochastic binary output from a single-photon detector.
\textbf{b}, Forward and backward propagation through the SPD activation function. 
The optical energy ($\lambda$) incident on an SPD is a function of $z_i$ that depends on the encoding scheme used. Forward propagation uses the stochastic binary activation function $f_{\text{SPD}}$, while backpropagation involves the mean-field function of the probability $P_{\text{SPD}}$.
\textbf{c}, Probability of an SPD detecting a click (output $a=1$) or not (output $a=0$), as a function of the incident light energy $\lambda$.
\textbf{d}, Optical inference using an SPDNN with $L$ layers. The activation values from the SPD array of each layer are passed to light emitters for the optical MVM of the next layer. The last layer uses a conventional photodetector (PD) array instead of an SPD array, and is operated with enough optical energy that the output of this layer has high SNR.
\textbf{e}, \textit{In silico} training of an SPDNN with $L$ layers. Each forward propagation is stochastic, and during backpropagation, the error vector is passed to the hidden layers using the mean-field probability function $P_{\text{SPD}}$ instead of the stochastic activation function $f_{\text{SPD}}$. In this figure, $\partial x$ is shorthand for $\partial C / \partial x$, where $C$ is the cost function.
}
\label{fig:diag}
\end{figure}

\section{Single-photon-detection neural networks: optical neural networks with stochastic activation from single-photon detection}
\label{subsec:spd_actv}

We consider ONNs in which one or more layers are each constructed from an optical matrix-vector multiplier followed by an array of SPDs (Figure \ref{fig:diag}a--c), and in which the optical powers used are sufficiently low that in each execution of the layer, each SPD has at most only a few photons impinging on it, leading to stochastic measurement outcomes of \textit{no click} or \textit{click}.

In our setting, we aim to perform \textit{inference} using the SPDNN---with its implementation in physical hardware---(Figure \ref{fig:diag}d) and to perform \textit{training} of the SPDNN \textit{in silico} (Figure \ref{fig:diag}e). That is, training is performed entirely using standard digital electronic computing.\footnote{It is not required that the training be done \textit{in silico} for it to succeed but is just a choice we made in this work. \textit{Hardware-in-the-loop} training, such as used in Ref.~\cite{wright2022deep}, is a natural alternative to purely \textit{in silico} training that even can make training easier by relaxing the requirements on how accurate the \textit{in silico} model of the physical hardware process needs to be.}

\subsection{Physics-aware stochastic training}

To train an SPDNN, we perform gradient descent using backpropagation, which involves a forward pass, to compute the current error (or loss) of the network, and a backward pass, which is used to compute the gradient of the loss with respect to the network parameters; our procedure is inspired by backpropagation-based training of stochastic and binary neural networks \cite{bengio2013estimating,hubara2016binarized}. We model the forward pass (upper part of Figure \ref{fig:diag}e) through the network as a stochastic process that captures the key physics of SPD of optical signals having Poissonian photon statistics \cite{gerry2005introductory}: the measurement outcome of SPD is a binary random variable (\textit{no click} or \textit{click}) that is drawn from the Bernoulli distribution with a probability that depends on the mean photon number of the light impinging on the detector. However, during the backward pass (lower part of Figure \ref{fig:diag}e), we employ a deterministic mean-field estimator to compute the gradients. This approach avoids the stochasticity and binarization of the SPD process, which typically pose difficulties for gradient estimation.

We now give a brief technical description of our forward and backward passes for training; for full details see Methods and Supplementary Notes 1A and 2A. We denote the neuron pre-activations of the $l$th stochastic layer of an SPDNN as $\textbf{z}^{(l)}=W^{(l)}\textbf{a}^{(l-1)}$, where $\textbf{a}^{(l-1)}$ is the activation vector from the previous layer ($\textbf{a}^{(0)}$ denotes the input vector $\textbf{x}$ of the data to be classified). In the physical realization of an SPDNN, $\textbf{z}^{(l)}$ is encoded optically (for example, in optical intensity) following an optical matrix-vector multiplier (optical MVM, which computes the product between the matrix $W^{(l)}$ and the vector $\textbf{a}^{(l-1)}$) but before the light impinges on an array of SPDs. We model the action of an SPD with a stochastic activation function, $f_{\text{SPD}}$ (Figure \ref{fig:diag}b; Eq.~\ref{eq:f_spd}). The stochastic output of the $l$th layer is then $\textbf{a}^{(l)}=f_{\text{SPD}}(\textbf{z}^{(l)})$.

For an optical signal having mean photon number $\lambda$ and that obeys Poissonian photon statistics, the probability of a \textit{click} event by an SPD is $P_{\text{SPD}}(\lambda) = 1-e^\lambda$ (Figure \ref{fig:diag}c). We define the stochastic activation function $f_{\text{SPD}}$ as follows:

\begin{equation}
\label{eq:f_spd}
  f_{\text{SPD}}(z) \coloneqq
  \begin{cases}
    1 & \text{with probability $p=P_{\text{SPD}}(\lambda(z))$}, \\
    0 & \text{with probability $1-p$,}
  \end{cases}
\end{equation}

\noindent where $\lambda(z)$ is a function mapping a single neuron's pre-activation value to a mean photon number. For an incoherent optical setup where the information is directly encoded in intensity, $\lambda(z) = z$; for a coherent optical setup where the information is encoded in field amplitude and the SPD directly measures the intensity, $\lambda(z) = |z|^2$. In general, the form of $\lambda(z)$ is determined by the signal encoding used in the optical MVM, and the detection scheme following the MVM. We use $f_{\text{SPD}}$ in modeling the stochastic behavior of an SPDNN layer in the forward pass. However, during the backward pass, we make a deterministic mean-field approximation of the network: instead of evaluating the stochastic function $f_{\text{SPD}}$, we evaluate $P_{\text{SPD}}(\lambda(z))$ when computing the activations of a layer: $\textbf{a}^{(l)}=P_{\text{SPD}}(\lambda(\textbf{z}^{(l)}))$ (Figure \ref{fig:diag}b). This is an adaptation of a standard machine-learning method for computing gradients of stochastic neural networks \cite{bengio2013estimating}.

\subsection{Inference}

When performing inference (Figure \ref{fig:diag}d), we can run just a single shot of a stochastic layer or we can choose to take the average of multiple shots---trading greater energy and/or time usage for reduced stochasticity. For a single shot, a neuron activation takes on the value $a^{[1]}=a\in\{0,1\}$; 
for $K$ shots, $a^{[K]}=\frac{1}{K}\sum^K_{k=1} a_k\in\{0,1/K,2/K,\ldots,1\}$. In the limit of infinitely many shots, $K\rightarrow\infty$, the activation $a^{[\infty]}$ would converge to the expectation value, $a^{[\infty]}=\mathbb{E}[a]=P_{\text{SPD}}(\lambda(z))$.
In this work we focus on the single-shot ($K=1$) and few-shot 
($K \leq 5$) regime, since the high-shot ($K \gg 100$) regime is very similar to the high-photon-count-per-shot regime that has already been studied in the ONN literature (e.g., in Ref.~\cite{wang2022optical}). 
An important practical point is that averaging for $K>1$ shots can be achieved by counting the clicks from each SPD. This was done in a clocked fashion by defining a discrete time window during which data was input and the SPDs were monitored for whether they clicked or not (the single-shot, $K=1$, case); to average for $K>1$ shots, we kept inputting the data for $K$ times the single-shot discrete time window and summed the number of click events for each SPD. We can think of $K$ as a discrete integration time, so averaging need not involve any data reloading or sophisticated control.

\begin{figure}[htp]
\includegraphics[width=.83\textwidth]{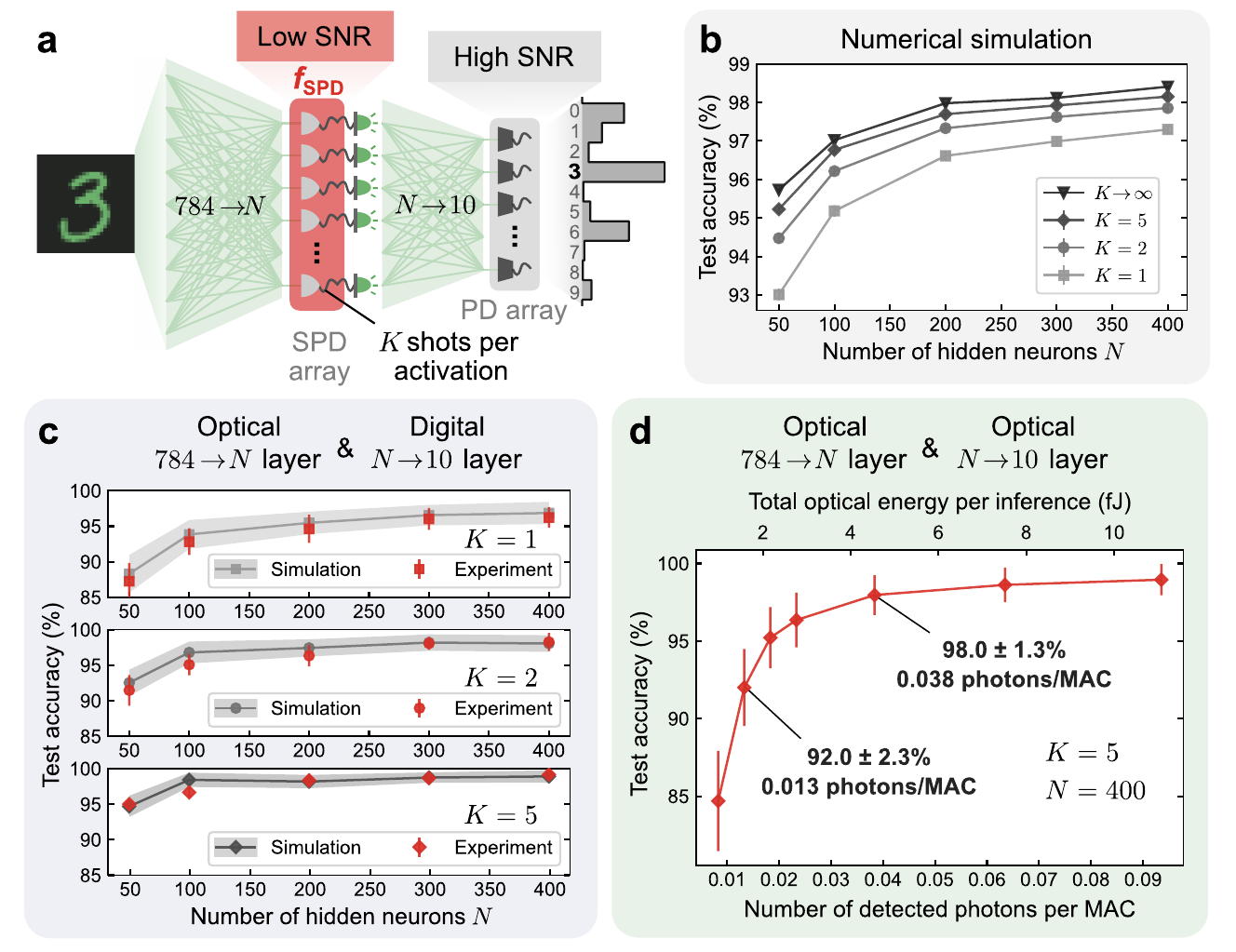}
\caption{\textbf{Performance of a single-photon-detection neural network (SPDNN) on MNIST handwritten-digit classification.}
\textbf{a}, An SPDNN realizing a multilayer perceptron (MLP) architecture of $N$ neurons in the hidden layer. The hidden layer ($784 \rightarrow N$) was computed using an incoherent optical matrix-vector-multiplier (MVM) followed by a single-photon-detector (SPD) array. Each SPD realized a stochastic activation function for a single hidden-layer neuron. During a single inference, the hidden layer was executed a small number of times ($1 \leq K \leq 5$), yielding averaged activation values. The output layer ($N \rightarrow 10$) was realized either optically---using an optical MVM and high photon budget to achieve high readout SNR, as in conventional ONNs, or with a digital electronic processor, yielding a result with full numerical precision.
\textbf{b}, Simulated test accuracy of MNIST handwritten-digit classification for models with different numbers of hidden neurons $N$ and shots per activation $K$.
Error bars, representing standard deviations from 100 repeated stochastic implementations with identical inputs and weights, are plotted but are too small to be easily visible.
\textbf{c},  Experimental evaluation of the SPDNN, with the output layer performed with full numerical precision on a digital computer. Results are presented for both $K=1$ (single-shot, i.e., no averaging; top), $K=2$ (middle), and $K=5$ (bottom) shots per activation. 
Mean values and standard deviations (shown as error bars) were calculated from repeated stochastic implementations using identical inputs and weights (see Supplementary Note 8 for details).
\textbf{d}, Experimental evaluation of the SPDNN, with both the hidden and the output layer executed using the optical experimental apparatus. The average number of detected photons used per inference in the hidden layer was kept fixed and the number used per inference in the output layer was varied.
Mean values and standard deviations (shown as error bars) were calculated from repeated stochastic implementations using identical inputs and weights (see Supplementary Note 9 for details).}
\label{fig:incoh}
\end{figure}

\section{MNIST handwritten-digit classification with a single-photon-detection multilayer perceptron}

We evaluated the performance---both in numerical simulations and in optical experiments---of SPDNNs on the MNIST handwritten-digit-classification benchmark task with a simple, $784\rightarrow N \rightarrow10$ multilayer perceptron (MLP) architecture (Figure \ref{fig:incoh}a).
The activation values in the hidden layer were computed by SPDs. The optical power was chosen so that the SNR of the SPD measurements was $\sim 1$, falling in the low-SNR regime (Figure \ref{fig:intro}b). The output layer was implemented either with full numerical precision on a digital electronic computer, or optically with an integration time set so that the measured signal comprised enough photons that a high SNR (Figure \ref{fig:intro}b) was achieved, as in conventional ONNs. Our use of a full-precision output layer is consistent with other works on binary neural networks \cite{hubara2016binarized,rastegari2016xnor,zhou2016dorefa}. In a shallow neural network, executing the output layer at high SNR substantially limits the overall energy efficiency gains from using small photon budgets in earlier layers, but in larger models, the relatively high energy cost of a high-SNR output layer is amortized. Nevertheless, as we will see, even with just a single-hidden-layer network, efficiency gains of \textgreater40$\times$ are possible by performing the hidden layer in the low-SNR regime. 

The models we report on in this section used non-negative weights in the hidden layers and real-valued weights in the output layers. This allows the hidden layers to be straightforwardly realized with optical MVMs using incoherent light.\footnote{A high-SNR layer with real-valued weights can be realized with an incoherent optical MVM if some digital-electronic postprocessing is allowed \cite{hayasaki1992optical,wang2022optical}---which is the approach we take for the optical output layer executions in our experiments. However, the postprocessing strategy doesn't directly apply in the low-SNR regime because readout becomes inseparable from the application of a nonlinear activation function, so we are constrained to non-negative weights and activations in the hidden layers.} In Section \ref{sec:coh} and Supplementary Note 2, we report on extensions to the case of real-valued weights in coherent optical processors.

\subsection{Simulation results}
\label{subsec:classif}

First, we digitally simulated the SPDNN models shown in Figure \ref{fig:incoh}a. We report the simulated test accuracies in Figure \ref{fig:incoh}b for the full test dataset of 10,000 images, as a function of the number of hidden neurons $N$ and the number of shots $K$ of binary SPD measurements integrated to compute each activation.

Due to the stochastic nature of the model, the classification output for a fixed input varies from run to run. We repeated inferences on fixed inputs from the test set 100 times; we report the mean and standard deviation of the test accuracy as data points and error bars, respectively. The standard deviations of the test accuracies are around $0.1\%$.

The accuracy achieved by the SPDNN is substantially higher than for linear models ($<93\%$ classification accuracy on MNIST \cite{lecun1998gradient}). This both shows that despite the hidden layer being stochastic, high-accuracy determistic classification is possible, and that the SPD activation function serves as a suitable nonlinearity in the neural network. 
The sizes of the models we simulated (in number of neurons $N$) are similar to those of traditional deterministic neural networks for MNIST classification \cite{hamerly2019large}, so the high accuracies achieved are not a simple consequence of averaging over many noisy neurons \cite{laydevant2021training}.

If we integrated an infinite number of SPD measurements for each activation ($K \rightarrow \infty$)---which is infeasible in experiment, but can be simulated---then the SPDNN output would become deterministic. The test accuracy achieved in this limit can be considered as an upper bound, as the classification accuracy improves monotonically with $K$.
Notably, even with just a single SPD measurement ($K=1$) for each activation, the mean test accuracy is around $97\%$.
The accuracy is substantially improved with just a few more shots of averaging, and approaches the deterministic upper bound when $K\gtrsim5$. The mean single-photon-detection probability, averaged over all neurons, is $\approx 0.5$, so the simulated number of detected photons per shot is very small: $\approx 0.5 N$. As we will quantify in the next section reporting the results of optical experiments, this means high accuracy can be achieved using much less optical energy than in conventional ONNs.

\subsection{Optical experimental results}

In our experimental demonstrations, we based our SPDNN on a free-space optical matrix-vector multiplier (MVM) that we had previously constructed for high-SNR experiments \cite{wang2022optical}, and replaced the detectors with SPDs so that we could operate it with ultra-low photon budgets (see Methods). The experiments we report were, in part, enabled by the availability of cameras comprising large arrays of pixels capable of detecting single photons with low noise \cite{dhimitri2022scientific}.
We encoded neuron values in the intensity of incoherent light; as a result, the weights and input vectors were constrained to be non-negative. However, this is not a fundamental feature of SPDNNs---in the next section (Section~\ref{sec:coh}), we present simulations of coherent implementations that lift this restriction.
A single-photon-detecting camera measured the photons transmitted through the optical MVM, producing the stochastic activations as electronic signals that were input to the following neural-network layer (see Methods and Supplementary Note 3 and 4).

In our first set of optical experiments, the hidden layer was realized optically and the output layer was realized \textit{in silico} (Figure \ref{fig:incoh}c): the output of the SPD measurements after the optical MVM was passed through a linear classifier executed with full numerical precision on a digital electronic computer. We tested using $K=1$ (no averaging), $K=2$, and $K=5$ shots of averaging the stochastic binary activations in the hidden layer. In Figure~3c, we can see that the experimental results agree well with results from simulations that additionally modeled the imperfections in our experimental setup (see Methods, Supplementary~Note~7), in contrast to the simulation results shown in Figure 3b, which did not account for these imperfections. The test accuracies were calculated using 100 test images, with inference for each image repeated 30 times to enable the computation of error bars. The hidden layer (the one computed optically in these experiments) used approximately 0.0008 detected photons per MAC ($K=1$), which is $\geq 6$ orders of magnitude lower than is typical in ONN implementations \cite{feldmann2021parallel,zhou2021large,sludds2022delocalized,bernstein2023single} and $\geq 3$ orders of magnitude lower than the lowest photons-per-MAC numbers reported to date \cite{wang2022optical,sludds2022delocalized}.

We then performed experiments in which both the hidden layer and the output layer were computed optically (Figure \ref{fig:incoh}d). In these experiments, we implemented a neural network with 400 hidden neurons and used 5 shots per inference ($N=400$, $K=5$). 
To execute the linear operations with signed weights (matrix elements) on our incoherent setup, we separated the weights into those that were positive and those that were negative. We first performed the matrix-vector multiplication with the negative weights only, then performed the matrix-vector multiplication with the positive weights only, and then obtained the final output by subtracting the results of the first from the second matrix-vector multiplication.
The total optical energy was varied by changing the number of photons used in the output layer; the number of photons used in the hidden layer was kept fixed.

The average value of the stochastic binary activations $a_i$ in the hidden layer was $\approx 0.522$. This corresponds to a total of $0.522 \times N \times K = 1044$ photons being detected in the hidden layer per inference. The total detected optical energy per inference comprises the sum of the detected optical energy in the hidden ($784\rightarrow 400$) layer and in the output ($400\rightarrow 10$) layer (see Methods, Supplementary Table 6 and Supplementary Note 9).

The results show that even though the output layer was operated in the high-SNR regime (Figure \ref{fig:intro}b), the full inference computation achieved high accuracy yet used only a few femtojoules of optical energy in total (equivalent to a few thousand photons). By dividing the optical energy by the number of MACs performed in a single inference, we can infer the per-MAC optical energy efficiency achieved: with an average detected optical energy per MAC of approximately 0.005 attojoules (or 0.014 attojoules) at a photon wavelength of 532 nm, equivalent to 0.013 photons (or 0.038 photons), the mean and standard deviation of test accuracy achieved $92.0\pm 2.3\%$ (or $98.0\pm 1.3\%$).
The optical energy per MAC is a metric of the energy efficiency of the system while the data is in the optical domain, quantifying a fundamental physical cost of computation, and allows for direct comparison with previous works that report this metric. In practical use of an SPDNN, the end-to-end system energy consumption per inference, covering both optical energy and energy used by the surrounding electronics, would be the preferred metric to optimize for. Engineering end-to-end system advantage is beyond the scope of this paper, but it is an interesting open question to what extent and in which situations reducing the optical energy from a few hundred photons per neuron to just 1 photon per neuron leads to a meaningful decrease in end-to-end system energy usage. Earlier analyses \cite{wang2022optical,anderson2024optical} in the context of ONNs with few-bit neuron values requiring analog-to-digital and digital-to-analog conversion between layers have concluded that once the optical energy is reduced to be on the order of 1 photon per MAC, the dominant energy cost with currently available electronics is from the electronic components. However, the use of binary stochastic neurons instead of few-bit deterministic neurons involves different hardware and could lead to different conclusions.

We now compare our results with what has been published previously. Our experiments with $N=50$ hidden neurons and $K=5$ shots of SPD measurements per activation (see Supplementary Figure 21) achieved a test accuracy of $90.6\%$ on MNIST handwritten-digit recognition while using only an average of $1390$ detected photons per inference (corresponding to $\sim$0.5~fJ of detected optical energy per inference). This represents a \textgreater40$\times$ reduction in the number of photons per inference to achieve \textgreater90\% accuracy on this task versus the previous state-of-the-art \cite{wang2022optical,sludds2022delocalized}.

While these numbers are already very low, most photons per inference in our experiments were used in the high-SNR output layer, so there is potential for further reduction with a more optimized experimental setup (see Supplementary Note 9). Furthermore, as we discuss in the next section, as the model size is increased, the fraction of optical energy used by a single high-SNR output layer becomes negligible.

\begin{figure}[htp]
\includegraphics [width=.98\textwidth]{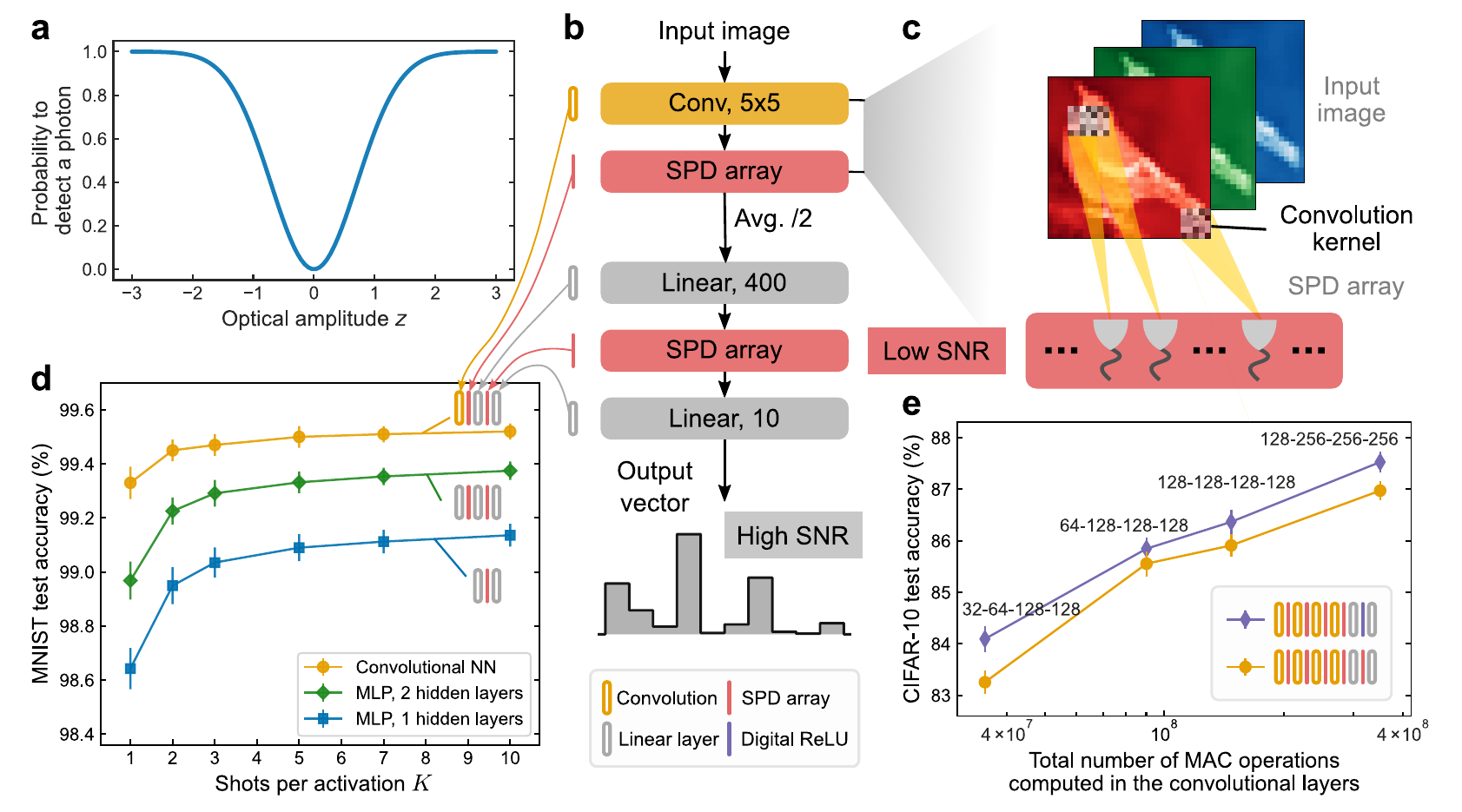}
\caption{\textbf{Simulation study predicting the performance of proposed \textit{coherent} single-photon-detection neural networks (SPDNNs).} 
\textbf{a}, The probability of detecting a photon as a function of the input light amplitude in a coherent SPDNN. Real-valued numbers are encoded in coherent light with either 0 phase (positive numbers) or $\pi$ phase (negative numbers). Measurement by a single-photon detector (SPD) results in the probabilistic detection of a photon that is proportional to the square of the encoded value $z$, in comparison to intensity encodings with incoherent light.
\textbf{b}, Structure of a convolutional SPDNN with a kernel size of $5\times5$. Single-shot SPD measurements ($K=1$) are performed after each layer (by an SPD array), except for the output layer. Average $2\times2$ pooling is applied after each convolutional operation. 
A digital rectified linear unit (ReLU) \cite{agarap2018deep} activation function can also be used in the linear layer as an alternative.
\textbf{c}, Schematic of a convolutional layer with SPD activations. 
\textbf{d}, Simulated test accuracy of coherent SPDNNs with varying architecture performing MNIST handwritten-digit classification. The multilayer perceptron (MLP) models had 400 neurons in each hidden layer.
The convolutional model consisted of a convolutional layer with 16 output channels, followed by two linear layers with an SPD activation inbetween.
\textbf{e}, Simulated test accuracy of coherent SPDNNs with varying architecture performing CIFAR-10 image classification. The models have four convolutional layers, each followed by SPD activation functions. The two linear layers can either be implemented in full-precision with a ReLU activation function (in purple) or using the SPD activation function. The number of output channels for each convolutional layer is indicated above the corresponding data point.
}
\label{fig:coh}
\end{figure}

\section{Simulation study of possible future deeper, coherent single-photon-detection neural networks}
\label{sec:coh}

We have successfully experimentally demonstrated a two-layer SPDNN, but can SPDNNs be used to implement deeper and more sophisticated models? One of the limitations of our experimental apparatus was that it used an intensity encoding with incoherent light and as a result could natively only perform operations with non-negative numbers. In this section we will show that SPDNNs capable of implementing signed numbers can be used to realize multilayer models (with up to 6 layers), including models with more sophisticated architectures than multilayer perceptrons---such as models with convolutional layers.

ONNs based on coherent light can naturally encode sign information in the phase of the light and have been realized in many different physical platforms \cite{shen2017deep,chang2018hybrid,lin2018all,spall2020fully,miscuglio2020massively,feldmann2021parallel,xu202111}. We propose---and study in simulation---SPDNNs using coherent light. Neuron values are encoded in optical amplitudes that are constrained to have phases that are either $0$ (positive values) or $\pi$ (negative values). With this encoding, detection by an SPD---which measures intensity and is hence insensitive to phase---results in a stochastic nonlinear activation function that is symmetric about zero (Figure \ref{fig:coh}a; see Methods). Alternative detection schemes could be employed that would modify the activation function, but we have focused on demonstrating the capabilities of this straightforward case, avoiding introducing additional experimental complexity.

We performed two sets of simulation experiments: one on coherent SPDNNs trained to perform MNIST handwritten-digit classification, and one on coherent SPDNNs trained to performed CIFAR-10 image classification. Figures \ref{fig:coh}d shows the architectures tested and simulation results for the MNIST benchmark (see Methods, Supplementary Note 2B). The accuracy achieved by MLPs with either one or two hidden layers was higher than that of the single-hidden-layer MLP simulated for the incoherent case (Figure \ref{fig:incoh}b), and an architecture with a single convolutional layer followed by two linear layers achieved \textgreater99\% accuracy even in the single-shot ($K=1$) regime.

Figure \ref{fig:coh}e shows the results of simulating variants of a 6-layer convolutional SPDNN (comprising 4 convolutional layers and 2 fully connected, linear layers) on CIFAR-10 image classification. All these simulation results were obtained in the single-shot ($K=1$) regime. The number of channels in each convolution layer was varied, which affects the total number of MACs used to perform an inference. We observed that the test accuracy increased with the size of the SPDNN, with accuracies approaching those of conventional convolutional neural networks of comparable size \cite{lee2016generalizing}, as well as of binarized convolutional neural networks \cite{hubara2016binarized,esser2016convolutional,qin2020binary}. 
In the models we simulated that only used SPD as the activation function (i.e., the ones in which there are no `Digital ReLU' blocks), the high-SNR linear output layer had only 4000 MAC operations, so the number of MACs in the high-SNR layer comprises less than 0.01\% of the total MACs performed during an inference. The models we simulated are thus sufficiently large that the total optical energy cost would be dominated by the (low-SNR) layers prior to the (high-SNR) output layer. Equivalently, the optical energy cost per MAC would be predominantly determined by the cost of the low-SNR layers.
These simulation results illustrate the ability of SPDNNs to scale to larger and deeper models, enabling them to perform more challenging tasks. The symmetric stochastic activation function that is realized by SPD of coherently encoded real values yields good accuracies on both MNIST and CIFAR-10 benchmarks and is straightforward to implement experimentally.

\section{Discussion}
\label{sec:discussion}  

Our research is an example of realizing a neural network using a stochastic physical system. Beyond optics, our work is related and complementary to recent investigations in electronic, spintronic, and quantum neuromorphic computing \cite{torrejon2017neuromorphic,grollier2020neuromorphic,cai2020power,harabi2023memristor,markovi2020physics,islam2023hardware,markovic2020quantum,cerezo2022challenges,roques2023biasing}, including in training physical systems to perform neural-network inference \cite{prezioso2015training,romera2018vowel,mitarai2018quantum,hughes2019wave,chen2020classification,cramer2022surrogate,wright2022deep,ross2023multilayer}. 

Neural networks are well-suited for execution on analog devices due to their inherent tolerance to low-precision operations \cite{wetzstein2020inference}. While previous works (e.g., Refs.~\cite{hamerly2019large,tait2022quantifying,anderson2024optical}) have studied energy consumption in analog optical systems, they have focused on the regime where the physical system effectively performs low-precision arithmetic in a way that is quite similar to in digital devices. Our research has instead investigated the regime where the analog system is operating at such a low signal-to-noise ratio (SNR) that it is no longer effective to treat the system simply as a low-precision approximation of a perfect arithmetic engine. Instead, we treat the system as fundamentally stochastic and train neural networks for it by modeling the stochastic physical processes that the system undergoes. Our results show that this \textit{physics-aware stochastic training} can enable optical neural networks (ONNs) to be operated at substantially lower SNR than what has previously been demonstrated, and in turn have much lower optical energy consumption.

Noise is a fundamental feature of analog machines and ultimately limits the energy efficiency in computing with any analog physical system. It has long been realized that stochasticity is not always detrimental: not only does it not necessarily prevent accurate computation, but can in some cases even enable fundamentally new and more efficient algorithms or types of computation. Our work shows that using a quantum physical model of a particular hardware's inherent stochastic response at the software level can enable surprisingly large gains in energy efficiency.  

While there are many reasons computer science has traditionally favored the abstraction of hardware from software, our work is part of a broad trend, spanning many different physical platforms \cite{berggren2020roadmap,christensen20222022,finocchio2023roadmap}, in which researchers engineer computations in a physics-aware manner. 
By short-circuiting the abstraction hierarchy---in our case, going from a physics-aware software description of a stochastic neural network directly to a physical optical realization of the constituent operations---it is possible to achieve orders-of-magnitude improvements in energy efficiency \cite{wetzstein2020inference,anderson2024optical} versus conventional CMOS computing. \textit{Physics-aware software}, in which software directly incorporates knowledge of the physics of the underlying computing hardware---such as in the \textit{physics-aware stochastic training} we used in this work---is understudied compared to purely software-level or hardware-level innovations (i.e., ``at the top'' or ``at the bottom'' of the hierarchy \cite{leiserson2020there}). It is thus ripe for exploration: within the domain of neural networks, there are a multitude of emerging physical platforms that could be more fully harnessed if the physical devices were not forced to conform to the standard abstractions in modern computer architecture \cite{wright2022deep}. Beyond neural-network accelerators, communities such as computational imaging \cite{kellman2020physics} have embraced the opportunity to improve system performance through co-optimizing hardware and software in a physics-aware manner. We believe there is an opportunity to make gains in even more areas and applications of computing technology by collapsing abstractions and implementing physics-aware software with physical hardware that could be orders of magnitude faster or more energy efficient than current digital CMOS approaches but that doesn't admit a clean, digital, deterministic abstraction.

\section*{Data and code availability}

All the simulation and experimental data presented in the paper, demonstration data for data gathering, as well as training data for the SPDNN models, are available at \url{https://doi.org/10.5281/zenodo.8188270}.
An expandable demonstration code to train SPDNNs as well as other stochastic physical systems is available at \url{https://github.com/mcmahon-lab/Single-Photon-Detection-Neural-Networks}. 

\section*{Acknowledgements}

We wish to thank NTT Research for their financial and technical support (S.-Y.M., P.L.M., T.W. and L.G.W.). Portions of this work were supported by the National Science Foundation (award no. CCF-1918549; J.L., P.L.M. and T.W.), a Kavli Institute at Cornell instrumentation grant (P.L.M. and T.W.), and a David and Lucile Packard Foundation Fellowship (P.L.M.). P.L.M. acknowledges membership of the CIFAR Quantum Information Science Program as an Azrieli Global Scholar. T.W. acknowledges partial support from an Eric and Wendy Schmidt AI in Science Postdoctoral Fellowship. We acknowledge valuable discussions with M.~Anderson, F.\,J.~Chen, R.~Hamerly, T.~Onodera, S.~Prabhu, M.\,M.~Sohoni and R.~Yanagimoto. We also acknowledge Z.~Eslami, V.~Kremenetski, F.~Presutti, C.~Wan and F.~Wu for helpful suggestions regarding the manuscript.

\section*{Author Contributions}

S.-Y.M., L.G.W., T.W., and P.L.M. conceived the project. S.-Y.M. and T.W. designed the experiments and built the experimental setup. S.-Y.M. and J.L. performed the neural-network training. S.-Y.M. performed the experiments, the data analysis, and the numerical simulations. All authors contributed to preparing the manuscript. T.W., L.G.W. and P.L.M. supervised the project.

\bibliographystyle{mcmahonlab}
\bibliography{references}

\newpage
\setcounter{equation}{0}

\section*{Methods}
\label{methods}

\subsection*{Stochastic optical neural networks using single-photon detection as the activation function}
In the single-photon-detection neural networks (SPDNNs), the activation function is directly determined by the stochastic physical process of single-photon detection (SPD). 
Each SPD measurement produces a binary output of either 0 or 1, with probabilities determined by the incident light intensity. Consequently, each SPD neuron activation, which corresponds to an SPD measurement in experiments, is considered as a binary stochastic process \cite{neal1990learning,lee2017energy,liu2018stochastic}.

Following the Poisson distribution, the probability of an SPD detecting a photon click is given by $P_{\text{SPD}}(\lambda)=1-e^{-\lambda}$ when exposed to an incident intensity of $\lambda$ photons per detection. Note that these photon statistics may vary based on the state of light (e.g. squeezed light), but here we only consider Poissonian light.
Therefore, the SPD process can be viewed as a Bernoulli sampling of that probability, expressed as $f_{\text{SPD}}(z) = \textbf{1}_{t<P_{\text{SPD}}(\lambda(z))}$, where $t$ is a uniform random variable $t\sim U[0,1]$ and $\textbf{1}_x$ is the indicator function that evaluates to 1 if $x$ is true. This derivation leads to Equation \ref{eq:f_spd} in the main text.
In our approach, the pre-activation value $z$ is considered as the direct output from an optical matrix-vector multiplier (MVM) that encodes the information of a dot product result. For the $i$th pre-activation value in layer $l$, denoted as $z^{(l)}_i$, the expression is given by:
\begin{equation}
z^{(l)}_i=\sum_{j=1}^{N_{l-1}} w^{(l)}_{ij}\cdot a^{(l-1)}_{j},
\end{equation}
where $N_{l-1}$ is the number of neurons in layer $l-1$, $w^{(l)}_{ij}$ is the weight between the $i$th neuron in layer $l$ and the $j$th neuron in layer $l-1$, $a^{(l-1)}_j$ is the activation of the $j$th neuron in layer $l$. The intensity $\lambda(z)$ is a function of $z$ that depends on the detection scheme employed in the optical MVM.
In optical setups using incoherent light, the information is directly encoded in the intensity, resulting in $\lambda=z$. 
If coherent light were used in a setup where 0 and $\pi$ phases represent the sign of the amplitude, the intensity would be determined by squaring the real-number amplitude if directly measured, resulting in $\lambda=z^2$. 
While more sophisticated detection schemes can be designed to modify the function of $\lambda(z)$, we focused on the simplest cases to illustrate the versatility of SPDNNs.
 
During the inference of a trained model, in order to regulate the level of uncertainty inherent in stochastic neural networks, we can opt to conduct multiple shots of SPD measurements during a single forward propagation. In the case of a $K$-shot inference, each SPD measurement is repeated $K$ times, with the neuron's final activation value $a^{[K]}$ being derived from the average of these $K$ independent stochastic binary values. Consequently, for a single shot, $a^{[1]}=a\in\{0,1\}$; for $K$ shots, $a^{[K]}=\frac{1}{K}\sum^K_{k=1} a_k\in \{0,1/K,2/K,\ldots,1\}$. By utilizing this method, we can mitigate the model's stochasticity, enhancing the precision of output values. Ideally, with an infinite number of shots ($K\rightarrow\infty$), the activation $a^{[\infty]}$ would equate to the expected value without any stochasticity, that is, $a^{[\infty]}=\mathbb{E}[a]=P_{\text{SPD}}(\lambda(z))$. The detailed process of inference of SPDNNs is described in Algorithm 2 in Supplementary Note 1A.

The training of our stochastic neuron models takes inspiration from recent developments in training stochastic neural networks. We have created an effective estimator for training our SPDNNs while accounting for the stochastic activation determined by the physical SPD process.
To train our SPDNNs, we initially adopted the idea of the ``straight-through estimator'' (STE) \cite{bengio2013estimating,yin2019understanding}, which enables us to bypass the stochasticity and discretization during neural network training. However, directly applying STE to bypass the entire SPD process led to subpar training performance. To address this, we adopted a more nuanced approach by breaking down the activation function and treating different parts differently.
The SPD process can be conceptually divided into two parts: the deterministic probability function $P_{\text{SPD}}$ and the stochasticity introduced by the Bernoulli sampling. For a Bernoulli distribution, the expectation value is equal to the probability, making $P_{\text{SPD}}$ the expectation of the activation. Instead of applying the ``straight-through'' method to the entire process, we chose to bypass only the Bernoulli sampling process. At the same time, we incorporate the gradients induced by the probability function, aligning them with the expectation values of the random variable. In this way, we obtained an unbiased estimator \cite{williams1992simple} for gradient estimation, thereby enhancing the training of our SPDNNs.

In the backward propagation of the $l$th layer, the gradients of the pre-activation $z^{(l)}$ can be computed as (the gradient with respect to any parameter $x$ is defined as $g_x=\partial C/\partial x$ where $C$ is the cost function): 
\begin{equation}
    g_{z^{(l)}} 
    = \frac{\partial a^{(l)}}{\partial \lambda^{(l)}}\circ \frac{\partial\lambda^{(l)}}{\partial z^{(l)}} \circ  g_{a^{(l)}}
    = P_{\text{SPD}}'(\lambda^{(l)})\circ \frac{\partial\lambda^{(l)}}{\partial z^{(l)}} \circ  g_{a^{(l)}},
\end{equation}
where $a^{(l)}=f_{\text{SPD}}(z^{(l)}) = \textbf{1}_{t<P_{\text{SPD}}(\lambda(z^{(l)}))}$ and the gradients $g_{a^{(l)}}$ are calculated from the next layer (previous layer in the backward propagation).
Using this equation, we can evaluate the gradients of the weights $W^{(l)}$ as 
    $g_{W^{(l)}} = g_{z^{(l)}}^\top a^{(l-1)}$,
where $a^{(l-1)}$ are the activation values from the previous layer. 
By employing this approach, SPDNNs can be effectively trained using gradient-based algorithms (such as SGD \cite{bottou2012stochastic} or AdamW \cite{loshchilov2017decoupled}), regardless of the stochastic nature of the neuron activations.

For detailed training procedures, please refer to Algorithms 1 and 3 in Supplementary Notes 1A and 1B, respectively.

\subsection*{Simulation of incoherent SPDNNs for deterministic classification tasks}
The benchmark MNIST (Modified National Institute of Standards and Technology database) \cite{de2015comparison} handwritten digit dataset consists of 60,000 training images and 10,000 testing images. Each image is a grayscale image with $28 \times 28 = 784$ pixels. To adhere to the non-negative encoding required by incoherent light, the input images are normalized so that pixel values range from 0 to 1.

To assess the performance of the SPD activation function, we investigated the training of the MLP-SPDNN models with the structure of $784\xrightarrow{W^{(1)}} N \xrightarrow{W^{(2)}} 10$, where $N$ represents the number of neurons in the hidden layer, $W^{(1)}$ ($W^{(2)}$) represents the weight matrices of the hidden (output) layer. The SPD activation function is applied to the $N$ hidden neurons, and the resulting activations are passed to the output layer to generate output vectors (Figure \ref{fig:incoh}a).
To simplify the experimental implementation, biases within the linear operations were disabled, as the precise control of adding or subtracting a few photons poses significant experimental challenges. We have observed that this omission has minimal impact on the model's performance.

In addition, after each weight update, we clamped the elements of $W^{(1)}$ in the positive range in order to comply with the constraint of non-negative weights of an incoherent optical setup. Because SPD is not required at the output layer, the constraints on the last layer operation are less stringent. Although our simulations indicate that the final performance is only marginally affected by whether the elements in the last layer are also restricted to be non-negative, we found that utilizing real-valued weights in the output layer provided increased robustness against noise and errors during optical implementation. As a result, we chose to use real-valued weights in $W^{(2)}$.

During the training process, we employed the LogSoftmax function on the output vectors and used cross-entropy loss to formulate the loss function. Gradients were estimated using the unbiased estimator described in the previous section and Algorithm 1.

For model optimization, we found that utilizing the SGD optimizer with small learning rates yields better accuracy compared to other optimizers such as AdamW, albeit at the cost of slower optimization speed. Despite the longer total training time, the SGD optimizer leads to a better optimized model. The models were trained with a batch size of 128, a learning rate of 0.001 for the hidden layer and 0.01 for the output layer, over 10,000 epochs to achieve optimized parameters. To prevent gradient vanishing in the plateau of the probability function $P_{\text{SPD}}$, pre-activations were clamped at $\lambda_{\text{max}}=3$ photons. 

It should be noted that due to the inherent stochasticity of the neural networks, each forward propagation generates varying output values even with identical weights and inputs. 
However, we only used one forward propagation in each step. This approach effectively utilized the inherent stochasticity in each forward propagation as an additional source of random search for the optimizer. Given the small learning rate and the significant noise in the model, the number of epochs exceeded what is typically required for conventional neural network training processes.
The training is performed on a GPU (Tesla V100-PCIE-32GB) and takes approximately eight hours for each model.

We trained incoherent SPDNNs with a varying number of hidden neurons $N$ ranging from 10 to 400. The test accuracy of the models improved as the number of hidden neurons increased (see Supplementary Note 1B for more details). During inference, we adjusted the number of shots per SPD activation $K$ to tune the SNR of the activations within the models.

For each model configuration with $N$ hidden neurons and $K$ shots of SPD readouts per activation, we repeated the inference process 100 times to observe the distribution of stochastic output accuracies. Each repetition of inference on the test set, which comprises 10,000 images, yielded a different test accuracy. The mean values and standard deviations of these 100 repetitions of test accuracy are plotted in Figure \ref{fig:incoh}b (see Supplementary Table 1 for more details). It was observed that increasing either $N$ or $K$ led to higher mean values of test accuracy and reduced standard deviations.

\subsection*{Experimental implementation of SPDNNs}
The optical matrix-vector multiplier setup utilized in this work is based on the design presented in \cite{wang2022optical}. The setup comprises an array of light sources, a zoom lens imaging system, an light intensity modulator, and a photon-counting camera. For encoding input vectors, we employed an organic light-emitting diode (OLED) display from a commercial smartphone (Google Pixel 2016 version). The OLED display features a $1920 \times 1080$ pixel array, with individually controllable intensity for each pixel. In our experiment, only the green pixels of the display were used, arranged in a square lattice with a pixel pitch of \SI{57.5}{\micro m}.
To perform intensity modulation as weight multiplication, we combined a reflective liquid-crystal spatial light modulator (SLM, P1920-500-1100-HDMI, Meadowlark Optics) with a half-wave plate (HWP, WPH10ME-532, Thorlabs) and a polarizing beamsplitter (PBS, CCM1-PBS251, Thorlabs). The SLM has a pixel array of dimensions $1920 \times 1152$, with individually controllable transmission for each pixel measuring $9.2\times9.2$ \SI{}{\micro m}. The OLED display was imaged onto the SLM panel using a zoom lens system (Resolv4K, Navitar).
The intensity-modulated light field reflected from the SLM underwent further de-magnification and was focused onto the detector using a telescope formed by the rear adapter of the zoom lens (1-81102, Navitar) and an objective lens (XLFLUOR4x/340, Olympus). 

We decompose a matrix-vector multiplication in a batch of vector-vector dot products that are computed optically, either by spatial multiplexing (parallel processing) or temporal multiplexing (sequential processing). To ensure a more accurate experimental implementation, we chose to perform the vector-vector dot products in sequence in most of the data collection. 
For the computation of an optical vector-vector dot product,  the value of each element in either vector is encoded in the intensity of the light emitted by a pixel on the OLED and the transmission of an SLM pixel. 
The imaging system aligned each pixel on the OLED display with its corresponding pixel on the SLM, where element-wise multiplication occurred via intensity modulation. The modulated light intensity from pixels in the same vector was then focused on the detector to sum up the element-wise multiplication values, yielding the vector-vector dot product result. 
Since the light is incoherent, only non-negative values can be allowed in both of the vectors.
For more details for the incoherent optical MVM, please refer to Supplementary Note 3. The calibration of the vector-vector dot products on the optical MVM is detailed in Supplementary Note 5.

In this experiment, we used a scientific CMOS camera (Hamamatsu ORCA-Quest qCMOS Camera C15550-20UP) \cite{dhimitri2022scientific} to measure both conventional light intensity measurement and SPD. 
This camera, with $4096 \times 2304$ effective pixels of $4.6\times4.6$ \SI{}{\micro m} each, can perform SPD with ultra-low readout noise in its photon counting mode.
This scientific CMOS camera is capable of carrying out the SPD process with ultra-low readout noise. 
When utilized as an SPD in the photon-counting mode, the camera exhibits an effective photon detection efficiency of $68\%$ and a dark count rate of approximately 0.01 photoelectrons per second per pixel (Supplementary Note 4). 
We typically operate with an exposure time in the millisecond range for a single shot of SPD readout.
For conventional intensity measurement that integrates higher optical energy for the output layer implementation, we chose another operation mode that used it as a common CMOS camera.

Further details on validating the stochastic SPD activation function measured on this camera are available in Supplementary Note 6. We also adapted our SPDNNs training methods to conform to the real-world constraints of our setup, ensuring successful experimental implementation (see Supplementary Note 7).
 
First, we conducted the implementation of the hidden layers and collect the SPD activations experimentally by the photon-counting camera as an SPD array. Then we performed the output layer operations digitally on a computer. 
This aims to verify the fidelity of collecting SPD activations from experimental setups. 
Supplementary Figure 16 provides a visual representation of the distribution of some of the output vectors.
For the experiments with 1 shot per activation ($K=1$), we collected 30 camera frames from the setup for each fixed input images and weight matrix, which are regarded as 30 independent repetitions of inference. They were then used to compute 30 different test accuracies by performing the output linear layer on a digital computer.
For the experiments with 2 shots per activation ($K=2$), we divided the 30 camera frames into 15 groups, with each group containing 2 frames. The average value of the 2 frames within each group serves as the activations, which are used to compute 15 test accuracies. For additional results and details, please refer to Supplementary Note 8.

Second, to achieve the complete optical implementation of the entire neural networks, we utilized our optical matrix-vector multiplier again to carry out the last layer operations. 
For example, we first focused on the data from the model with 400 hidden neurons and performed 5 shots per inference.
In this case, for the 30 binary SPD readouts obtained from 30 frames, we performed an averaging operation on every 5 frames, resulting in 6 independent repetitions of the inference. These activation values were then displayed on the SLM as the input for the last layer implementation. For the 5-shot activations, the possible values included 0, 0.2, 0.4, 0.6, 0.8, and 1.
When the linear operation were performed on a computer with full precision, the mean test accuracy was approximately $99.17\%$. 
To realize the linear operation with real-valued weight elements on our incoherent optical setup, we divided the weight elements into positive and negative parts. Subsequently, we projected these two parts of the weights onto the OLED display separately and performed them as two different operations. The final output value was obtained by subtracting the results of the negative weights from those of the positive weights.
This approach requires at least double the photon requirement for the output layer and offers room for optimization to achieve higher energy efficiency. Nevertheless, even with these non-optimized settings, we demonstrated a photon budget that is lower than any other ONN implementations known to us for the same task and accuracy.
For additional data and details, please refer to Supplementary Note 9.

\subsection*{Deeper SPDNNs operating with coherent light}
Optical processors with coherent light have the ability to preserve the phase information of light and have the potential to encode complex numbers using arbitrary phase values. In this work, we focused on coherent optical computing utilizing real-number operations. In this approach, positive and negative values are encoded in the light amplitudes corresponding to phase 0 and $\pi$, respectively.

As the intensity of light is the square of the amplitude, direct detection of the light amplitude, where the information is encoded, would involve an additional square operation, i.e., $\lambda(z)=|z|^2$.
This leads to a ``V-shape'' SPD probability function with respect to the pre-activation $z$, as depicted in Figure \ref{fig:coh}a. We chose to focus on the most straightforward detection case to avoid any additional changes to the experimental setup. Our objective is to demonstrate the adaptability and scalability of SPDNN models in practical optical implementations without the need for complex modifications to the existing setup.

\subsubsection*{Coherent SPDNNs for MNIST classification}

\paragraph*{MLP-SPDNNs}
Classifying MNIST using coherent MLP-SPDNNs was simulated utilizing similar configurations as with incoherent SPDNNs. The only difference was the inclusion of the coherent SPD activation function and the use of real-valued weights.
Contrary to the prior scenario with incoherent light, the input values and weights do not need to be non-negative.
The models were trained using the SGD optimizer \cite{bottou2012stochastic} with a learning rate of 0.01 for the hidden layers and 0.001 for the last linear layer, over a period of 10,000 epochs.

\paragraph*{Convolutional SPDNNs}
The convolutional SPDNN model used for MNIST digit classification, illustrated in Figure \ref{fig:coh}b, consists of a convolutional layer with 16 output channels, a kernel size of $5 \times 5$, a stride size of 1, and padding of 2. The SPD activation function was applied immediately after the convolutional layer, followed by average pooling of $2 \times 2$. The feature map of $14\times 14 \times 16=3136$ was then flattened into a vector of size 3136. After that, the convolutional layers were followed by a linear model of $3136 \rightarrow 400 \rightarrow 10$, with the SPD activation function applied at each of the 400 neurons in the first linear layer.

The detailed simulation results of the MNIST test accuracies of the coherent SPDNNs can be found in Supplementary Table 2 with varying model structures and shots per activation $K$. For additional information, see Supplementary Note 2B.

\subsubsection*{Coherent convolutional SPDNNs for Cifar-10 classification}
The CIFAR-10 dataset \cite{krizhevsky2009learning} has 60,000 images, each having $3\times 32\times 32$ pixels with 3 color channels, that belong to 10 different categories, representing airplanes, automobiles, birds, cats, deers, dogs, frogs, horses, ships and trucks. The dataset is partitioned into a training set with 50,000 images and a test set with 10,000 images. The pixel values have been normalized using the mean value of $(0.4914, 0.4822, 0.4465)$ and standard deviation of $(0.2471, 0.2435, 0.2616)$ for each of the color channels.
To boost performance, data augmentation techniques including random horizontal flips (50\% probability) and random $32\times 32$ crops (with 4-pixel padding) were implemented during training.

The convolutional SPDNN models for Cifar-10 classification have deeper structures. 
Same as the convolutional models trained for MNIST, the convolutional layers use a kernel size of $5 \times 5$, a stride size of 1 and padding of 2. Each convolutional layer is followed by the SPD activation function, average pooling of $2 \times 2$, as well as batch normalization.
After $N_\text{conv}$ convolutional layers ($N_\text{conv}=4$ in Figure \ref{fig:coh}e) with the number of output channels of the last one to be $N_\text{chan}^\text{last}$, the feature map of $(32/2^{N_\text{conv}})^2\times N_\text{chan}^\text{last}$ is flattened to a vector, followed by two linear layers of $(32/2^{N_\text{conv}})^2 N_\text{chan}^\text{last} \rightarrow 400 \rightarrow 10$.
In the first linear layer, either SPD or ReLU \cite{agarap2018deep} activation function were used for each of the 400 neurons, as depicted in Figure \ref{fig:coh}e.
We vary the number of convolutional layers and number of output channels of them to change the different model size (Figure \ref{fig:coh}e and Supplementary Figure 5). 
In these results, we only used a single shot of SPD measurement ($K=1$) to compute the SPD activations in the models, including the convolutional and linear layers.
For additional information, please refer to Supplementary Note 2C.

\end{document}


\clearpage
Supplementary Materials for 

\title{Quantum-limited stochastic optical neural networks operating at a few quanta per activation}

\author{Shi-Yuan~Ma}
\email{sm2725@cornell.edu}
\affiliation{School of Applied and Engineering Physics, Cornell University, Ithaca, NY 14853, USA}

\author{Tianyu~Wang}
\affiliation{School of Applied and Engineering Physics, Cornell University, Ithaca, NY 14853, USA}

\author{Jérémie~Laydevant}
\affiliation{School of Applied and Engineering Physics, Cornell University, Ithaca, NY 14853, USA}
\affiliation{USRA Research Institute for Advanced Computer Science, Mountain View, CA 94035, USA}

\author{Logan~G.~Wright} 
\email{Present address: Department of Applied Physics, Yale University, New Haven, Connecticut 06511, USA}
\affiliation{School of Applied and Engineering Physics, Cornell University, Ithaca, NY 14853, USA}
\affiliation{NTT Physics and Informatics Laboratories, NTT Research, Inc., Sunnyvale, CA 94085, USA}

\author{Peter~L.~McMahon}
\email{pmcmahon@cornell.edu}
\affiliation{School of Applied and Engineering Physics, Cornell University, Ithaca, NY 14853, USA}
\affiliation{Kavli Institute at Cornell for Nanoscale Science, Cornell University, Ithaca, NY 14853, USA}

\maketitle

\begin{spacing}{1.3}
\tableofcontents 
\end{spacing}

\clearpage
\newpage

\setcounter{page}{1}

 \part{Simulation results}
In this part, we introduce the details of simulation of the single-photon-detection neural networks (SPDNNs). Each neuron activation in an SPDNN, corresponding to a readout on a single-photon detector (SPD) in an experiment, is modeled as a binary stochastic process \cite{neal1990learning,lee2017energy,liu2018stochastic}. For each SPD measurement, the single-shot output is either 0 or 1, with probabilities determined by the incident optical energy. The exact form of the activation function is defined by the actual physical process of single-photon detection. For an incident beam with optical energy of $\lambda$ photons per detection, due to Poissonian photon statistics, the probability for an SPD to detect a click is $P_{\text{SPD}}(\lambda)=1-e^{-\lambda}$, as shown in Figure 2 in the main text. The detected binary results are used to compute the activation values. However, due to the stochasticity and discretization in the single-photon-detection process, estimating the gradients in the loss function is challenging, and conventional backpropagation algorithms fail to train these models. 

Training stochastic neuron models has been investigated for many years. One of the major families of algorithms dependent on such neurons is the Boltzmann machine \cite{hinton1984boltzmann, ackley1985learning}. REINFORCE algorithms (RA) \cite{williams1992simple} update the weights in the direction of the gradients of expected reinforcement without explicitly computing them. These algorithms have been investigated and applied in different tasks to train stochastic neural networks
effectively \cite{weaver2013optimal,fiete2006gradient}. In \cite{bengio2013estimating}, many methods of estimating gradients through stochastic neurons are studied. They found that the fastest training in their experiments was achieved by the ``straight-through estimator" (STE), which was previously introduced in Hinton's course, lecture 15b \cite{hinton2012}.  
In our simulation of SPDNNs, we were inspired by both methods and found an estimator that trained our SPDNNs effectively, with the activation induced by the physical single-photon detection process. When using STE in a binary neural network, the binarization process, either deterministic or stochastic, is regarded as identity function during back propagation. However, if we directly use the STE to go ``straight through'' the entire SPD process, the training performance is not very good. This is because the STE is a biased estimator of the gradients \cite{bengio2013estimating,yin2019understanding}, meaning the expectation value of the estimator is not the same as the true expectation value of the real random variable. The biased estimation of gradients harms training accuracy over a large number of epochs. Then we looked for an unbiased estimator inspired by RA \cite{williams1992simple,bengio2013estimating}. We can conceptually break the single-photon detection process into two parts, a determinisic probability function $P_{\text{SPD}}$, and the Bernoulli sampling that introduces the stochasticity. For a Bernoulli distribution, the expectation value is the probability of 1 itself, so $P_{\text{SPD}}$ is also the expectation value of the activation. Instead of going ``straight through'' the entire SPD process, we only skip the Bernoulli sampling process to avoid the uncertainty in backpropagation and include the gradients induced by the probability function to meet the expectation values of the random variable. 

To enhance training effectiveness in certain cases, we introduced a slope variable, $\eta$, which modifies the intensity value within the SPD activation function: $P_{\text{SPD}}^\eta(\lambda)=P_{\text{SPD}}(\eta\lambda)$. The incorporation of a technique called ``slope annealing'' \cite{chung2016hierarchical} allows controlled alteration of the gradients of the activation function, leading to more efficient navigation of the model's parameter space.
Additionally, we impose an upper limit on the intensity by clamping it to a maximum value $\lambda_{\text{max}}$. This prevents the occurrence of vanishing gradients resulting from excessively large values and the plateauing probability function. 
Both the application of the slope variable and intensity clamping can be utilized exclusively during the training phase.
In optical implementation, the annealing factor can be absorbed in the mapping from the trained weights to the controlled parameters on the experimental setup. 

The details of the backpropagation and training process are shown in Algorithms 1 and 3, with the exact activation functions of incoherent and coherent optical setups, respectively. In the following sections, we introduce the two SPDNN setups in detail and test their performance with different tasks and architectures. 

\section{Modeling and training}
\subsection{SPDNNs with incoherent optical setups}
\label{subsec:incoh_model}
When an optical neural network (ONN) operates with incoherent light, the values of the vector elements are encoded as the intensity of light. The encoded values are non-negative, and the operations are performed by modulating the intensity of light. Thus, for an optical matrix-vector multiplier (optical MVM) operating with incoherent light, the values in an output vector $z$ are readily the intensity to be measured by the detector, i.e., $\lambda=z$. The probability of having the SPD measurement of 1 is then $P_{\text{SPD}}(\lambda(z))=P_{\text{SPD}}(z)$. This probability $P_{\text{SPD}}$ is determined by the pre-activation value $z$. Thus, the SPD activation is a Bernoulli sampling of the probability $P_{\text{SPD}}$, $f^{\text{Incoh}}_{\text{SPD}}(z)=\textbf{1}_{t<P_{\text{SPD}}(z)}$, where $t$ is a uniform random variable $t\sim U[0,1]$ and $\textbf{1}_x$ is the indicator function on the true value of $x$, i.e. 
\begin{equation}
  f^{\text{Incoh}}_{\text{SPD}}(z) =
  \begin{cases}
    1 & \text{with probability $p=P_{\text{SPD}}(z)$,} \\
    0 & \text{with probability $1-p$,}
  \end{cases}
\end{equation}
where the probability function $P_{\text{SPD}}(z)=1-e^{-z}$. The activation in the forward propagation is calculated as $a=f^{\text{Incoh}}_{\text{SPD}}(z)$.

\vspace{20pt}
In the backward propagation, the stochastic Bernoulli sampling process is regarded as an identity function (``straight through''), so that the gradients can propagate through the whole model as it were deterministic. In this way, the SPDNNs with stochastic binary neurons can be efficiently trained. The training procedure of incoherent SPDNNs is detailed in Algorithm 1. 
With $L$ layers in the neural network, the SPD activation function is applied after every layer except the output layer. 
In the $l$th layer ($l\neq L$), $z^{(l)} = a^{(l-1)} W^{(l)}$ is the direct output of the optical MVM that encodes the information of the dot product results. In an incoherent optical setup, the output values are directly encoded in light intensity,
$\lambda^{(l)} = z^{(l)}$. 
In the training process, we clamp the intensity to a maximum value $\lambda_{\text{max}}$ to avoid vanishing gradients with large values. Meanwhile, the clamped intensity vector $\lambda^{(l)}$ is multiplied by the slope variable $\eta$ to compute the probability of detecting a click on the SPDs according to $P_{\text{SPD}}$: $p^{(l)}= P_{\text{SPD}}(\eta\lambda^{(l)})$. Then the activation values are the Bernoulli sampling of the computed probabilities: $a^{(l)}=\textbf{1}_{t<p^{(l)}}$, which are sent to the next layer in the forward propagation. 
In backward propagation, our gradient estimator assumes the gradients of the stochastic sampling process are 1: $\partial a^{(l)}/\partial p^{(l)}=1$. Thus, during the backward pass of the $l$th layer, given the gradient with respect to $a^{(l)}$, $g_{a^{(l)}}=\partial C/\partial a^{(l)}$, calculated from the next layer (previous layer in backward propagation), the gradient with respect to the pre-activation $z^{(l)}$ is calculated as follows:
\begin{equation}
    g_{z^{(l)}} = \frac{\partial a^{(l)}}{\partial z^{(l)}} \circ  g_{a^{(l)}} \\
    = \frac{\partial a^{(l)}}{\partial p^{(l)}}\circ \frac{\partial p^{(l)}}{\partial \lambda^{(l)}}\circ\frac{\partial\lambda^{(l)}}{\partial z^{(l)}} \circ  g_{a^{(l)}}\\
    = 1 \circ P_{\text{SPD}}'(\lambda^{(l)})\circ 1 \circ  g_{a^{(l)}}\\
    = P_{\text{SPD}}'(z^{(l)}) \circ g_{a^{(l)}},
\end{equation}
so the gradients with respect to the weights $W^{(l)}$ are  $g_{W^{(l)}} = g_{z^{(l)}}^\top a^{(l-1)}$. In this way, the gradients can be efficiently calculated to optimize the weights using a gradient-based optimizer with a learning rate. 

{\centering
\begin{minipage}{.96\linewidth}
  \begin{algorithm}[H]
    \caption{Physics-aware stochastic training of an SPDNN with an incoherent optical setup. $N_\text{batch}$ is the batch size, $N_l$ denotes the number of neurons in layer $l$ and $N_0$ is the input size. $C$ is the loss function. $L$ is the number of layers. $P_{\text{SPD}}(\lambda)$ is the function of the probability to detect a click on the single-photon detector (SPD) with respect to the incident light intensity $\lambda$ (in number of photons). Sample() is a probabilistic sampling of the probability. In SPDNNs, it refers to Bernoulli sampling, Sample($p$) has a probability of $p$ to be 1 and a probability of $1-p$ to be 0 (i.e. $\text{Sample}(p)\equiv \textbf{1}_{t<p}$, $t\sim U[0,1]$). In experiments, an SPD detection intrinsically consists both of the process, the SPD activation function $f_{\text{SPD}}(\lambda) =\textbf{1}_{t<P_{\text{SPD}}(\lambda)}$, $t\sim U[0,1]$. The $\lambda$ is equivalent to the pre-activation $z$ in an incoherent setup. Output() determines the function applied to the pre-activation right before the final output, such as Softmax or LogSoftmax. Update() specifies how to update the parameters given the calculated gradients, using optimizers such as SGD \cite{bottou2012stochastic}, Adam \cite{kingma2014adam} or AdamW \cite{loshchilov2017decoupled}.}
    \begin{algorithmic}[1]
    
        \Require A batch of inputs $a^{(0)}$ ($N_\text{batch}\times N_0$) with corresponding targets $y$, current weights $W^{(l)}$ ($N_{l}\times N_{l-1}$, $l \in \{0,1,\ldots,L\}$), current slope variable $\eta$, slope annealing factor $\theta$, current learning rate $\alpha$, decay coefficient $\gamma$ and the clamped photon number $\lambda_{\text{max}}$.
        \Ensure Updated weights $W^{(l)}$ ($l \in \{0,1,\ldots,L\}$), slope $\eta$ and learning rate $\alpha$ . 
        \State \underline{\textbf{\textit{I. Forward pass}}}
        \For{$l = 1$ to $L$}
        \State $z^{(l)} \gets a^{(l-1)} W^{(l)\top}$  
        \Comment{Linear operation to compute the pre-activation values}
        \State $\lambda^{(l)} \gets z^{(l)}$      \Comment{For incoherent light, intensity is directly modulated}
        \State $\lambda^{(l)} \gets \text{min}(\lambda^{(l)},\lambda_{\text{max}})$      \Comment{Clamp the maximum intensity}
        \If{$l<L$}
        \State $p^{(l)} \gets P_{\text{SPD}}(\eta\cdot\lambda^{(l)})$    \Comment{The probability of detecting a click, with the slope $\eta$ applied}
        \State $a^{(l)} \gets \text{Sample}(p^{(l)})$      \Comment{SPD activation values}
        \EndIf
        \EndFor
        \State $a^{(L)} \gets \text{Output}(\lambda^{(L)})$ 
        \Comment{Final output function} 
        \State \underline{\textbf{\textit{II. Backward pass}}}
        \State Compute $g_{a^{(L)}}=\frac{\partial C}{\partial a^{(L)}}$ knowing $a^{(L)}$ and $y$.
        \State $g_{z^{(L)}} \gets \frac{\partial a^{(L)}}{\partial z^{(L)}} \circ g_{a^{(L)}}$
        \For{$l = L$ to 1}
        \If{$l<L$}
        \State $g_{p^{(l)}} \gets g_{a^{(l)}}$      \Comment{``Straight-through'' here, skip the Bernoulli process}
        \State $g_{z^{(l)}} \gets P_{\text{SPD}}'(z^{(l)}) \circ g_{p^{(l)}}$ 
        \Comment{$\frac{\partial p^{(l)}}{\partial z^{(l)}} =\frac{\partial p^{(l)}}{\partial \lambda^{(l)}}\circ\frac{\partial\lambda^{(l)}}{\partial z^{(l)}}=P_{\text{SPD}}'(\lambda^{(l)})\circ 1 =P_{\text{SPD}}'(z^{(l)})$}
        \EndIf
        \State $g_{a^{(l-1)}} \gets g_{z^{(l)}} W^{(l)}$ 
        \State $g_{W^{(l)}} \gets g_{z^{(l)}}^\top a^{(l-1)}$   \Comment{The gradients for $W^{(l)}$}
        \EndFor
        \State \underline{\textbf{\textit{III. Parameter update}}}
        \For{$l = 1$ to $L$}
        \State $W^{(l)} \gets \text{Update}(W^{(l)}, g_{W^{(l)}},\alpha)$   \Comment{Update the weights}
        \State $W^{(l)} \gets \text{max}(W^{(l)},0)$       \Comment{Clip the weights to be non-negative for the incoherent setup}
        \EndFor
        \State $\eta \gets \theta\eta$            \Comment{Update the slope}
        \State $\alpha \gets \gamma\alpha$            \Comment{Update the learning rate}
        \end{algorithmic}
  \end{algorithm}
\end{minipage}
\par
}
\vspace{20pt}

Note that for an incoherent optical setup, the elements in the weights (realized by intensity modulations) are also non-negative, so the updated weights need to be clamped to non-negative values after each optimization step. After each optimization step, the slope variable is updated by multiplying by a factor $\theta$, as the ``slope annealing'' trick \cite{chung2016hierarchical} to improve the training performance when necessary.

During the inference of a trained model, the forward pass of test inputs is similar to the training process, except that the maximum clamping $\lambda_{\text{max}}$ is not applied. Additionally, to control the level of uncertainty in the stochastic neural networks, we can choose to use multiple shots of SPD measurements during each inference. In a ``$K$-shot'' inference, we use $K$ shots of binary SPD readouts, and the final activation value of the neuron, denoted as $a^{[K]}$,
is the average of the $K$ independent stochastic binary values. 
This process essentially involves integrating a few more photons using the SPD, as is usually done for conventional ONN implementation \cite{sludds2022delocalized}.

For a single shot of SPD measurement per activation, $a^{[1]} = a \in \{0,1\}$, while for $K$ shots, $a^{[K]} = \frac{1}{K}\sum_{k=1}^{K} a_k \in \{0,1/K,2/K,\ldots,1\}$. This approach reduces the uncertainty in the models, resulting in more precise output values. In the ideal case where an infinite number of shots are integrated ($K\rightarrow\infty$), the activation $a^{[\infty]}$ would converge to the expectation value without stochasticity, denoted as $a^{[\infty]} = \mathbb{E}[a] = P_{\text{SPD}}(z)$.
As we will see in \ref{subsec:incoh_mnist}, the SPDNN models have higher test accuracy as the shots per activation $K$ increases.
The detailed inference procedure is explained in Algorithm 2.

{\centering
\begin{minipage}{.96\linewidth}
  \begin{algorithm}[H]
    \caption{Inference of an SPDNN with an incoherent optical setup. $L$ is the number of layers. $N_\text{batch}$ is the batch size, $N_l$ denotes the number of neurons in layer $l$ and $N_0$ is the input size. $K$ is the number of shots used in one inference. Sampling(), experiment. The predictions of an inference is based on the label of the output node with the maximum output value.}
    \begin{algorithmic}[1]
        \Require A batch of test inputs $a^{(0)}$ ($N_\text{batch}\times N_0$) and trained weights $W^{(l)}$ ($N_{l}\times N_{l-1}$, $l \in \{0,1,\ldots,L\}$), slope annealing factor $\eta$.
        \Ensure The output $a^{(L)}$. 
        \For{$l = 1$ to $L$}
        \State $z^{(l)} \gets a^{(l-1)} W^{(l)\top}$    \Comment{Linear operation to compute the pre-activation}
        \State $\lambda^{(l)} \gets z^{(l)}$        \Comment{For incoherent light, intensity is directly modulated}
        \If{$l<L$}                                  \Comment{SPD activation process}
        \State $p^{(l)} \gets P_{\text{SPD}}(\eta\cdot\lambda^{(l)})$    \Comment{The probability of detecting a click, with the slope $\eta$ applied}
        \For{$k = 1$ to $K$}                                      \Comment{$K$ shots in one inference}
        \State $a^{(l),k} \gets \text{Sampling}(p^{(l)})$       \Comment{SPD output for each shot}
        \EndFor
        \State $a^{(l)} \gets \frac{1}{K}\sum_{k=1}^K a^{(l),k}$  
        \Comment{Average over all $K$ shots for the activation values}
        \EndIf
        \EndFor
        \State $a^{(L)} \gets \lambda^{(L)}$        \Comment{Use the output intensity directly in the inference}
    \end{algorithmic}
  \end{algorithm}
\end{minipage}
\par
}
\vspace{20pt}

\subsection{SPDNNs with coherent optical setups}

In coherent optical MVMs \cite{shen2017deep,miscuglio2020massively,bogaerts2020programmable,lin2018all,spall2020fully,feldmann2021parallel,xu202111}, the information is conveyed through both the amplitude and phase of light states. These multipliers have the potential to encode complex numbers using arbitrary phase, but in most applications, only phases of 0 and $\pi$ are used for positive and negative real-number values, to align with conventional machine learning models. Our work focuses on real-valued coherent optical MVMs. 
Now that the information is encoded in the amplitude and phase instead of the intensity, the photon detection process involves measuring the square modulus of the complex number, which adds an extra square function to the pre-activation values.
Thus, the coherent SPD activation function is $f^{\text{Coh}}_{\text{SPD}}(z)=\textbf{1}_{t<P_{\text{SPD}}(z^2)}$, where $t$ is a uniform random variable $t\sim U[0,1]$ and $\textbf{1}_x$ is the indicator function on the true value of $x$, i.e. 
\begin{equation}
  f^{\text{Coh}}_{\text{SPD}}(z) =
  \begin{cases}
    1 & \text{with probability $p=P_{\text{SPD}}(z^2)$,} \\
    0 & \text{with probability $1-p$,}
  \end{cases}
\end{equation}
where $P_{\text{SPD}}(z^2)=1-e^{-z^2}$. The activation in the forward propagation is calculated by $a=f^{\text{Coh}}_{\text{SPD}}(z)$. The expectation of the coherent SPD activation is $\mathbb{E}[f_{\text{SPD}}^{\text{Coh}}]=P_{\text{SPD}}(z^2)$.

{\centering
\begin{minipage}{.96\linewidth}
  \begin{algorithm}[H]
    \caption{Physics-aware stochastic training of an SPDNN with coherent light. $N_\text{batch}$ is the batch size, $N_l$ denotes the number of neurons in layer $l$ and $N_0$ is the input size. $C$ is the loss function. $L$ is the number of layers. $P_{\text{SPD}}(\lambda)$ is the function of the probability to detect a click on the single-photon detector (SPD) with respect to the incident light intensity $\lambda$ (in number of photons).  Sample() is a probabilistic sampling of the probability. In SPDNNs, it refers to Bernoulli sampling, Sample($p$) has a probability of $p$ to be 1 and a probability of $1-p$ to be 0 (i.e. $\text{Sample}(p)\equiv \textbf{1}_{t<p}$, $t\sim U[0,1]$). For a coherent setup, $\lambda=z^2$ where $z$ is the pre-activation, output of a matrix-vector multiplier. Output() determines the function applied to the pre-activation right before the final output, such as Softmax or LogSoftmax. Update() specifies how to update the parameters given the calculated gradients, using optimizers such as SGD \cite{bottou2012stochastic}, Adam \cite{kingma2014adam} or AdamW \cite{loshchilov2017decoupled}. 
    }
    \begin{algorithmic}[1]
        \Require A batch of inputs $a^{(0)}$ ($N_\text{batch}\times N_0$) with corresponding targets $y$, current weights $W^{(l)}$ ($N_{l}\times N_{l-1}$, $l \in \{0,1,\ldots,L\}$), current slope variable $\eta$, slope annealing factor $\theta$, current learning rate $\alpha$, decay coefficient $\gamma$ and the clamped photon number $\lambda_{\text{max}}$.
        \Ensure Updated weights $W^{(l)}$ ($l \in \{0,1,\ldots,L\}$), slope $\eta$ and learning rate $\alpha$. 
        \State \underline{\textbf{\textit{I. Forward pass}}}
        \For{$l = 1$ to $L$}
        \State $z^{(l)} \gets a^{(l-1)} W^{(l)\top}$    \Comment{Linear operation to compute the pre-activation}
        \State $\lambda^{(l)} \gets (z^{(l)})^2$      \Comment{For coherent light, intensity is the square of the amplitude}\State $\lambda^{(l)} \gets \text{min}(\lambda^{(l)},\lambda_{\text{max}})$      \Comment{Clamp the maximum intensity}
        \If{$l<L$}
        \State $p^{(l)} \gets P_{\text{SPD}}(\lambda^{(l)})$    \Comment{The probability of detecting a click on the SPDs}
        \State $a^{(l)} \gets \text{Sample}(p^{(l)})$          \Comment{SPD output for each shot}
        \EndIf
        \EndFor
        \State $a^{(L)} \gets \text{Output}(\lambda^{(L)})$    \Comment{Final output function}   
        \State \underline{\textbf{\textit{II. Backward pass}}}
        \State Compute $g_{a^{(L)}}=\frac{\partial C}{\partial a^{(L)}}$ knowing $a^{(L)}$ and $y$.
        \State $g_{z^{(L)}} \gets \frac{\partial a^{(L)}}{\partial z^{(L)}} \circ g_{a^{(L)}}$ 
        \For{$l = L$ to 1}
        \If{$l<L$}
        \State $g_{p^{(l)}} \gets g_{a^{(l)}}$      \Comment{``Straight-through'' here, skip the Bernoulli process}
        \State $g_{z^{(l)}} \gets 2 z^{(l)}\circ P_{\text{SPD}}'\left((z^{(l)})^2\right) \circ g_{p^{(l)}}$ 
        \Comment{$\frac{\partial p^{(l)}}{\partial z^{(l)}} =\frac{\partial p^{(l)}}{\partial \lambda^{(l)}}\circ\frac{\partial\lambda^{(l)}}{\partial z^{(l)}}=2 z^{(l)}\circ P_{\text{SPD}}'\left((z^{(l)})^2\right)$}
        \EndIf
        \State $g_{a^{(l-1)}} \gets g_{z^{(l)}} W^{(l)}$    
        \State $g_{W^{(l)}} \gets g_{z^{(l)}}^\top a^{(l-1)}$    \Comment{The gradients with respect to $W^{(l)}$}
        \EndFor
        \State \underline{\textbf{\textit{III. Parameter update}}}
        \For{$l = 1$ to $L$}
        \State $W^{(l)} \gets \text{Update}(W^{(l)}, g_{W^{(l)}},\alpha)$   \Comment{Update the weights}   
        \EndFor
        \State $\eta \gets \theta\eta$            \Comment{Update the slope}
        \State $\alpha \gets \gamma\alpha$            \Comment{Update the learning rate}
        \end{algorithmic}
  \end{algorithm}
\end{minipage}
\par
}
\vspace{20pt}

The coherent activation function, depicted in Figure 4a of the main text, exhibits a distinct ``V'' shape due to the additional square operation, which is symmetric about the y axis. It could be problematic as an activation function \cite{ramachandran2017searching}. One possible solution is to modify the information encoding and detection scheme to alter the exact form of $\lambda(z)$ (e.g. \cite{spall2020fully}).
However, in this section, we have chosen to employ the most straightforward intensity-detection scenario, which does not necessitate modifications to conventional ONN implementation. Remarkably, despite its simplicity, this activation function delivers comparable performance and demonstrates impressive results. By adopting this approach, we alleviate experimental complexities while ensuring reliable inference in our SPDNN models.

\section{Classification performance}
\subsection{MNIST classification task}
\label{subsec:mnist}
To illustrate the capability of the SPD activation function, we first use it in a simple multi-layer perceptron (MLP) architecture and train the models on the benchmark MNIST classification task. The models have a structure of $784\rightarrow N \rightarrow 10$ with $N$ neurons in the hidden layer, as discussed in the main text. The MNIST dataset has 60,000 images for training and 10,000 images for testing. Each image is grayscale and has $28 \times28 = 784$ pixels. To meet the non-negative encoding of incoherent light, the input images are normalized to have pixel values ranging from 0 to 1. 

\subsubsection{SPDNNs with incoherent optical setups}
\label{subsec:incoh_mnist}

The models consist of two linear layers: the $784\rightarrow N$ hidden layer has the weight matrix $W^{(1)}$ with a shape of $N\times 784$, and the $N\rightarrow 10$ output layer has the weight matrix $W^{(2)}$ with a shape of
$10\times N$. The SPD activation function is applied to each hidden neuron after the linear operation of $W^{(1)}$ to compute the neuron activations; then the computed activation values are passed to the output layer to produce the output vectors. The elements in the first linear operation, $W^{(1)}$, are clamped to be non-negative to meet the requirement of an incoherent optical setup. In general, real-valued weights can be realized with an incoherent optical MVM if some digital-electronic post-processing is available. In our case, where the activations are measured by SPDs, the activation function is directly applied in the single-photon detection process, which makes digital post-processing impossible. Similarly, biases of the linear operations are also disabled. If we want to get away with digital post-processing by applying a bias term directly to the optical intensity, at the level of a few photons, the approach is also challenging in experiments. 
However, since the output layer is implemented using conventional optical computing with a higher signal-to-noise ratio (SNR), we can effectively implement real-valued weights of $W^{(2)}$.  In optical implementation, this would involve extra operations to map these values onto the incoherent setup. 

During the training process, we apply the LogSoftmax function to the output vectors and use the cross-entropy loss to construct the loss function. To avoid the issue of vanishing gradients, we clamp the pre-activation values at $\lambda_{\text{max}} = 3$ photons. 
It is important to note that due to the stochastic nature of the neural networks, each forward pass generates different output values, even with the same weights and inputs. However, we only use a single forward pass in each training epoch,
which has been shown to be the most efficient training approach. The stochasticity introduced in each forward propagation could add to the random search of the stochastic optimizer itself, helping with the training process.

We have found that using the SGD optimizer \cite{bottou2012stochastic} with small learning rates leads to better accuracy compared to other optimizers, such as AdamW \cite{loshchilov2017decoupled}. Although training with SGD takes longer overall, it helps us achieve a better-optimized model in the end. For our final results, we used a batch size of 128 and a learning rates of 0.001 for the hidden layer and 0.01 for the output layer in the SGD optimizer. We trained each SPDNN model for 10,000 epochs to obtain optimized parameters, and an even higher number of epochs may be needed to achieve better accuracy.
Given the small learning rate and the significant amount of noise in the model, the number of epochs required is much larger than what is typically seen in common neural network training processes. The training and test errors for an incoherent MLP SPDNN with a structure of $784 \rightarrow 400 \rightarrow 10$ are shown in Supplementary Figure \ref{suppfig:training}. The training process was performed on a GPU (Tesla V100-PCIE-32GB) and took approximately eight hours to complete.

\begin{figure}[htp]
\includegraphics [width=0.98\textwidth] {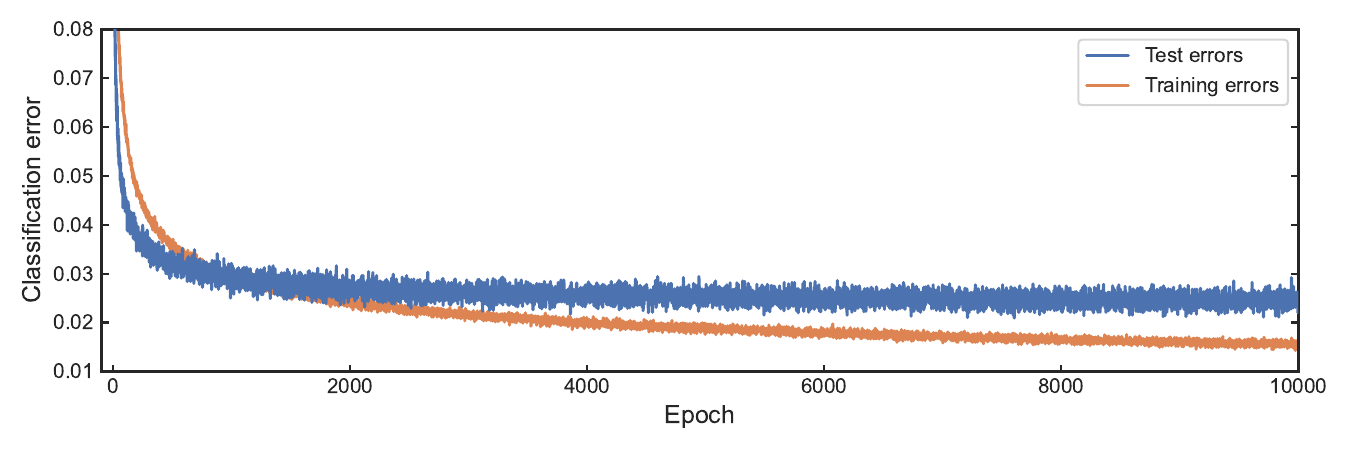}
\caption{\textbf{Training curves of an incoherent SPDNN model for MNIST classification.}
The plot illustrates the progression of test and training errors throughout the training process of an incoherent SPDNN model with an MLP architecture of $784\rightarrow 400\rightarrow 10$. The optimization is conducted using an SGD optimizer with learning rates of 0.001 for the hidden layer and 0.01 for the output layer. The final trained model is obtained at 10,000 epochs.}
\label{suppfig:training}
\end{figure}

\begin{figure}[htp]
\centering
\includegraphics [width=0.999\textwidth] {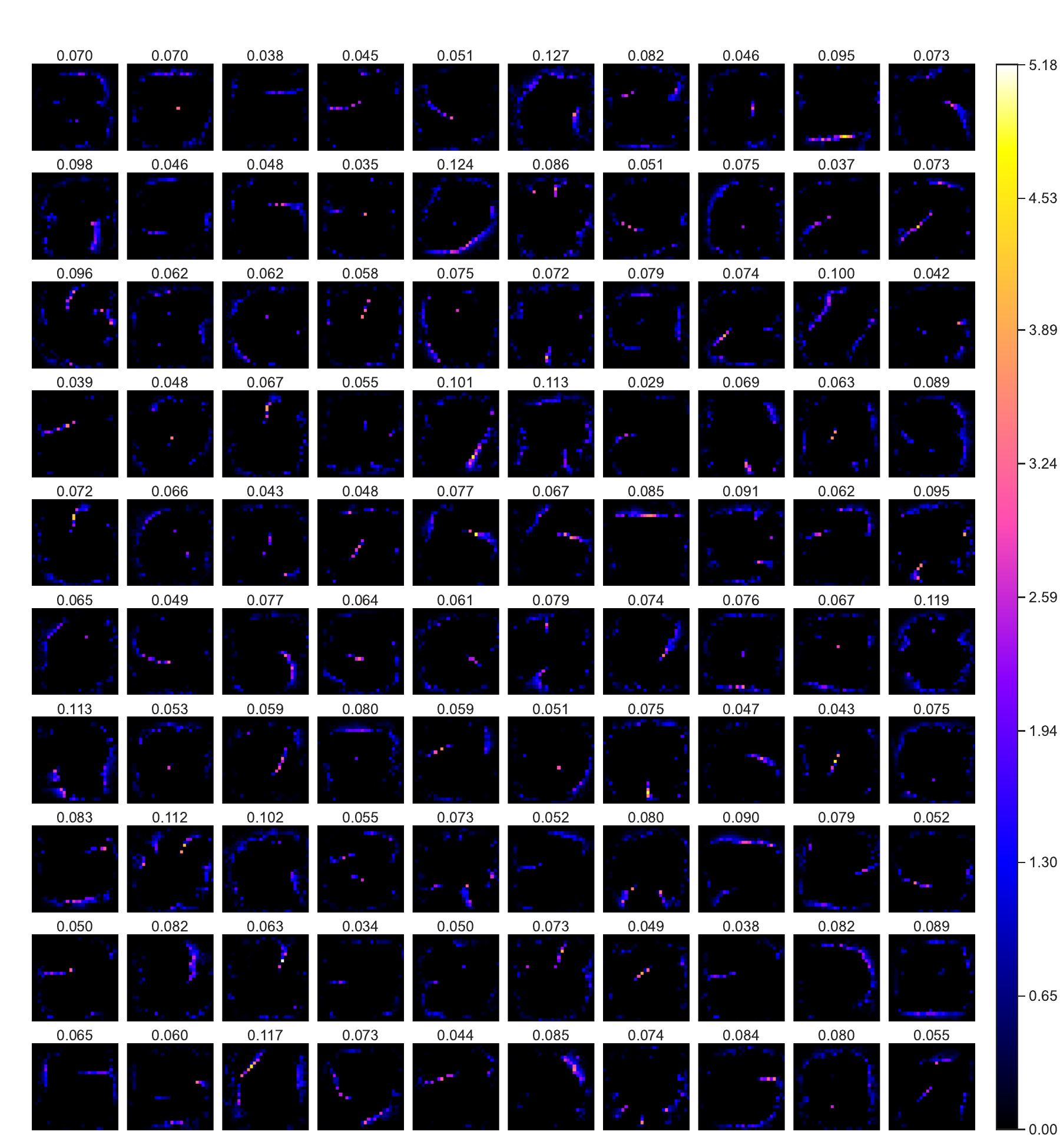}
\caption{\textbf{Visualization of weight elements in the first linear layer of an incoherent SPDNN.} The architecture of this model is $784\rightarrow 100\rightarrow 10$, and we display the weight matrix $W^{(1)}$ of the first layer (with dimensions $100\times 784$). Each block represents a row vector in $W^{(1)}$ containing 784 elements. These column vectors are rearranged to form a 2D block with dimensions $28\times28$, matching the original shape of the MNIST input images. The $100$ rows in $W^{(1)}$, corresponding to the $100$ hidden neurons in the neural network, are arranged in a $10\times10$ grid to be visualized. The average value of each block is indicated at the top, and the overall average value of the weight matrix is approximately $\sim 0.07$.}
\label{suppfig:incoh_w_1}
\end{figure}

The magnitude of the weight element values in the first linear layer, $W^{(1)}$, is influenced by the range of values in the input vectors, $a^{(0)}$, and the specific form of the SPD activation function, $f^{\text{Incoh}}_{\text{SPD}}(z)$. 
In the forward pass of an incoherent SPDNN, the pre-activation values, $z^{(1)}$, are computed as $z^{(1)} = a^{(0)} W^{(1)\top}$. The activation function, $f^{\text{Incoh}}_{\text{SPD}}(z)$, is defined as $\textbf{1}_{t<P_{\text{SPD}}(z)}$, where $t$ is a random variable uniformly distributed between 0 and 1, and $P_{\text{SPD}}(z)$ represents the probability of photon detection.
When the input vectors, $a^{(0)}$, are normalized to the range of 0 to 1, the weight elements in $W^{(1)}$ are optimized based on the specific form of $P_{\text{SPD}}$ because it depends on the exact value of pre-activations $z$. 
In our simulation of an incoherent SPDNN, the elements of $z^{(1)}$ are represented in terms of photon numbers, where the value 1 corresponds to 1 photon.
When $z \gtrsim 3$, $P_{\text{SPD}}$ reaches the plateau part of the probability function. Thus, we have to make sure the value of the pre-activation to be around 1 photon to ensure an effective forward pass.
When considering a uniform bright image where each element has the maximum value of 1, and with an input vector size of $28\times 28 = 784$, if we aim for an output value of approximately 1 photon, the average value of the weight elements in $W^{(1)}$ should be around $1/784 \approx 0.0013$.
The average pixel value in the MNIST dataset is approximately 0.13 (when each pixel value is normalized to the range of 0 to 1). Based on this, we can estimate that to achieve an output value of approximately 1 photon, the average weight element value should be around 0.01.
Taking into account that both the input images and weight matrices tend to be sparse, this estimation may be slightly lower than the actual scenario.
Supplementary Figure \ref{suppfig:incoh_w_1} illustrates the matrix elements of $W^{(1)}$ for a model with $N=100$ hidden neurons. The weight elements range from 0 to 5.18, with an average value of 0.07. Each block represents a row vector of size 784, rearranged in the form of $28\times 28$. The average value of $W^{(1)}$ may vary slightly in different network structures, ranging from 0.06 to 0.08.

During the inference of SPDNNs, the pixel values of the test images are normalized to the range of 0 to 1 as well.
This will be correspond to the dynamic range on the optical setup. We trained incoherent MLP-SPDNN models with varying numbers of hidden neurons, $N$, ranging from 10 to 400. As discussed in Section \ref{subsec:incoh_model}, we can adjust the number of SPD measurements per activation, denoted as $K$, to control the level of stochasticity in the models.

\begin{table}[htp] 
    \centering
    \begin{tabular}{|c||c|c|c|c|c|c|c|}
        \hline 
        Model & $K=1$  & $K=2$  & $K=3$  & $K=5$  & $K=7$  & $K=10$  & $K\rightarrow\infty$ \\
        \hhline{|========|}
        784--10--10 & $78.03 \pm 0.32$ & $83.18 \pm 0.26$ & $84.79 \pm 0.22$ & $86.13 \pm 0.17$ & $86.65 \pm 0.17$ & $87.08 \pm 0.16$ & $87.91 \pm 0.00$ \\
        784--20--10 & $86.74 \pm 0.24$ & $89.98 \pm 0.18$ & $90.96 \pm 0.15$ & $91.71 \pm 0.13$ & $92.00 \pm 0.13$ & $92.22 \pm 0.13$ & $92.66 \pm 0.00$ \\
        784--50--10 & $93.04 \pm 0.16$ & $94.49 \pm 0.15$ & $94.92 \pm 0.12$ & $95.24 \pm 0.11$ & $95.38 \pm 0.10$ & $95.47 \pm 0.09$ & $95.73 \pm 0.00$ \\
        784--100--10 & $95.20 \pm 0.16$ & $96.24 \pm 0.11$ & $96.53 \pm 0.10$ & $96.75 \pm 0.09$ & $96.85 \pm 0.07$ & $96.91 \pm 0.07$ & $97.02 \pm 0.00$ \\
        784--200--10 & $96.62 \pm 0.12$ & $97.33 \pm 0.10$ & $97.54 \pm 0.08$ & $97.70 \pm 0.08$ & $97.75 \pm 0.08$ & $97.80 \pm 0.06$ & $97.98 \pm 0.00$ \\
        784--300--10 & $97.00 \pm 0.12$ & $97.61 \pm 0.08$ & $97.80 \pm 0.08$ & $97.93 \pm 0.07$ & $97.97 \pm 0.06$ & $98.01 \pm 0.05$ & $98.12 \pm 0.00$ \\
        784--400--10 & $97.31 \pm 0.11$ & $97.85 \pm 0.10$ & $98.01 \pm 0.09$ & $98.15 \pm 0.06$ & $98.20 \pm 0.06$ & $98.27 \pm 0.05$ & $98.41 \pm 0.00$ \\
        \hline
    \end{tabular}
    \caption{\textbf{Test accuracy (\%) of incoherent MLP-SPDNNs on MNIST with varying hidden layer size $N$ and shots per activation $K$.} These models have an MLP structure of $784\rightarrow N\rightarrow 10$, where $N$ represents the number of hidden neurons. Each hidden neuron uses $K$ shots of binary SPD readouts to compute its activation value. The reported test accuracy values are obtained by calculating the mean value and standard deviation over 100 repetitions of inferences on the MNIST test set, which comprises 10,000 images.}
    \label{tab:MNIST_N_K}
\end{table}

The results of the MNIST test accuracy for different combinations of $N$ and $K$ are summarized in Supplementary Table \ref{tab:MNIST_N_K}. The values of $N$ include 10, 20, 50, 100, 200, 300, and 400, while $K$ takes on the values of 1, 2, 3, 5, 7, 10, and $\infty$. In the case of $K\rightarrow\infty$, we use the expectation of the activation values, $P_{\text{SPD}}$, as the activation function, which is equivalent to integrating an infinite number of shots per SPD detection. This serves as an upper bound that is approached as $K$ increases.
Due to the stochastic nature of SPDNNs, the output vectors vary across different repetitions of inference. To capture the overall behavior of the models, we repeated the full inference process 100 times for each structure with $N$ hidden neurons and $K$ shots per activation. This allows us to calculate the mean test accuracy and standard deviation, representing the distribution of test accuracies. Each independent repetition of inference uses the MNIST test dataset, consisting of 10,000 images. We observe that as either $N$ or $K$ increases, the mean test accuracy tends to improve while the standard deviation decreases.

\subsubsection{SPDNNs with coherent optical setups}
\label{subsec:coh_mnist}
The MNIST handwritten-digit classification task was performed using the same simulation configurations as with incoherent SPDNNs, but but using the coherent SPD activation function and real-number operations. Unlike the previous case, no clamping of the weights was necessary. The models were trained using the SGD optimizer with a learning rate of 0.01 for the hidden layers and 0.001 for the last linear layer, for a period of 10,000 epochs.
To evaluate the impact of model size, we trained models with both one and two hidden layers. The training curves of the model with the structure of $784\rightarrow 400\rightarrow 400 \rightarrow 10$ are shown in Supplementary Figure \ref{suppfig:training_coh}. The results for models with different structures and shots of SPD measurements per activation can be found in Supplementary Table \ref{tab:MNIST_N_K_coh}, and the weights of a model with the structure of $784\rightarrow 100\rightarrow 10$ are illustrated in Supplementary Figure \ref{suppfig:coh_w1}.

\begin{figure}[htp]
\includegraphics [width=0.98\textwidth] {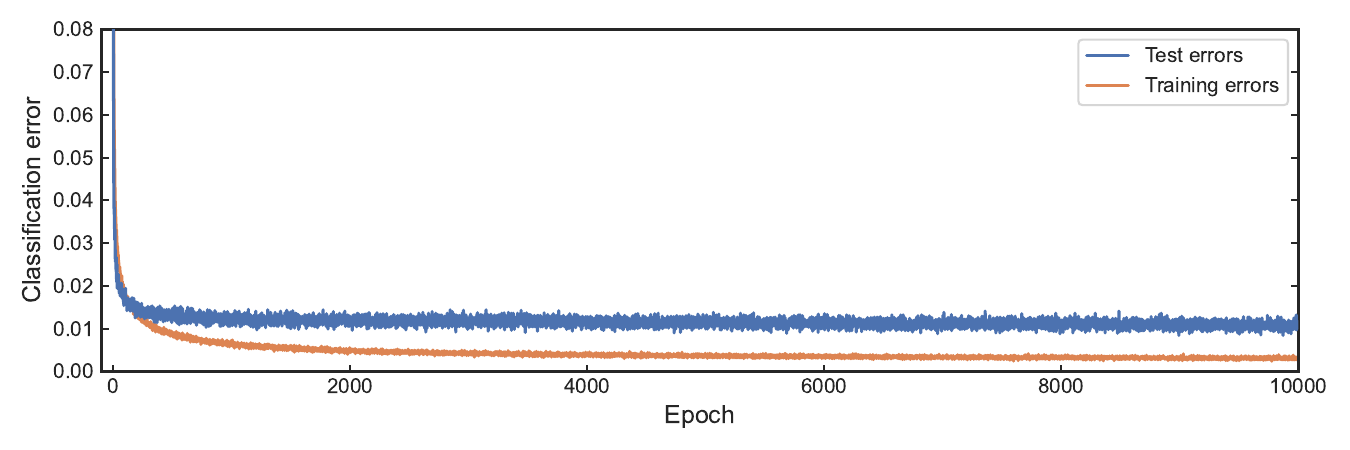}
\caption{\textbf{Training curves of a coherent SPDNN model on MNIST classification.} 
 The plot shows the evolution of the test and training errors during the training process of a coherent SPDNN model with an MLP architecture of $784\rightarrow 400\rightarrow 400 \rightarrow 10$ neurons. The optimization is performed using SGD with different learning rates for the hidden (0.001) and output (0.01) layers. The end result of the training is represented by the final values of the 10,000 epochs.}
\label{suppfig:training_coh}
\end{figure}

\begin{figure}[htp]
\includegraphics [width=0.999\textwidth] {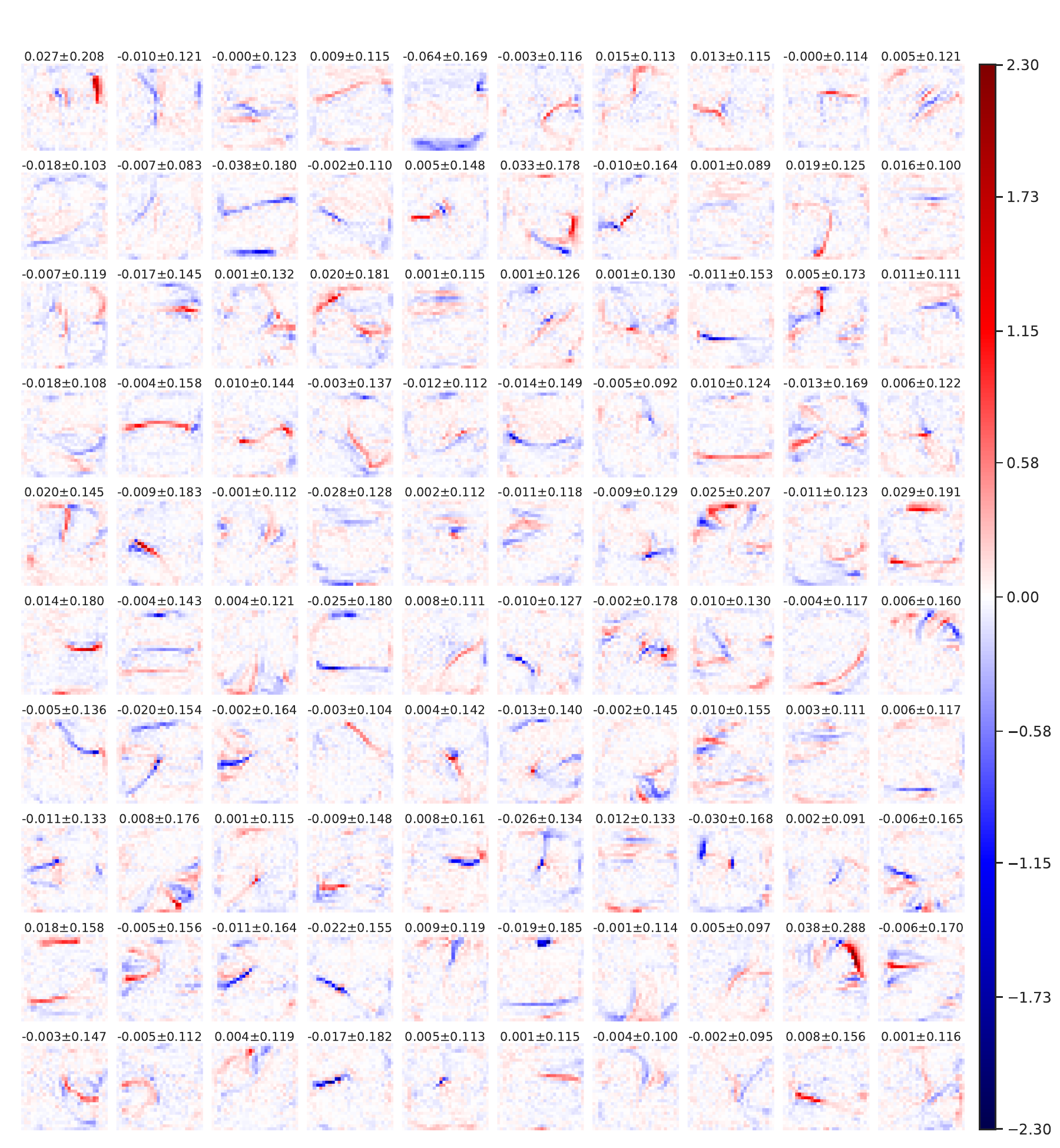}
\caption{\textbf{Visualization of weight elements in the first linear layer of a coherent SPDNN.} The architecture of this model is $784\rightarrow 100\rightarrow 10$, and we display the weight matrix $W^{(1)}$ of the first layer (with dimensions $100\times 784$). Each block represents a row vector in $W^{(1)}$ containing 784 elements. These column vectors are rearranged to form a 2D block with dimensions $28\times28$, matching the original shape of the MNIST input images. The $100$ rows in $W^{(1)}$, corresponding to the $100$ hidden neurons in the neural network, are arranged in a $10\times10$ grid to be visualized. The average value and standard deviation of the elements in each block are indicated at the top.}
\label{suppfig:coh_w1}
\end{figure}

\begin{table}[htbp] 
    \centering    
    \begin{tabular}{|c||c|c|c|c|c|c|c|}
        \hline 
        Model & $K=1$  & $K=2$  & $K=3$  & $K=5$  & $K=7$  & $K=10$  & $K\rightarrow\infty$ \\
        \hhline{|========|}
        784--10--10 & $80.21 \pm 0.27$ & $85.63 \pm 0.24$ & $87.49 \pm 0.19$ & $88.85 \pm 0.15$ & $89.42 \pm 0.15$ & $89.87 \pm 0.14$ & $90.70 \pm 0.00$ \\
        784--20--10 & $89.31 \pm 0.23$ & $92.43 \pm 0.17$ & $93.42 \pm 0.14$ & $94.10 \pm 0.13$ & $94.39 \pm 0.14$ & $94.60 \pm 0.11$ & $95.05 \pm 0.00$ \\
        784--25--10 & $91.51 \pm 0.20$ & $94.16 \pm 0.16$ & $94.89 \pm 0.14$ & $95.47 \pm 0.12$ & $95.70 \pm 0.10$ & $95.86 \pm 0.08$ & $96.10 \pm 0.00$ \\
        784--30--10 & $92.72 \pm 0.19$ & $94.92 \pm 0.13$ & $95.58 \pm 0.12$ & $96.05 \pm 0.10$ & $96.26 \pm 0.09$ & $96.41 \pm 0.08$ & $96.74 \pm 0.00$ \\
        784--50--10 & $95.50 \pm 0.15$ & $96.88 \pm 0.11$ & $97.26 \pm 0.09$ & $97.55 \pm 0.08$ & $97.67 \pm 0.07$ & $97.77 \pm 0.07$ & $97.93 \pm 0.00$ \\
        784--100--10 & $97.41 \pm 0.13$ & $98.16 \pm 0.08$ & $98.36 \pm 0.08$ & $98.52 \pm 0.07$ & $98.58 \pm 0.06$ & $98.61 \pm 0.05$ & $98.70 \pm 0.00$ \\
        784--200--10 & $98.34 \pm 0.10$ & $98.76 \pm 0.07$ & $98.88 \pm 0.07$ & $98.97 \pm 0.05$ & $99.00 \pm 0.05$ & $99.04 \pm 0.04$ & $99.12 \pm 0.00$ \\
        784--400--10 & $98.64 \pm 0.08$ & $98.95 \pm 0.06$ & $99.04 \pm 0.05$ & $99.09 \pm 0.05$ & $99.12 \pm 0.04$ & $99.14 \pm 0.03$ & $99.19 \pm 0.00$ \\
        784--400--400--10 & $98.95 \pm 0.08$ & $99.21 \pm 0.05$ & $99.29 \pm 0.04$ & $99.33 \pm 0.04$ & $99.35 \pm 0.04$ & $99.37 \pm 0.03$ & $99.40 \pm 0.00$ \\
        784--C16--400--10  & $99.33 \pm 0.06$  & $99.45 \pm 0.04$  & $99.47 \pm 0.04$  & $99.50 \pm 0.04$  & $99.51 \pm 0.03$  & $99.52 \pm 0.03$  & $99.54 \pm 0.00$ \\
        \hline
    \end{tabular}

    \caption{\textbf{Test accuracy (\%) of coherent SPDNN models on MNIST with different model structures and shots per activation $K$.} These models have an MLP structure with one or two hidden layers with the number of neurons denoted in the table, except for the last model, which has a convolutional layer denoted as ``C16'' with 16 output channels and followed by a $2\times2$ average pooling and the following linear layers. The mean accuracy and standard deviation are calculated based on 100 repetitions of inferences using the MNIST test set of 10,000 images.}
    \label{tab:MNIST_N_K_coh}
\end{table}

Furthermore, convolutional SPDNNs were also used for MNIST classification. The architecture included a convolutional layer with 16 output channels, a kernel size of $5\times 5$ and a stride of 1. An SPD activation function was immediately applied after each convolution layer, without batch normalization. Average pooling of $2\times 2$ was performed after each of the SPD activations. After the convolution layer, the total number of features was 3136, then the convolution layers were followed by a linear model of $3136\rightarrow 400\rightarrow10$, with the SPD activation function applied at each of the 400 hidden neurons as well. This structure is depicted in Figure 4b in the main text.
For optimization, we used an SGD optimizer with a learning rate of 0.01 for the entire model. The convolutional SPDNN model can be optimized easily without fine-tuning the parameters. After 200 epochs, the accuracy quickly reached $99.4\%$.

\vspace{40pt}
\begin{figure}[htp]
\includegraphics [width=0.67\textwidth] {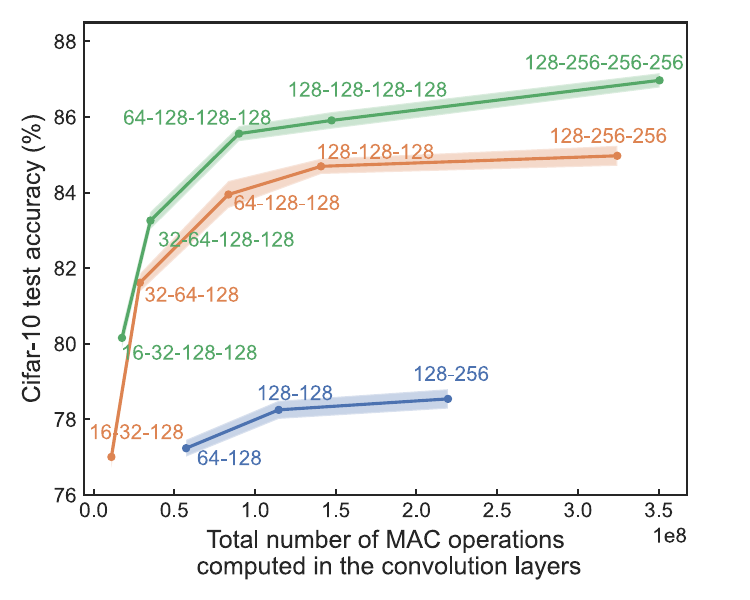}
\caption{\textbf{Test accuracy of convolutional SPDNN models on CIFAR-10 classification.} The plot shows the mean test accuracy (data points) and the corresponding standard deviation (shaded region), calculated by averaging the results of 100 repeated inference runs. The number of output channels in the convolutional layers is indicated near each data point, with different colors denoting different numbers of convolutional layers. The linear layers following the convolutional layers have a consistent structure of $400\rightarrow 10$, with SPD activation applied to the 400 neurons.}
\label{suppfig:cifar10_convs}
\end{figure}
\vspace{40pt}

\subsection{CIFAR-10 classification task}
The CIFAR-10 dataset \cite{krizhevsky2009learning} has 60,000 images, each having $3\times 32\times 32$ pixels with 3 color channels, that belong to 10 different categories, representing airplanes, automobiles, birds, cats, deer, dogs, frogs, horses, ships and trucks. The dataset is partitioned into a training set with 50,000 images and a test set with 10,000 images. The pixel values have been normalized using the mean value of $(0.4914, 0.4822, 0.4465)$ and standard deviation of $(0.2471, 0.2435, 0.2616)$ for each of the color channels.
To boost performance, data augmentation techniques including random horizontal flips (50\% probability) and random $32\times 32$ crops (with 4-pixel padding) were implemented during training.
We used the AdamW optimizer \cite{loshchilov2017decoupled} with a learning rate of 0.0005 and betas of $(0.99, 0.98)$. The models were trained for thousands of epochs.

The convolutional SPDNNs have a structure where the SPD activation function is applied after each convolution layer and before an average pooling of $2\times 2$. 
The final architecture consists of a series of convolution layers followed by a linear layer of 400 neurons, and a final layer of $400\rightarrow 10$ for the output.
Similar to the convolutional models trained for MNIST, the convolutional layers use a kernel size of $5 \times 5$, a stride size of 1 and padding of 2.
Batch normalization was used in the models after each convolutional layer.
Either SPD or ReLU activation function was applied to each of the 400 neurons in the first linear layer, as depicted in Figure 4e in the main text.

After $N_\text{conv}$ convolutional layers ($N_\text{conv}=2$, 3 or 4 in this case) with the number of output channels of the last one to be $N_\text{chan}^\text{last}$ (either 128 or 256 in this case), the feature map of $(32/2^{N_\text{conv}})^2\times N_\text{chan}^\text{last}$ is flattened to a vector, followed by two linear layers of $(32/2^{N_\text{conv}})^2 N_\text{chan}^\text{last} \rightarrow 400 \rightarrow 10$.
In addition to the results presented in Figure 4e in the main text, we experimented with more architectures ranging from 2 to 4 convolution layers, and the results are displayed in Supplementary Figure \ref{suppfig:cifar10_convs}. 
In these models, only the SPD activation function was used.
The x-axis of the plot represents the number of multiply--accumulate (MAC) operations in the convolutional layers. The layout of the number of channels for each convolution layer is noted around each data point for each model. For example, ``64--128" indicates that there are two convolution layers each with 64 and 128 output channels, respectively. 
These mean values (data points) and standard deviations (shaded area) of the test accuracies are obtained from 100 repeated inference, and the activations only involved a single shot of SPD measurement ($K=1$) in all the neurons, including the convolutional and linear layers.

We further investigated the effects of multiple shots of SPD measurements per activation in the SPD activations in convolutional ($K_{\text{conv}}$) and linear ($K_{\text{lin}}$) layers, respectively. We chose to test a model with four convolutional layers of 128, 256, 256, and 256 output channels and varied the $K_{\text{lin}}$ and $K_{\text{conv}}$ to see the test accuracies. The results are summarized in Supplementary Table \ref{tab:cifar_K_K_coh}. 

\begin{table}[htbp] 
    \centering    
    \begin{tabular}{|c||c|c|c|c|c|c|}
        \hline 
        & $K_{\text{lin}}=1$  & $K_{\text{lin}}=2$  & $K_{\text{lin}}=3$  & $K_{\text{lin}}=5$  & $K_{\text{lin}}=10$  & $K_{\text{lin}}\rightarrow\infty$ \\
        \hhline{|=======|}
        $K_{\text{conv}}=1$  & 86.94 $\pm$ 0.23 & 87.11 $\pm$ 0.21 & 87.17 $\pm$ 0.21 & 87.23 $\pm$ 0.19 & 87.26 $\pm$ 0.20 & 87.28 $\pm$ 0.21 \\
        $K_{\text{conv}}=2$  & 88.65 $\pm$ 0.18 & 88.77 $\pm$ 0.18 & 88.83 $\pm$ 0.18 & 88.88 $\pm$ 0.16 & 88.86 $\pm$ 0.18 & 88.90 $\pm$ 0.16 \\
        $K_{\text{conv}}=3$  & 89.16 $\pm$ 0.15 & 89.26 $\pm$ 0.15 & 89.31 $\pm$ 0.16 & 89.33 $\pm$ 0.15 & 89.35 $\pm$ 0.14 & 89.39 $\pm$ 0.14 \\
        $K_{\text{conv}}=5$  & 89.55 $\pm$ 0.14 & 89.68 $\pm$ 0.15 & 89.71 $\pm$ 0.14 & 89.74 $\pm$ 0.13 & 89.73 $\pm$ 0.13 & 89.77 $\pm$ 0.13 \\
        $K_{\text{conv}}=10$  & 89.82 $\pm$ 0.12 & 89.91 $\pm$ 0.11 & 89.95 $\pm$ 0.10 & 90.00 $\pm$ 0.11 & 90.00 $\pm$ 0.12 & 90.02 $\pm$ 0.13 \\
        $K_{\text{conv}}\rightarrow\infty$  & 90.09 $\pm$ 0.09 & 90.20 $\pm$ 0.08 & 90.21 $\pm$ 0.07 & 90.25 $\pm$ 0.07 & 90.26 $\pm$ 0.05 & 90.31 $\pm$ 0.00\\
        \hline
    \end{tabular}

    \caption{\textbf{Test accuracy (\%) of the convolutional SPDNN on CIFAR-10 with varying shots per activation $K$ in the convolutional and linear layers.}
    The SPDNN model in this table consists of four convolutional layers with 128, 256, 256, and 256 output channels, respectively. The convolutional layers are followed by a linear layer with 400 neurons and an output layer with 10 neurons. The SPD activation function is applied to each of the 400 neurons in the first linear layer. $K_{\text{conv}}$ represents the number of shots of SPD readouts per activation in the convolutional layers, while $K_{\text{lin}}$ represents the shots per activation in the linear layer. The mean accuracy and standard deviation are calculated based on 100 repetitions of inferences using the CIFAR-10 test set of 10,000 images.}
    \label{tab:cifar_K_K_coh}
\end{table}

In these SPDNNs, the number of operations in the output layer is negligible compared to the entire models. 
In terms of the number of MAC operations (dot products, DPs), $N_\text{MAC}^\text{out}=4000$ ($N_\text{DP}^\text{out}=10$). The number of dot products, or the activation size, is directly related to the number of optical detections in ONN implementations. 
The output layer is the only layer that needs to be implemented with a ``high SNR'' (see Figure 1 in the main text). 
The small portion of operations in this layer indicates the capability of low-SNR stochastic layers in a deeper model. 
This further suggests the potential to leverage stochastic physical systems with low SNRs to perform reliable neural-network inference.

\begin{table}[htbp] 

    \centering 
    \begin{tabular}{|>{\centering\arraybackslash}p{3.8cm}||>{\centering\arraybackslash}p{1.8cm}|>{\centering\arraybackslash}p{2.2cm}||>{\centering\arraybackslash}p{1.8cm}|>{\centering\arraybackslash}p{2.2cm}|}
        \hline
        Model& $N_\text{MAC}^\text{total}$ & $N_\text{MAC}^\text{out}/{N_\text{MAC}^\text{total}}$ & $N_\text{DP}^\text{total}$ & $N_\text{DP}^\text{out}/{N_\text{DP}^\text{total}}$ \\
        \hhline{|=||=|=||=|=|}   
        C64--C128 & $5.73 \times 10^{7}$ & $6.60 \times 10^{-5}$ & $2.29 \times 10^{6}$ & $4.36 \times 10^{-6}$ \\
        C128--C128 & $1.15 \times 10^{8}$ & $3.40 \times 10^{-5}$ & $4.59 \times 10^{6}$ & $2.18 \times 10^{-6}$ \\
        C128--C256 & $2.20 \times 10^{8}$ & $1.80 \times 10^{-5}$ & $8.78 \times 10^{6}$ & $1.14 \times 10^{-6}$ \\\hline
        C16--C32--C128 & $1.11 \times 10^{7}$ & $3.37 \times 10^{-4}$ & $4.42 \times 10^{5}$ & $2.26 \times 10^{-5}$ \\
        C32--C64--C128 & $2.87 \times 10^{7}$ & $1.36 \times 10^{-4}$ & $1.15 \times 10^{6}$ & $8.72 \times 10^{-6}$ \\
        C64--C128--C128 & $8.36 \times 10^{7}$ & $4.70 \times 10^{-5}$ & $3.34 \times 10^{6}$ & $2.99 \times 10^{-6}$ \\
        C128--C128--C128 & $1.41 \times 10^{8}$ & $2.80 \times 10^{-5}$ & $5.64 \times 10^{6}$ & $1.77 \times 10^{-6}$ \\
        C128--C256--C256 & $3.24 \times 10^{8}$ & $1.20 \times 10^{-5}$ & $1.30 \times 10^{7}$ & $7.71 \times 10^{-7}$ \\\hline
        C16--C32--C128--C128 & $1.76 \times 10^{7}$ & $2.24 \times 10^{-4}$ & $7.05 \times 10^{5}$ & $1.42 \times 10^{-5}$ \\
        C32--C64--C128--C128 & $3.52 \times 10^{7}$ & $1.13 \times 10^{-4}$ & $1.41 \times 10^{6}$ & $7.10 \times 10^{-6}$ \\
        C64--C128--C128--C128 & $9.01 \times 10^{7}$ & $4.40 \times 10^{-5}$ & $3.60 \times 10^{6}$ & $2.77 \times 10^{-6}$ \\
        C128--C128--C128--C128 & $1.47 \times 10^{8}$ & $2.70 \times 10^{-5}$ & $5.90 \times 10^{6}$ & $1.70 \times 10^{-6}$ \\
        C128--C256--C256--C256 & $3.51 \times 10^{8}$ & $1.10 \times 10^{-5}$ & $1.40 \times 10^{7}$ & $7.13 \times 10^{-7}$ \\    
        \hline
    \end{tabular}
    \caption{\textbf{Number of operations in the convolutional SPDNN models.}
    The table displays the number of multiply--accumulate (MAC) operations and dot products (DPs) in the SPDNN models. 
    $N_\text{MAC}^\text{total}$ ($N_\text{DP}^\text{total}$) represents the total number of MAC operations (dot products) in the entire SPDNN models, including all the convolutional layers and two linear layers. 
    $N_\text{MAC}^\text{out}$ ($N_\text{DP}^\text{out}$) represents the number of MAC operations (dot products) in the output layer, which is the only layer implemented with a ``high SNR''. 
    For the $400\rightarrow 10$ output layer, $N_\text{MAC}^\text{out}=4000$ and $N_\text{DP}^\text{out}=10$.
    The portion of the high-SNR output layer's operations relative to the entire model's operations in terms of MAC operations (dot products) is presented in the third (fifth) column.
    ``C$N_\text{chan}$'' denotes an SPD convolutional layer with $N_\text{chan}$ output channels.}
    \label{tab:cifar_ns}
\end{table}

\newpage
\subsection{Additional classification tasks}

Apart from the benchmark MNIST and CIFAR-10 classification tasks, we added additional tests for KMNIST (``Kuzushiji MNIST'') \cite{clanuwat2018deep} and FashionMNIST \cite{xiao2017fashion} to further demonstrate the capabilities of our SPDNN models. To ensure fair and straightforward comparisons, we chose to test these tasks using the same structure we used for MNIST classification tasks, specifically the MLP with $784\rightarrow400\rightarrow10$. We compared the following four different models:

\begin{itemize}
\item A deterministic MLP with one hidden layer of 400 neurons, $784\rightarrow400\rightarrow10$. The weights are real-valued. The activation function is ReLU.
\item A coherent MLP-SPDNN with one hidden layer of 400 neurons, $784\rightarrow400\rightarrow10$. The weights are real-valued. This model is the same as the one shown in Figure 4d in the main text ($N=400$).
\item An incoherent MLP-SPDNN with one hidden layer of 400 neurons, $784\rightarrow400\rightarrow10$. The weights are non-negative. This model is the same as the one shown in Figure 3b in the main text (MLP, 1 hidden layer).
\item A linear classifier. The weights are real-valued.
\end{itemize}

\begin{figure}[htp]
\centering
\includegraphics [width=0.96\textwidth]{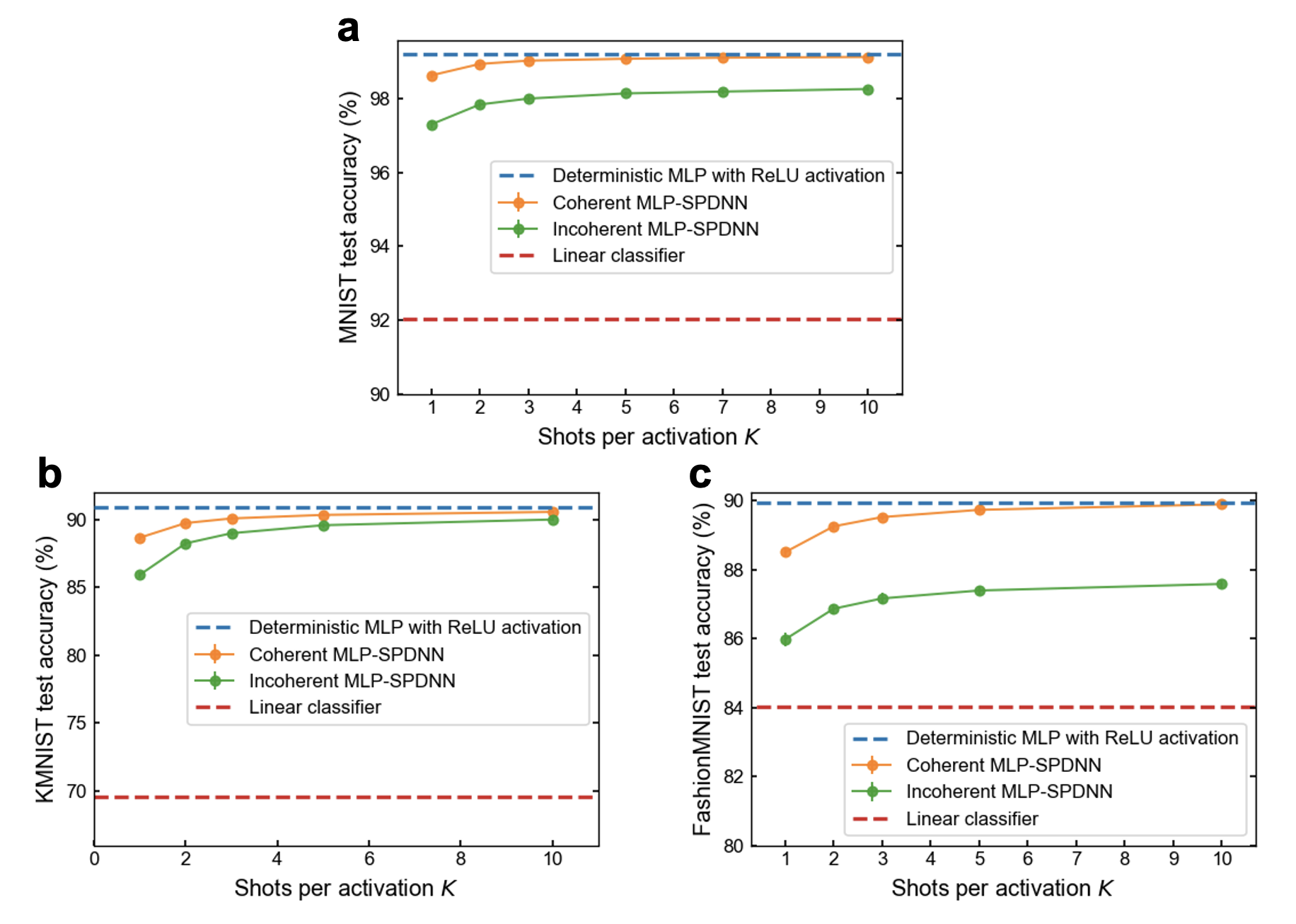}
\caption{\textbf{Comparison of performance of different models on three image classification tasks.} The deterministic model and both coherent and incoherent SPDNNs share the MLP architecture of $784\rightarrow400\rightarrow10$. The x-axes represent the number of shots per activation $K$ in the hidden layer of the corresponding SPDNN model. These models are trained for \textbf{a,} MNIST, \textbf{b,} KMNIST, and \textbf{c,} FashionMNIST classification tasks, respectively.}
\label{fig:3tasks}
\end{figure}

Using a training process similar to that for the MNIST classification task (\ref{subsec:mnist}), we optimized these SPDNN models with both incoherent and coherent optical setups. The blue dashed line represents the deterministic model with the same MLP architecture as the SPDNNs. The red dashed line represents the linear classifier. These lines set the upper and lower bounds of test accuracy, respectively, for each individual classification task.

By varying the number of shots per activation $K$ in these SPDNN models, we observed trends consistent with those seen in the MNIST task (Supplementary Figure \ref{fig:3tasks}), although some small differences exist among these tasks. For example, the test accuracy for coherent and incoherent MLP-SPDNNs tends to converge for the KMNIST classification task with increasing shots per activation $K$. However, this convergence is less pronounced for the other two tasks.

These additional results illustrate that our SPDNN models are effective across various tasks. They demonstrate that our highly stochastic SPDNN models can achieve reliable performance comparable to deterministic models.

\newpage
\part{Experimental setup}

\vspace{24pt}
\section{Incoherent optical matrix-vector multiplier}
\label{sec:setup}

\begin{figure}[htp]
\includegraphics [width=0.98\textwidth] {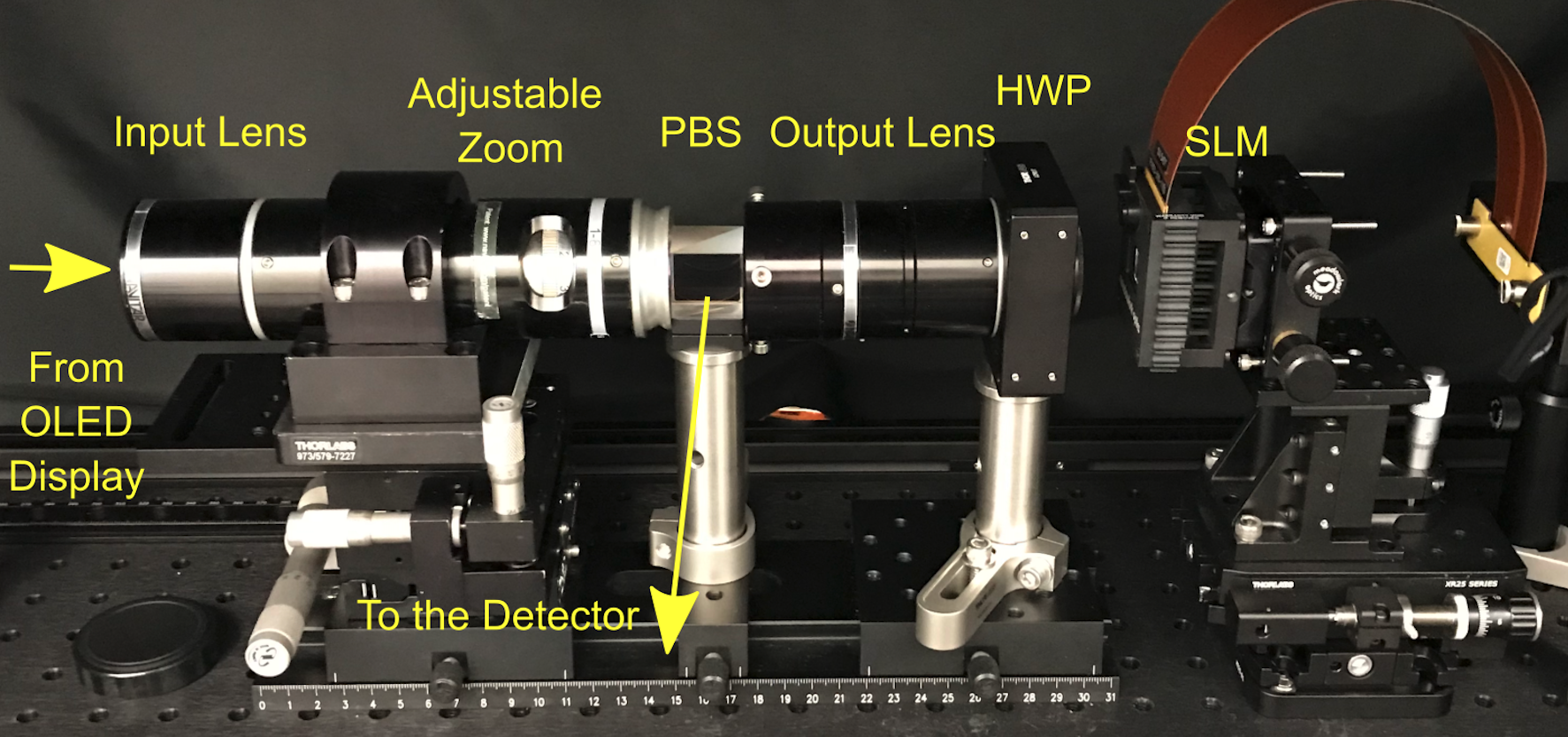}
\caption{\textbf{A photo of the experimental setup.} Input light source (OLED display) and detection parts are not included in this photo. PBS: polarizing beam splitter; HWP: half-wave plate; BPF: band-pass ﬁlter; SLM: spatial light modulator. }
\label{suppfig:setup}
\end{figure}

The optical matrix-vector multiplier (optical MVM) setup is based on the setup designed in \cite{wang2022optical}. It consists of an array of light sources, a zoom lens imaging system, an intensity modulator, and a photodetector. We used an organic light-emitting diode (OLED) display of a commercial smartphone (Google Pixel 2016 version) as the light source for encoding input vectors. The OLED display consists of a $1920 \times 1080$ pixel array, with individually controllable intensity for each pixel. There are pixels of three colors on the display, only the green pixels (light wavelength of $\sim532$ nm) are used in the experiment. The green pixels are arranged in a square lattice with a pixel pitch of 57.5 µm. A reflective liquid-crystal spatial light modulator (SLM, P1920-500-1100-HDMI, Meadowlark Optics) was combined with a half-wave plate (HWP, WPH10ME-532, Thorlabs) and a polarizing beam splitter (PBS, CCM1-PBS251, Thorlabs) to perform intensity modulation as weight multiplication. The SLM has a pixel array of dimensions $1920 \times 1152$, with individually controllable transmission for each pixel. Each pixel has the size of $9.2\times9.2$ µm. A zoom lens system (Resolv4K, Navitar) was used to image the OLED display onto the SLM panel (Supplementary Figure \ref{suppfig:setup}). 
The intensity-modulated light field reflected from the SLM was further de-magnified and imaged onto the detector, by a telescope formed by the rear adapter of the zoom lens (1-81102, Navitar) and an objective lens (XLFLUOR4x/340, Olympus). 
An additional band-pass filter (BPF, FF01-525/15-25, Semrock) and polarizer (LPVISE100-A, Thorlabs) were inserted into the telescope in order to reduce the bandwidth and purify the polarization of the light reflected by the PBS. A scientific CMOS camera (ORCA-Quest qCMOS Camera C15550-20UP) is used to measure the light intensity, as well as single-photon detection. The qCMOS camera has $4096\times2304$ effective pixels with the size of $4.6\times4.6$ µm.

\vspace{20pt}
\begin{figure}[htpb]
\includegraphics [width=0.8\textwidth] {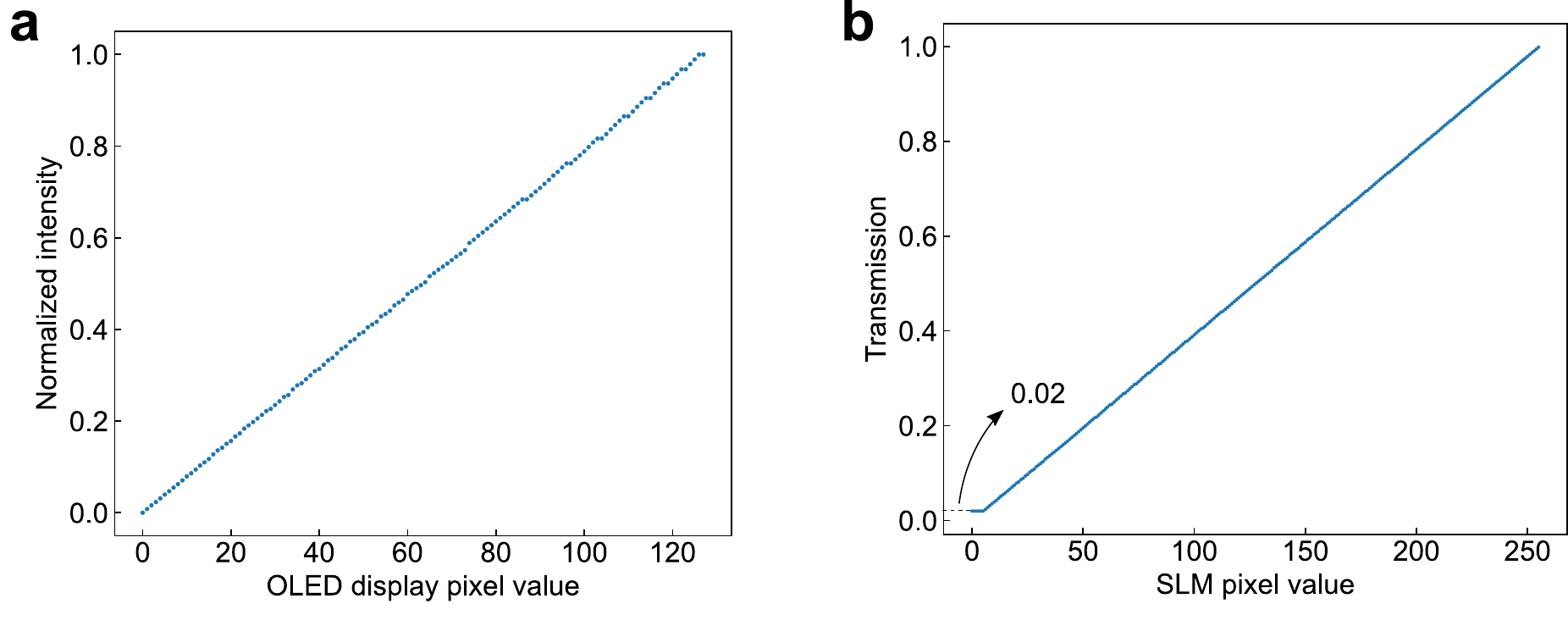}
\caption{\textbf{Look-up tables (LUTs) of the components in the incoherent optical matrix-vector multiplier (optical MVM) setup.} \textbf{a,} The 7-bit LUT of the OLED display to control the pixel intensity.  \textbf{b,} The 8-bit LUT of the SLM for intensity modulation. The minimum transmission was measured to be $\sim2\%$ of the maximum transmission.}
\label{suppfig:luts}
\end{figure}
\vspace{20pt}

During the computation of a vector-vector dot product, the value of each element in either vector is encoded in the intensity of the light emitted by a pixel on the OLED and the transmission of an SLM pixel. Via the imaging system, each pixel on the OLED display is aligned to a corresponding pixel on the SLM, where element-wise multiplication takes place by the intensity modulation. The modulated light intensity from pixels in the same vector is then focused on the detector to sum up the element-wise multiplication values to be the dot product result. Since the light is incoherent, only non-negative values can be represented. Matrix-vector multiplication is realized by doing a batch of this kind of vector-vector multiplications in parallel, either multiplexed in space or time.

OLED pixels feature a high extinction ratio and high dynamic range in intensity, which are ideal for characterizing the accuracy of vector-vector dot products. A commercial-grade integrated OLED panel with a high pixel count is readily available at a low cost, which made it possible to encode very large vectors that were essential for demonstrating vector-vector dot products on our setup. The true darkness of OLED pixels allowed us to achieve high dynamic range in intensity modulation and to reduce noise caused by background light pollution. The intensity of each individual pixel can be controlled independently with 256 (8-bit) control levels. However, since the actual output intensity was not linear with the pixel control level, we calibrated a linear look-up table (LUT) that contains 124 distinct intensity levels ($\sim$7 bits, Supplementary Figure \ref{suppfig:luts}a).
We converted a phase-only SLM into an intensity modulator with a half-wave plate (HWP) and a polarizing beam splitter (PBS). The SLM pixels are made of birefringent liquid crystal layers, whose refractive index can be tuned by applying voltage across them. By controlling the refractive index of extraordinary light, the SLM pixels introduce a phase difference between the extraordinary and ordinary light, whose polarizations are perpendicular to each other. When a PBS and HWP were placed in front of a reflective SLM, the light field passed the components twice, once during the trip towards the SLM and once after being reﬂected by the SLM (Supplementary Figure 6). One of the functions of PBS was to separate the output from the input light: the input light (incident to the SLM) was horizontally polarized and transmitted by the PBS, while the output light (reﬂected from the SLM) was vertically polarized, and therefore reﬂected by the PBS. The other function of the PBS is to convert the polarization state of the output light to its amplitude: the light modulated by the SLM was in general elliptically polarized, controlled by the phase difference. The amplitude of the light field (and intensity in this case too) was modulated by selecting only the vertical component of the SLM-modulated light at the output port of the PBS. The HWP was placed with its fast axis rotated 22.5 degrees from the extraordinary axis of the SLM such that the intensity transmission could be tuned from 0 to 100\%. 
In the experiment, each of the SLM pixels can be independently controlled for intensity modulation with a 256 (8-bit) LUT (Supplementary Figure \ref{suppfig:luts}b). The maximum extinction ratio of the transmission intensity was measured to be $\sim$50. Alternatively, instead of using a phase-modulation SLM, the intensity modulator can be more compactly implemented with a monolithic LCD panel in a transmission geometry.

\vspace{24pt}
\section{Single-photon detection by a scientific CMOS camera}
\label{sec:qcmos}
Single-photon detection is core to implementing an SPDNN. In our experiment, we use a scientific CMOS camera, the Hamamatsu ORCA-Quest qCMOS Camera, to realize the function of single-photon detectors \cite{dhimitri2022scientific}. CMOS cameras usually cannot detect single photons due to the relatively high readout noise compared to the signals induced by individual photons. The ORCA-Quest qCMOS camera, however, has well-controlled readout noise as low as $0.3$ equivalent photoelectrons. This makes viewing the individual spikes of photon response possible on the output of the camera. An example of the distribution of pixel values from the camera is shown in Supplementary Figure \ref{suppfig:qcmos_distr}a. These pixel values are from a sequence of frames collected with some intensity of input light. The output pixel values have a digital bias of $\approx200$, and the analog gain is $\sim7.94$ pixel values per photoelectron. We can see the individual spikes corresponding to different numbers of detected photons, with the first peak referring to no detected photons. Due to readout noise of the camera, we can still see a near-Gaussian distribution around the peak value of each detected photon number. To do single-photon detection, a threshold can be set to determine if there is a photon (or more photons) detected. If the pixel value is larger than the threshold, we record a click; otherwise, there is no click. In this way, although the camera has already completed analog-to-digital conversion (ADC) before thresholding, the qCMOS camera can still emulate the function of a single-photon detector. 

\begin{figure}[htp]
\includegraphics [width=0.78\textwidth] {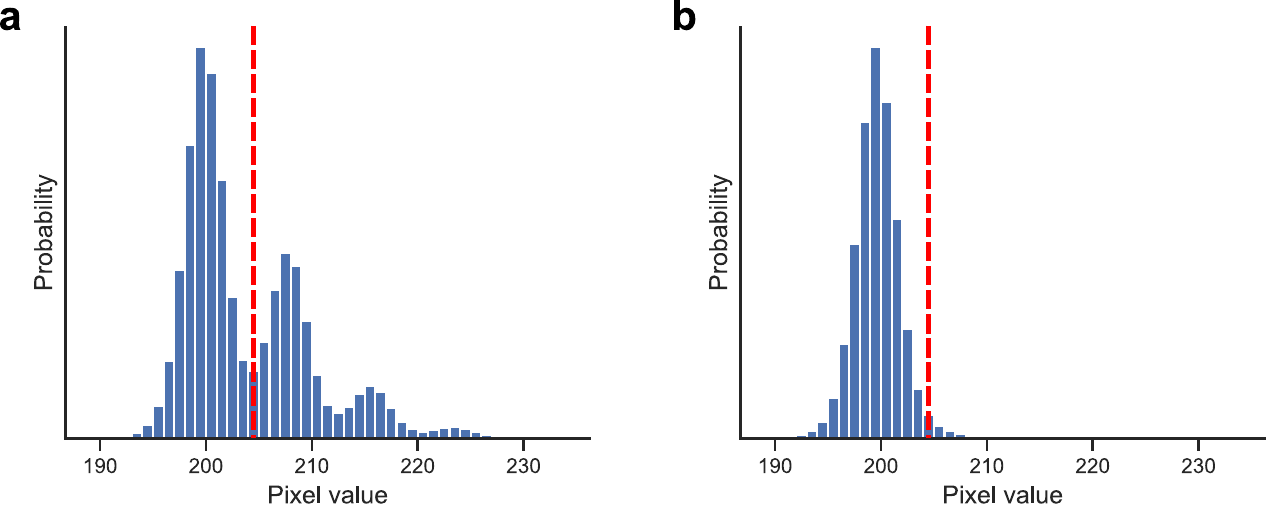}
\caption{\textbf{Pixel value distributions of the ORCA-Quest qCMOS camera.} \textbf{a,} An example of pixel value distribution in collected frames with input signal light of a few photons. The red dashed line indicates the threshold to tell if there is a photon click. We can see clearly that the discrete numbers of photons can be resolved in the individual peaks. \textbf{b,} Pixel value distribution of the dark frames without input light.}
\label{suppfig:qcmos_distr}
\end{figure}

This camera has an overall quantum efficiency of $\sim86\%$ at our working wavelength of 532 nm and a dark count rate of 0.006 photoelectrons per second per pixel at the working temperature of -35ºC with air cooling. 
Due to the readout noise,
the thresholding process may add additional errors because of the overlap of the signal peaks. Supplementary Figure \ref{suppfig:qcmos_distr}b shows the pixel value distribution of dark frames without input signals. We can see that using the same threshold, there is a small tail of the distribution on the right side of the threshold. This small portion of the pixel values from dark frames would trigger a photon click as well, which further adds to the dark count rate. Similarly, the output pixel values from detected photons also have a small probability to fall in the ``no click'' region, which makes the effective photon detection efficiency a bit lower. In our experiment, we calibrated the qCMOS camera in the single-photon detection mode and found that the effective dark count rate of around 0.01 photoelectrons per second per pixel, and the effective photon detection efficiency to be 68\%, on average. Note that there are also variations among different pixels. In the experiment implementation, only one single pixel is used for each vector-vector dot product. We will see that the photon detection efficiency and dark counts do not significantly influence the results, as discussed in \ref{sec:rob}.

\clearpage
\section{Validation of the optical vector-vector multiplications}
\label{sec:vvm}

\begin{figure}[htp]
\centering
\includegraphics [width=0.78\textwidth] {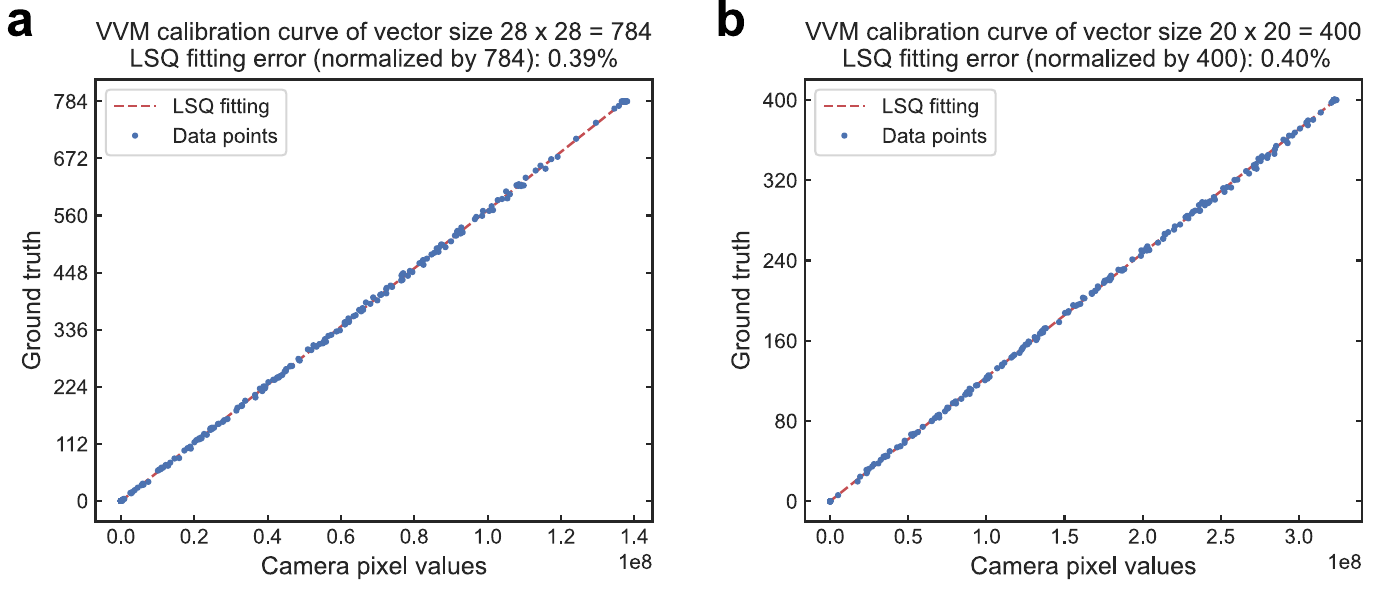}
\caption{\textbf{Calibration curves of vector-vector multiplication (VVM) precision in the setup.}
Panel \textbf{a} (\textbf{b}) shows the VVM calibration of the vector size of 784 (400), which were used for the hidden (output) layer in the optical implementation of the SPDNN model with a structure of $784\rightarrow400\rightarrow10$.
The calibration curves (red dashed line) were obtained using the least-squares regression method and the data points are collected with a long exposure time to eliminate the photon noise. }
\label{suppfig:vvm_cali}
\end{figure}

The major part of computation in the ONN implementation is the linear operations. The accuracy of matrix-vector multiplication is essential in a successful ONN inference. In this section, we calibrate the accuracy of our optical MVM. We use the setup with either single-photon detection or conventional intensity measurement that involves a much higher intensity. Focusing the lights to one pixel of $\sim5$ µm is challenging, which reduces the dot product precision slightly. However, as we will see in \ref{sec:rob}, the SPDNN models are very robust to this amount of errors. 

To generate a test dataset representative of general dot products, we randomly generated vector pairs $\vec{x}$ and $\vec{w}$ based on natural scene images from the STL10 dataset. Each vector was generated from a single color channel of one or more images patched together, depending on the target vector size (each image of size $L \times L$ contributes $N=L^2$ elements to the vector). We chose natural images since they are more representative of the inputs in image classification with globally inhomogeneous and locally smooth features. To adjust the sparsity of the vectors, different thresholds were applied to the image pixel values such that the dot product results cover a wider range of possible values. This was achieved by shifting the original pixel values (float point numbers normalized to the range $0\text{-}1$) in the entire image up or down by a certain amount, unless the value was already saturated at 1 (maximum) or 0 (dark). For example, a shift of -1 would make the whole image dark. A shift of +0.2 would make all the pixel values that were originally larger than 0.8 saturated, and would increase all other pixel values by 0.2. This method allowed us to tune the overall intensity of the modulated images without losing the randomness of the distribution.

Calibration curves of vector-vector dot product results are shown in Supplementary Figure \ref{suppfig:vvm_cali}. The results are averaged over a large number of repetitions to get rid of the photon noise to see the systematic errors in the optical MVM. The vectors are randomly generated to cover the full range of the light intensity from the minimum to maximum transmission, as discussed in \cite{wang2022optical}. The vector size is $28\times28$, which is equivalent to the size of the first layer in MNIST classification. 

\vspace{44pt}
\section{Validation of the SPD activation function}

To validate the SPD activation function in the SPDNN implementation, we need to consider not only the precision of the linear operations, but also the non-linear activation function. As the incident light onto the qCMOS camera is attenuated to just a few photons, the photon noise becomes significant and the measurement less accurate. To address this, we first measure a higher light intensity with long exposure times and estimate the exact light intensity with a shorter exposure time using the ratio of exposure times. We then use the shorter exposure time to perform single-photon detection to output a photon click (value 1) or no photon click (value 0). The probability of a photon click is estimated by averaging over a large number of repetitions. The intensity is tuned by adjusting both the exposure time and neutral density (ND) filters that attenuated the light. The expected theoretical curve is also plotted for comparison. The results are shown in Supplementary Figure \ref{suppfig:actv_valid}.

\vspace{20pt}
\begin{figure}[htp]
\includegraphics [width=0.4\textwidth] {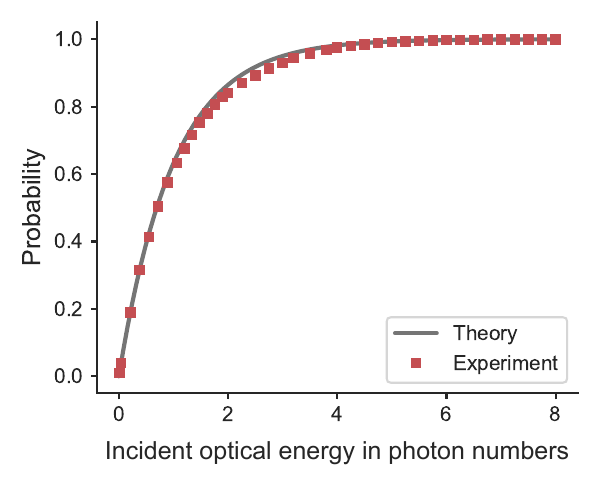}
\caption{\textbf{Validation of the SPD activation function.} The theory curve is the expected function of $f(\lambda)=1-e^{-\lambda}$ with the intensity $\lambda$ in photon numbers. The experiment data was taken by the Hamamatsu ORCA-Quest qCMOS camera.}
\label{suppfig:actv_valid}
\end{figure}

\clearpage
\part{Implementation of SPDNNs}

 \vspace{24pt}
\section{Adaptation to experimental limitations}
\label{sec:input_light}
The implementation of SPDNNs on an optical MVM can be challenged by experimental restrictions that affect the precision of the network inference. Some of these limitations include the non-negative encoding resulting from the use of an incoherent light source and limitations in the precision of the setup. In this section, we describe how these limitations can be addressed to successfully implement SPDNNs on our setup. As discussed in \ref{sec:setup}, our incoherent optical MVM has systematic errors in the dot-product results, even in the absence of photon noise. Additionally, the SLM used in the system has a finite extinction ratio of approximately 50 (Supplementary Figure \ref{suppfig:luts}b). These limitations present a significant challenge in the implementation of the SPDNNs because, in the models, both the input vectors and weights have many small values close to 0. This is problematic because, within the full range of 0 to 1, having a minimum value of 0.02 instead of 0 has a non-trivial effect on the accuracy of the dot product calculation. These small values are accumulated over many elements, leading to a relatively large value, compared to the final dot product result. As a result, the performance of the SPDNNs is severely impacted by these limitations.

\vspace{20pt}
\begin{figure}[htp]
\includegraphics [width=0.96\textwidth] {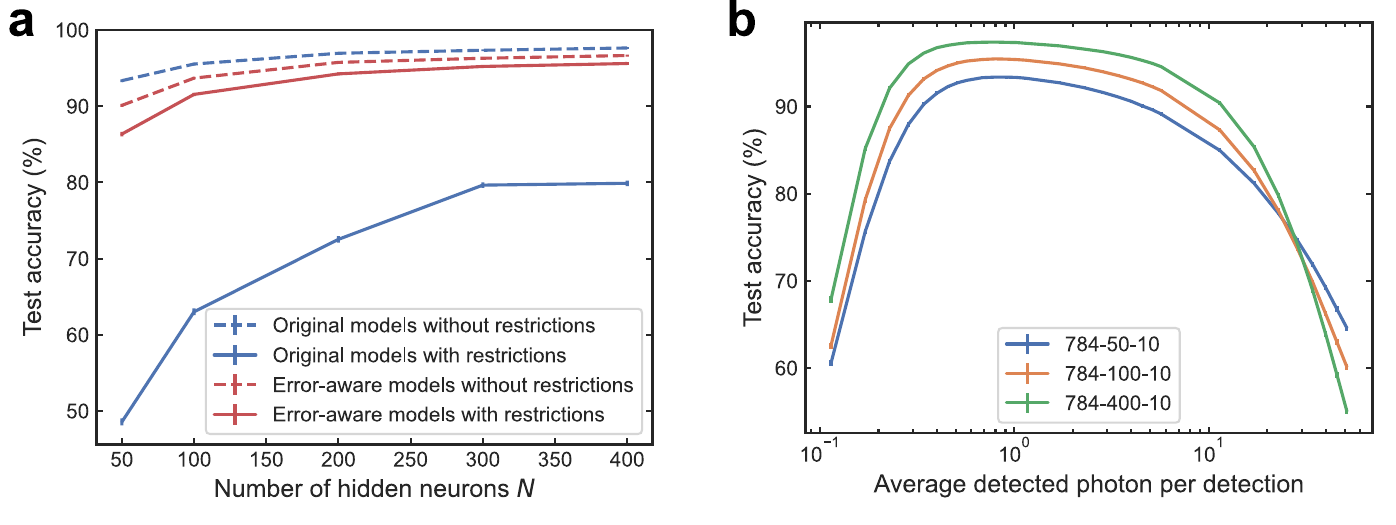}
\caption{\textbf{Simulation of SPDNN performance with different experimental settings.} \textbf{a,} MNIST test accuracy of SPDNN models under experimental restrictions. The models have a structure of $784\rightarrow400\rightarrow10$ ($N=400$, $K=1$). \textbf{b,} MNIST test accuracy as a function of the input light intensity. The intensity was varied by adjusting the range of input values using a constant factor. Both panels show results obtained with an incoherent setup and a single shot of SPD readout per activation ($K=1$). 
}
\label{suppfig:input_test}
\end{figure}

Supplementary Figure \ref{suppfig:input_test}a demonstrates the results of implementing the neural network models using the real LUTs from our setup. The test accuracy significantly drops, making the experimental implementation a failure. To address this issue, we used error-aware training techniques (as discussed in \cite{wu2021harnessing}) to train our models with an understanding of these experimental restrictions. During the error-aware training process, the real LUTs were used in the implementation of the models. The results of this error-aware training are shown in the red curves in Supplementary Figure \ref{suppfig:input_test}a. It can be seen that, with error-aware training, the SPDNN models are highly robust to changes in input range, especially with a relatively large number of hidden neurons.

Conventional ONN inferences can operate effectively at various light intensity levels, as long as the intensity is sufficiently high to suppress photon noise and maximize detection precision. These systems can, in principle, integrate arbitrarily high light intensities to enhance detection precision. 
However, in the optical implementation of SPDNN inferences, the SPD activation function relies on the precise number of photons detected. As a result, controlling the operating intensity in the setup becomes crucial to ensure accurate quantization of the detected optical energy. Calibrating the intensity to the appropriate level for SPD activation function presents a challenge, especially considering the inherent significant noise in intensity measurements at low photon counts.

Despite these challenges, our simulation results demonstrate the robust performance of SPDNNs even with slight variations in the input intensities. We systematically varied the input intensity across a range from 0.1 to 100 times the original expected intensity that was used during training. Supplementary Figure \ref{suppfig:input_test}b illustrates that the model's performance remains stable within a wide range of intensities. The test accuracy remains nearly consistent, even when the input energy deviates significantly from the original training intensity.
This observed stability highlights the resilience of SPDNNs to variations in input intensity levels. 
It further suggests that these SPDNN models can be successfully implemented with lower photon budgets, which is promising for practical applications where minimizing optical energy usage is desirable.

\vspace{24pt}
\section{Optical implementation of SPD activations}
\label{sec:exp_l1}
 
\begin{figure}[htp]
\centering
\includegraphics [width=0.999\textwidth] {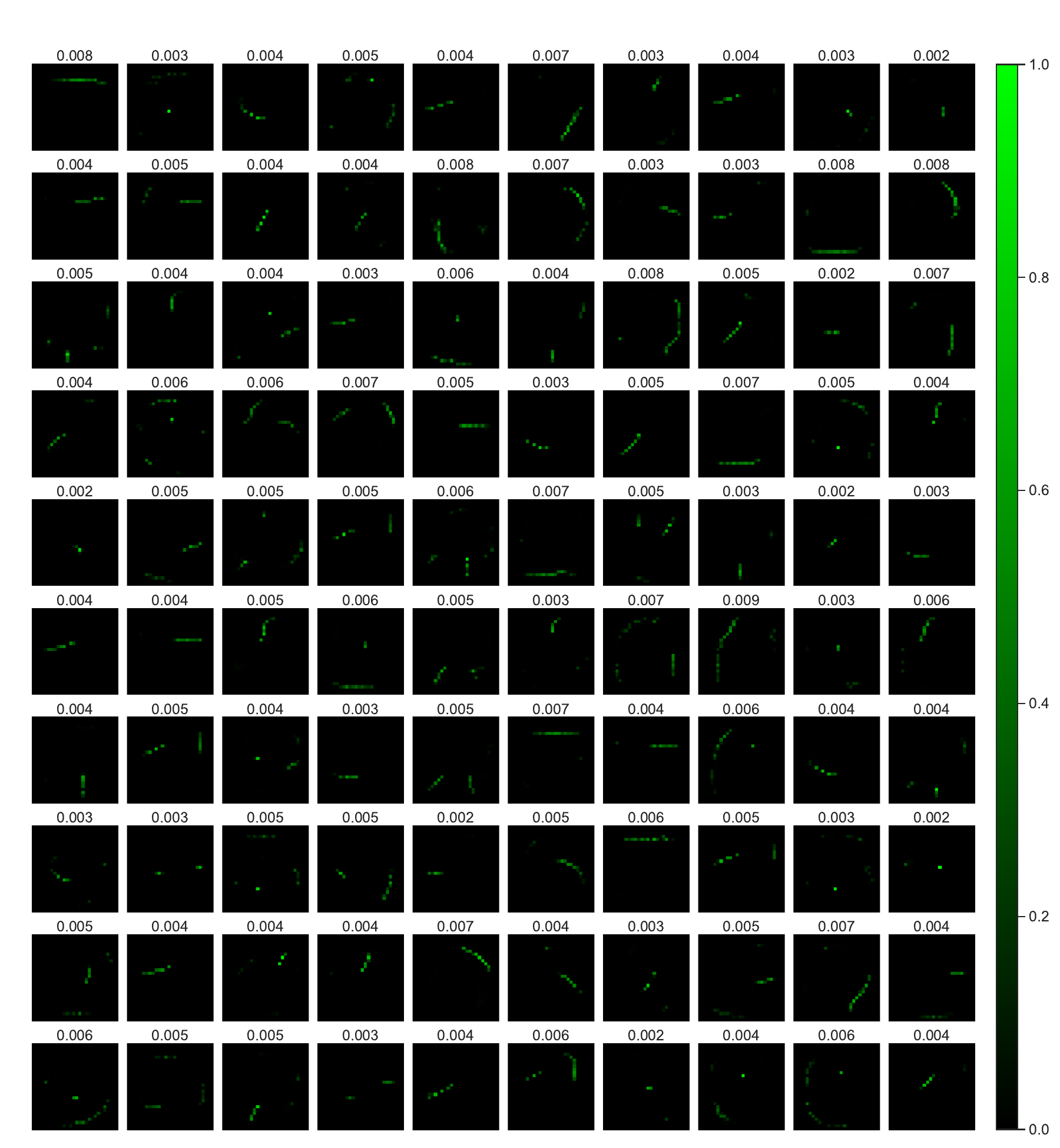}
\caption{\textbf{An example of weight values to be displayed on the OLED display during the experiment.} This model has $N=100$ hidden neurons, and we display the weight matrix $W^{(1)}$ of the first layer (with dimensions $100\times 784$). Each block represents a row vector in $W^{(1)}$ containing 784 elements. These column vectors are rearranged to form a 2D block with dimensions $28\times28$, matching the original shape of the MNIST input images. The $100$ rows in $W^{(1)}$, corresponding to the $100$ hidden neurons in the neural network, are arranged in a $10\times10$ grid to be visualized. The weight values have been normalized to a range of 0 to 1 to fit the intensity range of the OLED display. The average value of each block is indicated at the top. The color map used in the plot has been selected to emulate the actual color on the OLED display, as only green pixels are utilized ($\sim532$ nm), thereby presenting what could be observed on the OLED display in the experimental setup.}
\label{suppfig:exp_w1}
\end{figure}

\begin{figure}[htp]
\includegraphics [width=0.83\textwidth] {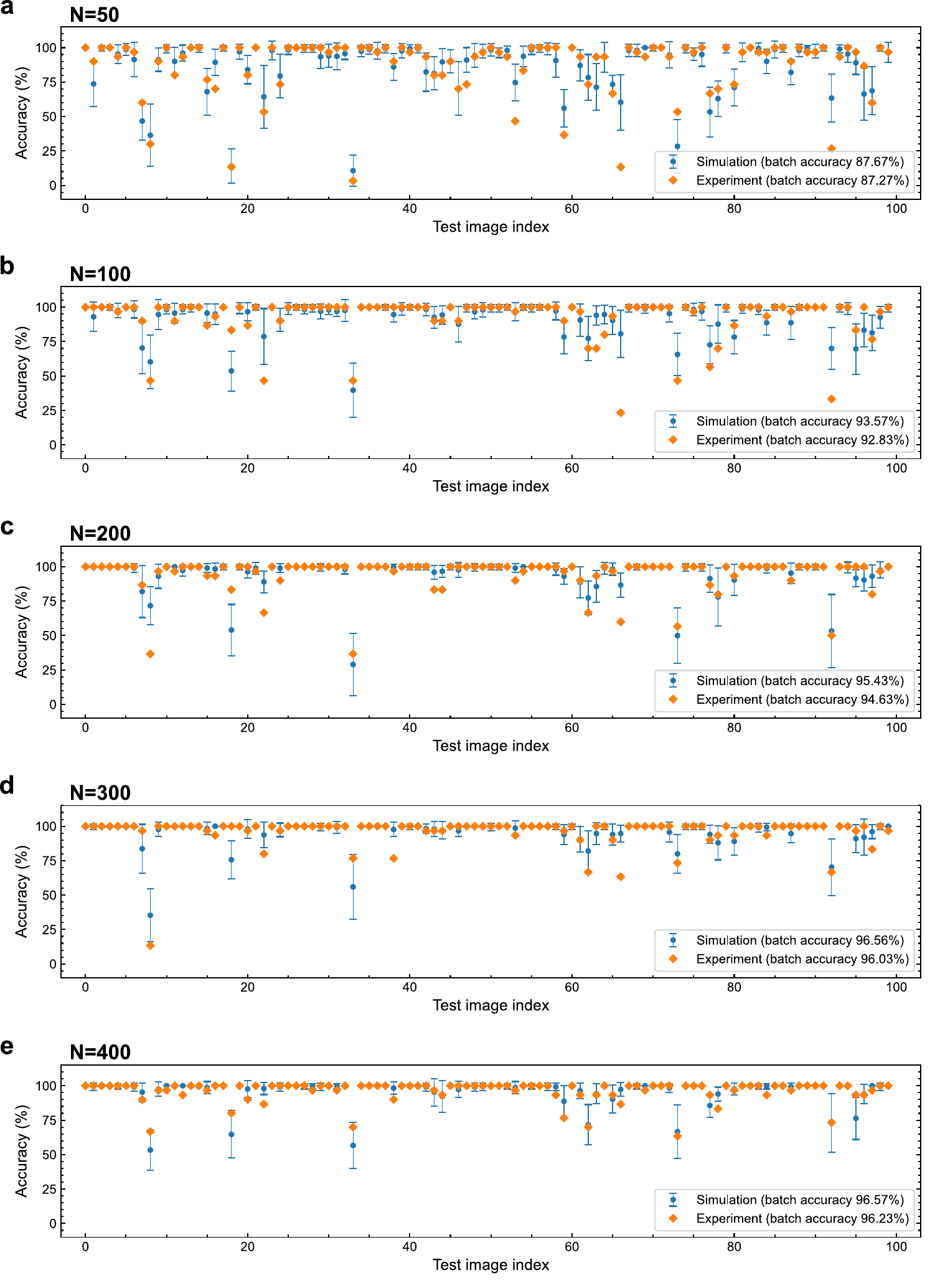}
\caption{\textbf{Test accuracy of individual images with 1 SPD measurement per activation ($K=1$).} 
The MLP-SPDNN models with a varying number of hidden neurons $N$ from 50 to 400 (panels \textbf{a}--\textbf{e}) were evaluated using a single shot per SPD activation ($K=1$). 
Test accuracy was evaluated for each individual image, and each data point represents the average accuracy obtained from 30 inferences. 
We have not plotted error bars for experimental data, but the variance is directly related to the data shown: since the outcome of each inference is a Bernoulli random variable with $p$, the variance is $p(1-p)$.
The accuracy values in the legends are averaged over all test images.
The simulation results used the same SPDNN models as in the experiment and considered the experimental restrictions. 
The error bars were calculated by repetitions of the whole process (average accuracy of 30 inferences).}
\label{suppfig:batch_acc}
\end{figure}

\begin{figure}[htp]
\includegraphics [width=0.83\textwidth]{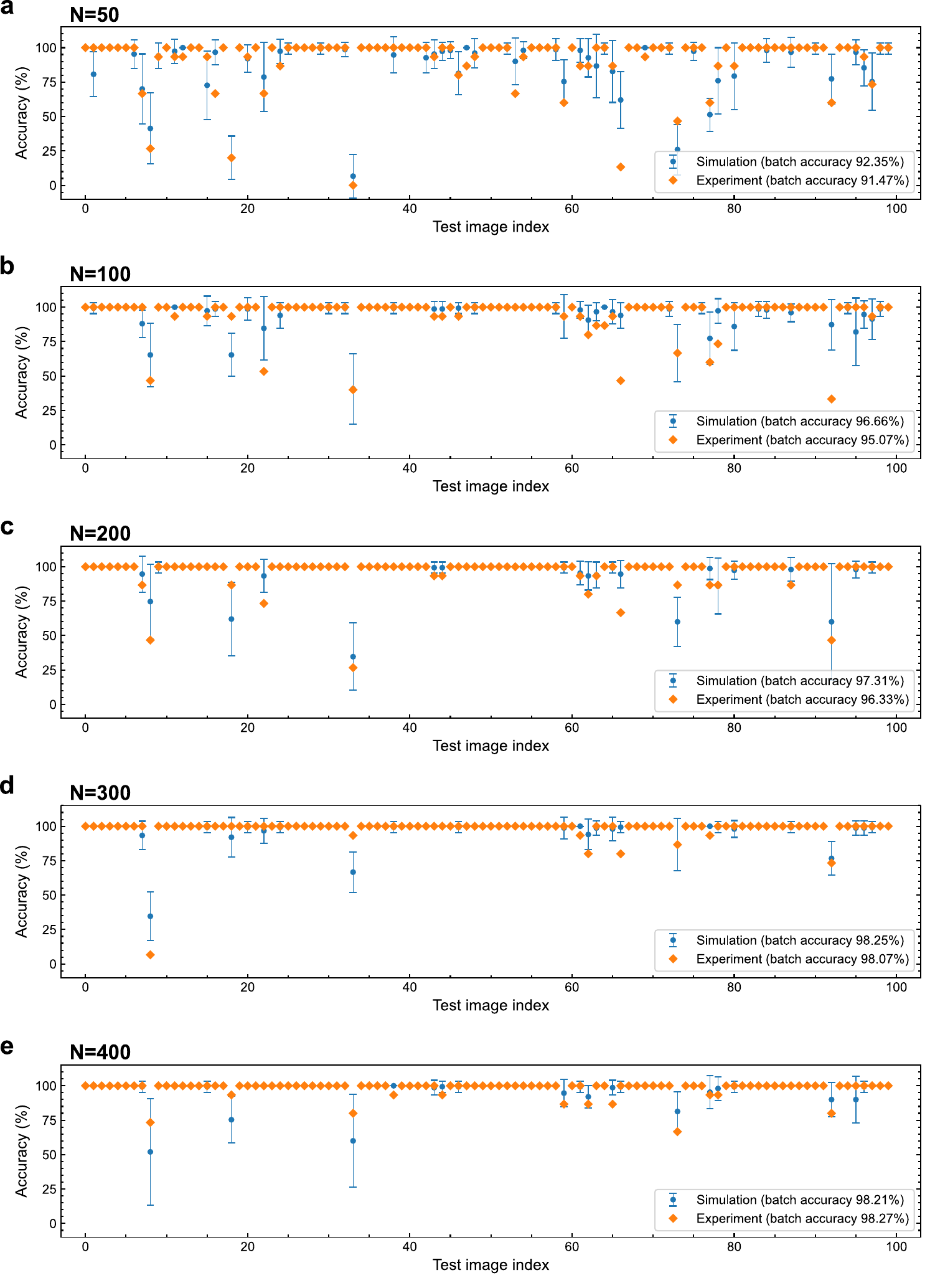}
\caption{\textbf{Test accuracy of individual images with 2 SPD measurements per activation ($K=2$).} 
The MLP-SPDNN models with a varying number of hidden neurons $N$ from 50 to 400 (panels \textbf{a}--\textbf{e}) were evaluated using 2 shots per SPD activation ($K=2$). 
Test accuracy was evaluated for each individual image, and each data point represents the average accuracy obtained from 15 inferences. 
We have not plotted error bars for experimental data, but the variance is directly related to the data shown: since the outcome of each inference is a Bernoulli random variable with $p$, the variance is $p(1-p)$.
The accuracy values in the legends are averaged over all test images.
The simulation results used the same SPDNN models as in the experiment and considered the experimental restrictions. 
The error bars were calculated by repetitions of the whole process (average accuracy of 15 inferences).}
\label{suppfig:batch_acc_2shots}
\end{figure}

\begin{figure}[htp]
\includegraphics [width=0.83\textwidth] {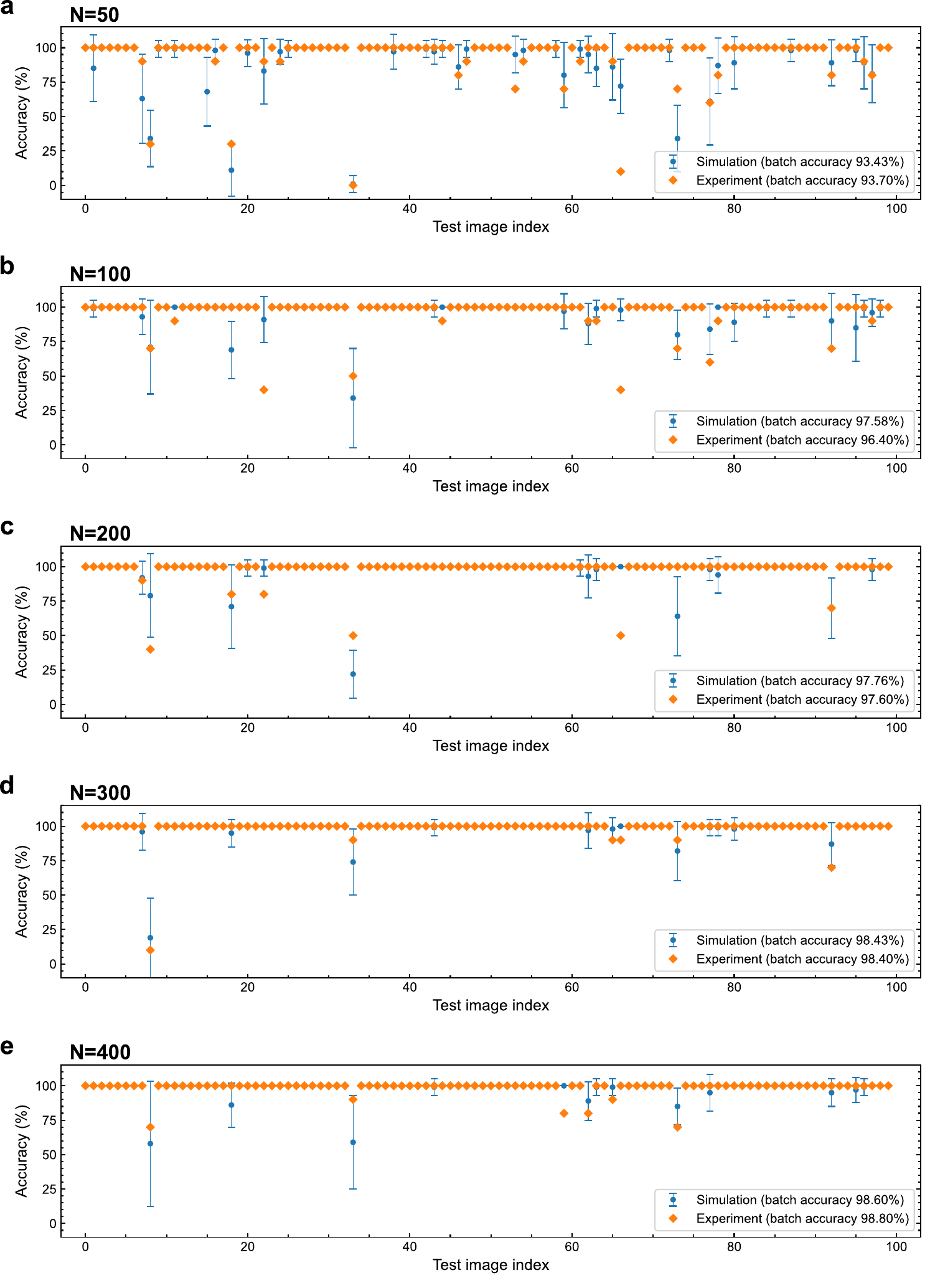}
\caption{\textbf{Test accuracy of individual images with 3 SPD measurements per activation ($K=3$).} 
The MLP-SPDNN models with a varying number of hidden neurons $N$ from 50 to 400 (panels \textbf{a}--\textbf{e}) were evaluated using 3 shots per SPD activation ($K=3$). 
Test accuracy was evaluated for each individual image, and each data point represents the average accuracy obtained from 10 inferences. 
We have not plotted error bars for experimental data, but the variance is directly related to the data shown: since the outcome of each inference is a Bernoulli random variable with $p$, the variance is $p(1-p)$.
The accuracy values in the legends are averaged over all test images.
The simulation results used the same SPDNN models as in the experiment and considered the experimental restrictions. 
The error bars were calculated by repetitions of the whole process (average accuracy of 10 inferences).}
\label{suppfig:batch_acc_3shots}
\end{figure}

In this section, we demonstrate the experimental implementation of SPD activation functions in SPDNN inferences. 
The weight matrix of the first layer $W^{(1)}$ in the SPDNNs was displayed on the OLED screen,
with each element encoded as the intensity of a corresponding pixel. The models used in experiments are trained to account for the limitations of the experimental setup, as discussed in \ref{sec:input_light}. In Supplementary Figure \ref{suppfig:exp_w1}, the weights of the model with $N=100$ hidden neurons are depicted. The two-dimensional arrangement of each weight vector block mirrors the visual representation on the OLED display, which matches each pixel of the MNIST input images.
The MNIST test images were displayed on the SLM, where the pixels were aligned with those on the OLED display. The transmission of each SLM pixel is determined by the values of the corresponding pixels in the input image. After passing through the SLM, the modulated intensity of each pixel is then combined through the imaging system to perform an optical fan-in process, and the dot product result is obtained by detecting the accumulated optical energy.

As described in \ref{sec:qcmos}, the qCMOS camera serves as a single-photon detector (SPD). The modulated light is recorded by the qCMOS camera, and the SPD activation function is applied during this single-photon detection process.
A single pixel on the qCMOS camera is used and emulates a single-photon detector by applying a threshold on the output pixel values. As shown in Supplementary Figure \ref{suppfig:qcmos_distr}, the output signal is primarily discrete in individual photon numbers, with only slight mixing due to readout noise from the electronic apparatus.
Robustness to the false clicks due to the mixing will be discussed in \ref{sec:rob}.
To control the input light intensity, either the exposure time is varied or light attenuation can be achieved through the use of neutral density filters, or a combination of both, depending on realistic experimental settings.
A typical exposure time value is $\sim 1$ ms for one shot of SPD measurement. 

Due to the inherent stochastic nature of SPDNN models, the output from different repetitions of the same test image and weight matrix varies. To assess the performance of the model, it is necessary to repeat the inference many times to represent the distribution of the output. This will give us a better understanding of the model's behavior and its accuracy.
In the experimental setup, multiple repetitions of the inference were obtained by capturing multiple frames with the camera. 
Each repetition is identified by a unique frame index number.
Note that this repetition includes the entire inference process. 
Each repetition of inference for a given input image involves $N \times K$ SPD measurements, with $K$ shots of SPD measurements for each of the $N$ neuron activations in the hidden layer.

In our experiments, we collected a set of 30 frames for each test image and weight matrix combination. Each frame provides us with a measurement of the stochastic binary output values produced by the SPD activation function, effectively serving as one-shot activations for each hidden neuron in the model. 
The activations for different frames, distinguished by different frame indices, are considered as distinct repetitions of the inference process.

As explained in \ref{subsec:incoh_model}, we can improve the precision of the inference by employing multiple shots per activation. 
With $K$ shots, we average the $K$ binary values to obtain the actual activation value during that inference.
In our experiment, every $K$ frames are averaged to obtain the activation value during a $K$-shot inference. 
As each frame collected during the experiment is independent and identically distributed, the specific sequence or arrangement of the frame indices has no impact on the resulting outcome.
Subsequently, the activation values were used as inputs to the output linear layer, performed digitally with full precision, to calculate the final output and make predictions for the label of the test image. The test accuracy results can be found in Supplementary Table \ref{tab:first_acc} and Figure 3 in the main text.

Our experimental results show that the collected SPD activations are capable of producing similar test accuracy results to those obtained from simulation. However, simply having similar accuracy values may not be a sufficient indicator of a faithful implementation. To validate this further, we compare the prediction accuracy of individual test images between the experimental and simulated results. 
The results obtained from 1-shot inferences $K=1$ are presented in Supplementary Figure \ref{suppfig:batch_acc}, where we use each frame as a separate repetition of the inference process, with a total of 30 repetitions performed. 
For each repetition, if the prediction made was accurate, it was recorded as a 1; otherwise, it was recorded as a 0. To visualize the distribution of the output accuracy, we then calculated and plotted the mean values and standard deviations of the test accuracy based on the 30 repetitions.

We conducted simulations of the same inference process on a digital computer using the same models and input images. To ensure a closer simulation to reality, we also incorporated realistic experimental restrictions, such as the limited extinguish value of the SLM, the dynamic range and precision of the LUT in both the SLM and the OLED display, and the systematic errors in the optical MVM.

Similarly, we examined the results of the inferences with $K=2$ ($K=3$) shots per activation, which are illustrated in Supplementary Figure \ref{suppfig:batch_acc_2shots} (\ref{suppfig:batch_acc_3shots}). In this setup, we combine every 2 (3) frames to be averaged to compute a neuron activation, and we repeated this process 15 (10) times to obtain the final results.

By comparing the simulation results with the experimental results obtained from the collected SPD activation values, we aimed to validate the performance of the latter. The comparison revealed that, for the majority of input images, the predictions are highly resilient to the inherent stochasticity in the model. Interestingly, the results are not as unpredictable as one might expect, as a closer examination shows that most of the errors stem from a limited number of specific input images (see Supplementary Figure \ref{suppfig:batch_acc}--\ref{suppfig:batch_acc_3shots}).

\begin{figure}[htp]
\includegraphics [width=0.83\textwidth] {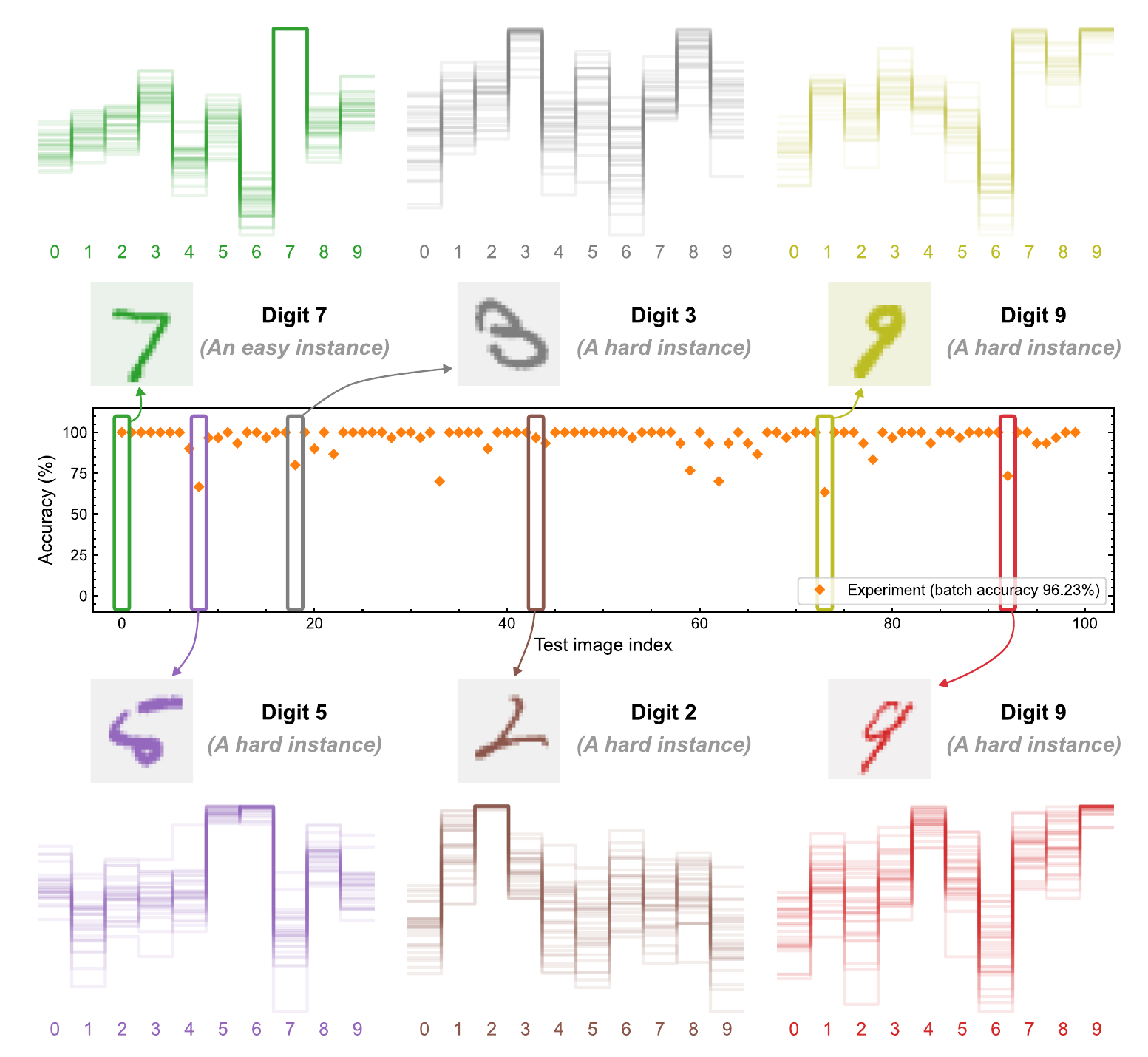}
\caption{\textbf{Visualization of the output vectors of the SPDNN with given input images.} In this figure, the SPDNN model has $N=400$ hidden neurons and $K=1$ shot of SPD measurement per activation (Supplementary Figure \ref{suppfig:batch_acc}e). 
The prediction of a single inference of a particular test image is stochastic and is either correct or incorrect. For each test image, we performed 30 inferences for each test image and report the average accuracy. We have not plotted error bars, but the variance is directly related to the data shown: since the outcome of each inference is a Bernoulli random variable with $p$, the variance is $p(1-p)$.
The output vectors from the 30 repetitions of inference with the fixed corresponding image are plotted together, with a 10\% transparency on the curves to show the density. 
These output vectors are computed from the experimentally collected SPD activations by performing the output layer digitally on a computer (Supplementary Table \ref{tab:first_acc}). }
\label{suppfig:expactv_outvecs}
\end{figure}

\begin{table}[htb] 
    \centering
    \begin{tabular}{c|ccccc} 
        \hline
        Model  & $K=1$  & $K=2$  & $K=3$  & $K=5$  & $K=10$  \\\hline
        $784-50-10$  & $\quad 87.3 \pm 2.5\%\quad$  & $\quad 91.5 \pm 2.2\%\quad$  & $\quad 93.7 \pm 1.8\%\quad$  & $\quad 95.0 \pm 0.0\%\quad$  & $\quad 95.0 \pm 0.8\%\quad$  \\
        $784-100-10$  & $\quad 92.8 \pm 1.9\%\quad$  & $\quad 95.1 \pm 1.5\%\quad$  & $\quad 96.4 \pm 1.2\%\quad$  & $\quad 96.7 \pm 0.7\%\quad$  & $\quad 97.7 \pm 0.5\%\quad$  \\
        $784-200-10$  & $\quad 94.6 \pm 2.0\%\quad$  & $\quad 96.3 \pm 1.5\%\quad$  & $\quad 97.6 \pm 1.3\%\quad$  & $\quad 98.3 \pm 0.9\%\quad$  & $\quad 98.3 \pm 0.9\%\quad$  \\
        $784-300-10$  & $\quad 96.0 \pm 1.5\%\quad$  & $\quad 98.1 \pm 0.9\%\quad$  & $\quad 98.4 \pm 1.1\%\quad$  & $\quad 98.7 \pm 0.5\%\quad$  & $\quad 98.7 \pm 0.5\%\quad$  \\
        $784-400-10$  & $\quad 96.2 \pm 1.5\%\quad$  & $\quad 98.3 \pm 1.3\%\quad$  & $\quad 98.8 \pm 0.9\%\quad$  & $\quad 99.2 \pm 0.9\%\quad$  & $\quad 99.7 \pm 0.5\%\quad$  \\
        \hline
    \end{tabular}
    \caption{\textbf{Test accuracy of the experimental SPD activations with different shots per activation $K$.} The results are based on 30 repetitions of one-shot binary SPD activations collected for each model structure. The test accuracy was calculated by averaging $K$ shots per activation, and the last layer was performed with full precision. The table displays the mean and standard deviation of the test accuracy obtained from $30/K$ repetitions of inference.}
    \label{tab:first_acc}
\end{table}

The close correspondence between the experimental and simulated results for these specific ``problematic'' input images further validates the reliability of our experimental implementation. Although the experimental results are slightly inferior to the simulation results, the distribution of accuracy per input image is highly comparable. 
In particular, input images that exhibit high sensitivity to the model's stochasticity tend to result in larger deviations in the experimental results, while input images that are robust to the model's stochasticity exhibit high accuracy both in simulations and in experiments. These results provide strong evidence of the reliability of the experimental implementation and demonstrate the robustness and noise resilience of SPDNN implementations.

To further understand the characteristics of the stochastic neural-network inference, we examined the output vectors of each input test image. 
As depicted in Supplementary Figure \ref{suppfig:expactv_outvecs}, the 30 output vectors from different repetitions of each input image are plotted together to demonstrate the stochasticity in the neural network. These output vectors were computed by the experimentally measured SPD activations and digitally implemented output layer, with $N=400$ hidden neurons and $K=1$ shot of SPD measurement per activation (Supplementary Figure \ref{suppfig:batch_acc}e). 
No additional operations were performed after the linear operation of the output layer (see Algorithm 2).

Each of the 10 values in the output vector corresponds to the classes in MNIST digit classification, ranging from 0 to 9, as indicated at the bottom. The curves of the 30 output vectors were plotted with 10\% transparency to show the distribution density.

As shown in Supplementary Figure \ref{suppfig:batch_acc}--\ref{suppfig:batch_acc_3shots}, most of the test images have very high accuracy and are predicted correctly by SPDNN with high certainty, such as image 0 of digit ``7'' (depicted in the upper left in Supplementary Figure \ref{suppfig:expactv_outvecs}). 
Despite the stochastic distribution of the output values among the 30 repetitions, the value of class ``7'' remains consistently higher than the other elements, resulting in a 100\% test accuracy for this image (see Figure 1a in the main text).

We also examined these ``problematic'' images, such as image 8 of digit ``5'' (lower left), which is predicted to be digit ``6'' nearly half of the chance. This misclassification is not surprising to human observers, as the image shares features with both digits ``5'' and ``6''. Interestingly, the output values for class 8 in this case are relatively high but not the highest, which also aligns with human intuition.

Similar phenomena can be found for the other ``problematic'' images as well, indicating that the model has indeed learned meaningful features from the dataset.
These findings solidify the fact that stochastic neural networks can perform reliable deterministic classification tasks, and the inherent stochasticity in the model does not compromise its ability to make accurate predictions.

\vspace{24pt}
\section{Full-optical implementation of the entire SPDNN models}
\label{sec:exp_l2}
 
 In this section, we showcase a full-optical implementation of a neural network by demonstrating the implementation of the last linear layer optically as well, using the SPD activation values obtained from the inference of the first layer. This provides a comprehensive illustration of the feasibility of optical implementation for the entire network. It is important to note that, in conventional binarized neural networks, the last layer is usually implemented using full precision, as demonstrated in previous studies such as \cite{hubara2016binarized,rastegari2016xnor,bulat2019matrix,qin2020binary}. Our results demonstrate that SPDNNs can be implemented entirely using optics with remarkably low energy requirements. This capability holds promise for further advancements, especially with the integration of coherent optical computing platforms, which will be discussed later.

 Similar to the first layer, we use the same setup to perform the optical matrix-vector multiplication. 
 The difference is that now we do not need to perform single-photon detection that has to control the light intensity at a few photons per detection. In fact, the inference of the last linear layer can be implemented just as the conventional ONNs, where we accumulate a sufficiently high number of photons to reach a high SNR of each detection.  
 The collected SPD activation values, as described in \ref{sec:exp_l1}, are used as inputs to the last linear layer. In the experimental implementation, we choose the data from the model with $N=400$ hidden neurons and $K=5$ shots per activation.  For the 30 frames of one-shot binary SPD activations, every 5 frames of them are averaged to obtain the 6 independent repetitions of the inference. 
 The input activation values to be displayed on the SLM are shown in Supplementary Figure \ref{suppfig:exp_actv}. The possible values for the 5-shot activations are 0, 0.2, 0.4, 0.6, 0.8, and 1.
 If the linear operation was performed in full-precision on a computer, the mean test accuracy would be approximately $99.2\%$. 
To perform the linear operation with real-valued weight elements on our incoherent setup, we divide the weight elements into positive and negative parts. We perform the operation separately for each part, and finally obtain the output value by subtracting the results with negative weights from those with positive weights. 
The two sets of weights to be projected onto the OLED display are shown in Supplementary Figure \ref{suppfig:exp_w2}, where the ten blocks of weights corresponds to the ten output nodes.
This approach at least double the photon budget required for the last layer and has the potential to be optimized for greater energy efficiency. However, even with these non-optimized settings, our results demonstrate that the optical energy budget is already several orders of magnitude lower than the start-of-the-art ONN implementations.

\begin{table}[htbp] 
    \centering
    \begin{tabular}{c|ccc|cc|c}    
        \hline
        \thead{Exposure\\ time} & \thead{Photons per \\detection (pos.)} & \thead{Photons per \\detection (neg.)} & \thead{Total photons\\ in output layer} & \thead{Total detected photons\\ in a full inference} & \;\;\;\thead{Detected photons \\ per multiplication}\;\;\;  & \;\,\thead{  Test accuracy  }\;\,\\ \hline
        1.03 ms & $79.9\pm0.10$ & $79.4\pm0.10$ & $1592.5\pm1.5$ & 2636.3 (0.98 fJ) & 0.008 (0.003 aJ) & $84.7\pm 3.2\%$ \\
        2.06 ms & $159.9\pm0.17$ & $158.2\pm0.20$ & $3184.7\pm2.5$ & 4228.4 (1.57 fJ) & 0.013 (0.005 aJ) & $92.0\pm 2.3\%$ \\
        3.09 ms & $239.7\pm0.17$ & $237.8\pm0.22$ & $4777.4\pm2.6$ & 5821.1 (2.17 fJ) & 0.018 (0.007 aJ) & $95.2\pm 2.0\%$ \\
        4.12 ms & $320.2\pm0.21$ & $317.0\pm0.20$ & $6369.4\pm2.5$ & 7413.2 (2.76 fJ) & 0.023 (0.009 aJ) & $96.4\pm 1.8\%$ \\
        7.21 ms & $560.1\pm0.30$ & $555.2\pm0.33$ & $11145.7\pm5.1$ & 12189.5 (4.54 fJ) & 0.038 (0.014 aJ) & $98.0\pm 1.3\%$ \\
        12.36 ms & $960.1\pm0.32$ & $950.2\pm0.40$ & $19107.3\pm4.1$ & 20151.1 (7.51 fJ) & 0.063 (0.024 aJ) & $98.2\pm 1.1\%$ \\
        18.54 ms & $1439.3\pm0.56$ & $1425.8\pm0.41$ & $28658.1\pm6.6$ & 29701.8 (11.08 fJ) & 0.094 (0.035 aJ) & $99.0\pm 1.0\%$ \\\hline
    \end{tabular}
    \caption{\textbf{Optical energy consumption in SPDNN inference with varying photon budgets in the optical implementation of the output layer.} 
    The first column displays the exposure time of the camera, which determines the number of detected photons. The average photons per detection for both positive (pos.) and negative (neg.) output are calculated from the 6000 dot products derived from 100 input images, 6 repetitions in the first layer inference, and 10 output nodes. The total photons in the output layer are determined by averaging 600 inferences of the last layer, each computing 10 output values. The total detected photons in a full inference are the sum of photons detected in both layers. The average photons per MAC is calculated by dividing the total number of MACs by the total detected photons. Standard deviations are calculated based on 30 repetitions of the last layer detection. The total detected number of photons in a full inference, along with the corresponding optical energy of photons at 532 nm, are displayed in the fifth column, with standard deviations omitted for simplicity. These results add the 1043.7 photons used in the first layer. The sixth column displays the average detected number of photons per MAC during a full inference, dividing the numbers in the fifth column by the total number of MACs, 317,600. The last column shows the test accuracy of the inferences at each photon budget.
    }
    \label{tab:last_acc}
\end{table}

\begin{figure}[htp]
\centering
\includegraphics [width=0.88\textwidth] {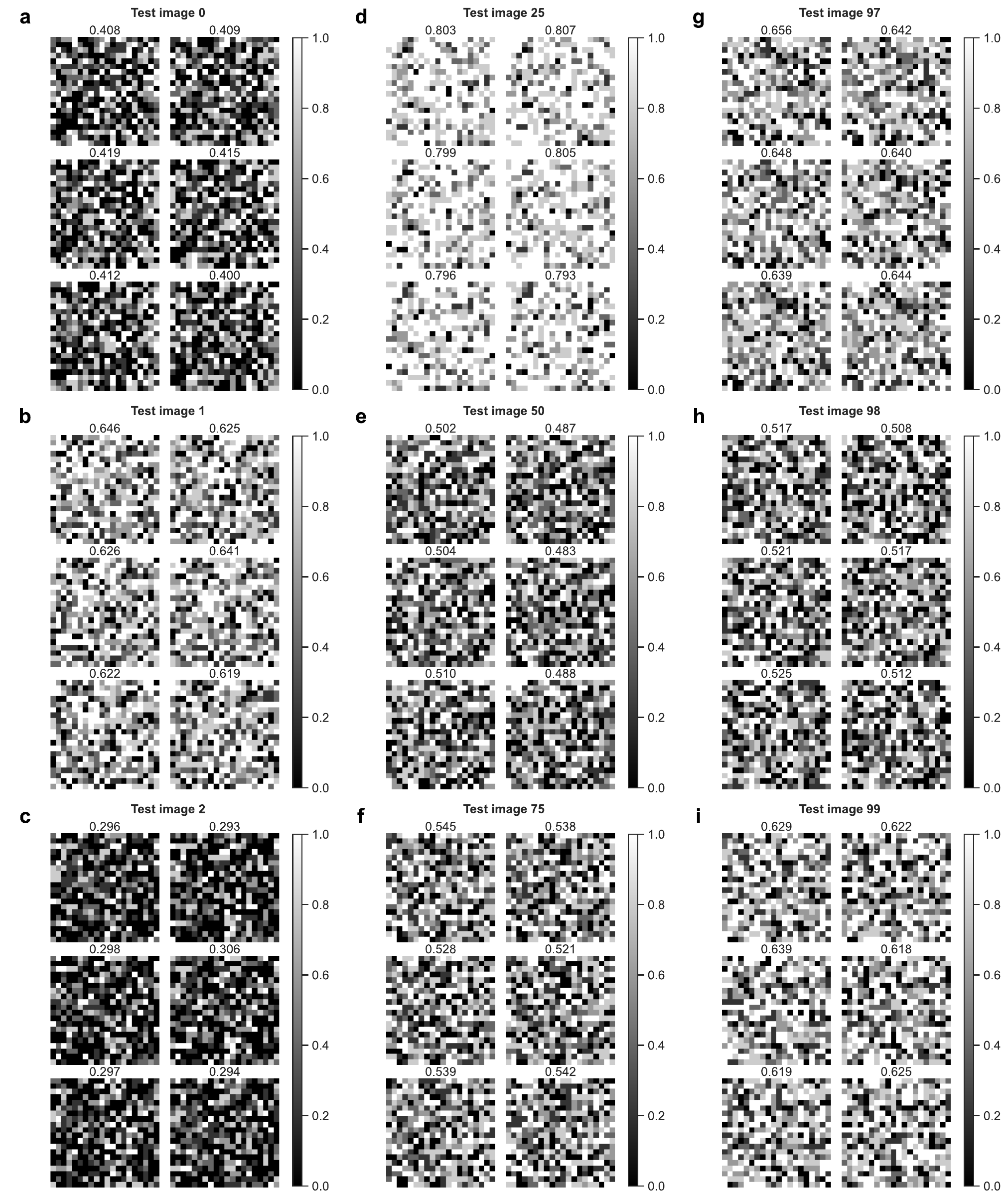}
\caption{\textbf{Visualization of activation values on the SLM during the last layer experiment.} 
This figure displays the activations obtained from the data collected for the model of $N=400$ hidden neurons and $K=5$ shots per activation. The possible values for the 5-shot activations are 0, 0.2, 0.4, 0.6, 0.8, and 1. The activations of size 400 are rearranged into a $20 \times 20$ shape, which corresponds to their physical layout on the SLM. Panels \textbf{a} to \textbf{i} display the activations of test images with indices 0, 1, 2, 25, 50, 75, 97, 98, and 99, respectively, each with 6 repetitions of inference. The average value of the activations in each block is indicated at the top. The overall average activation value of the 100 test images is $\sim 0.5219$.}
\label{suppfig:exp_actv}
\end{figure}

\begin{figure}[htp]
\centering
\includegraphics [width=0.9\textwidth] {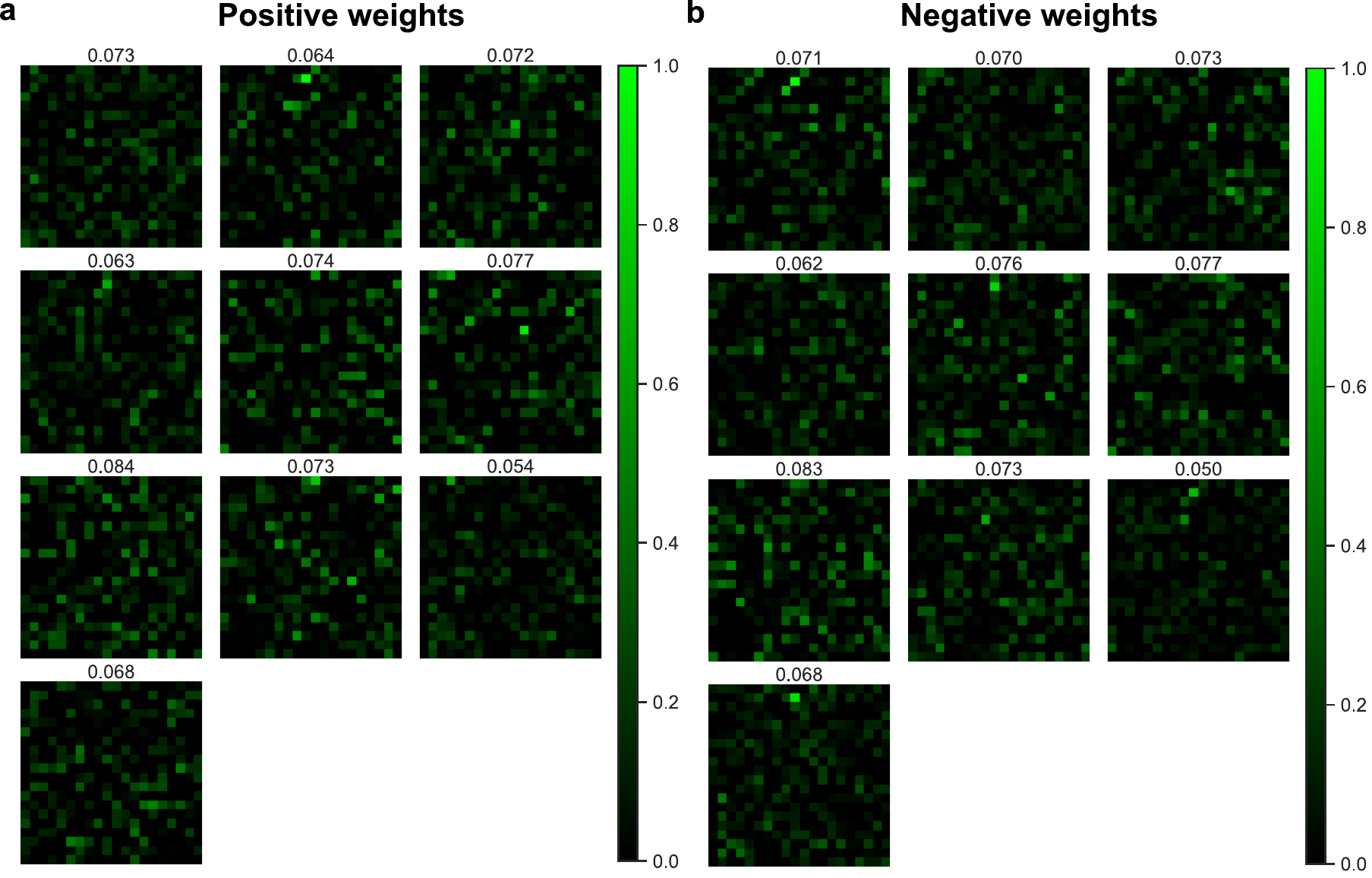}
\caption{\textbf{Visualization of output layer weights on the OLED display.} This figure presents the positive and negative weights from a model with $N=400$ hidden neurons and a weight matrix of shape $400 \times 10$. The positive and negative weights are depicted in panels \textbf{a} and \textbf{b}, respectively. To match the layout on the OLED display, each weight vector of size 400, corresponding to one of the 10 output nodes, was rearranged into a block with a shape of $20 \times 20$ and displayed using green pixels. The values were normalized to the range of 0 to 1 and the average value of each block is indicated at the top.}
\label{suppfig:exp_w2}
\end{figure}

\begin{figure}[htp]
\includegraphics [width=.98\textwidth] {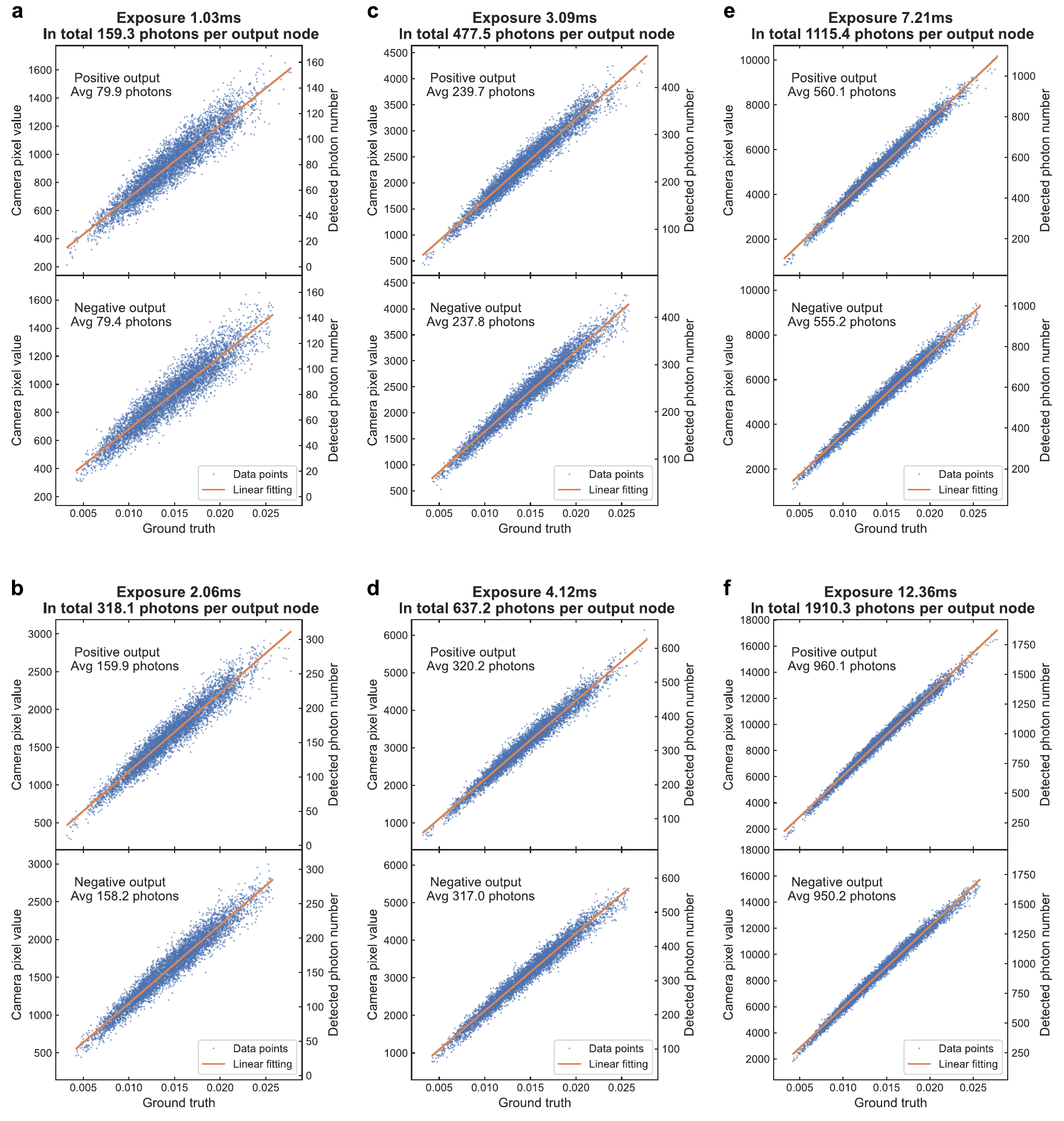}
\caption{\textbf{Calibration of collected experimental data of the last layer inference.} This figure shows the raw data of ``high-SNR'' optical measurement on the qCMOS camera with various exposure times, each depicted in a separate panel. For each exposure time, one output value was obtained by measuring the output from both positive and negative weights. Each plot includes 6,000 data points, representing 100 test images, 6 repetitions in the hidden layer activation, and 10 output nodes. The ground truth values were computed using full-precision operations on a digital computer. Both the raw camera pixel values and the corresponding detected photon numbers are displayed on the y-axis, with the average detected photon numbers for the 6,000 data points noted in each plot.}
\label{suppfig:l2_cali}
\end{figure}

In the implementation, we adjust the exposure time of the camera to control the optical energy per detection. In order to perform the inference on the 100 input images and 10 output nodes, along with 6 repetitions of the activation values and 2 sets of weights, we need to perform a total of $100\times 6\times 10\times 2=12000$ vector-vector dot products, each with a size of 400. Each vector-vector dot product detection is repeated 100 times.
The results are presented in Supplementary Table \ref{tab:last_acc}. The photons per detection of either positive or negative output are each averaged over $100\times 6\times 10=6000$ dot products. The total photons detected in the last layer per inference are averaged over the 100 input images and 6 repetitions, totaling  $100\times 6=600$ inferences. The standard deviation of the photon numbers are calculated based on the 100 repeated detections for each dot product.
The total detected photons in a full inference is the sum of those in the last layer and the first layer. The average value of the binary activations collected for the $N=400$ model is $\sim$0.52186, resulting in a total of $0.52186 \times 400 \times 5 \approx 1043.7$ detected photons per inference in the first layer, with 5 shots per activation. This number is then combined with the total detected photons in the last layer to obtain the overall photon count for a full inference.
We can see that the photon budget can be reduced by 5 folds if we only have one shot per inference. 
In a full inference with $N=400$ hidden neurons and $K=5$ shots per activation, the total number of vector-vector products in the first layer is 400 and that in the last layer is $10$ for the 10 output nodes. 
With dot products of size 784 in the first layer and 400 in the last layer, the total number of MACs in one inference process is equal to 317,600 ($400 \times 784 + 10 \times 400$). To calculate the number of detected photons per MAC, we divide the total number of detected photons in a full inference by the total number of MACs. 
The prediction of a given inference is made by directly evaluating the output values of each of the 10 output nodes. The output values are calculated as the difference between the positive and negative output intensity. The label of the node with the highest output value is then determined to be the predicted label.
The test accuracy on the 100 test images is presented with its mean and standard deviation in the final column of Supplementary Table \ref{tab:last_acc}. The standard deviation is determined by considering both the 6 repetitions of the first layer's inference and the 100 repetitions of detections in the last layer.

\begin{figure}[htp]
\includegraphics [width=.88\textwidth] {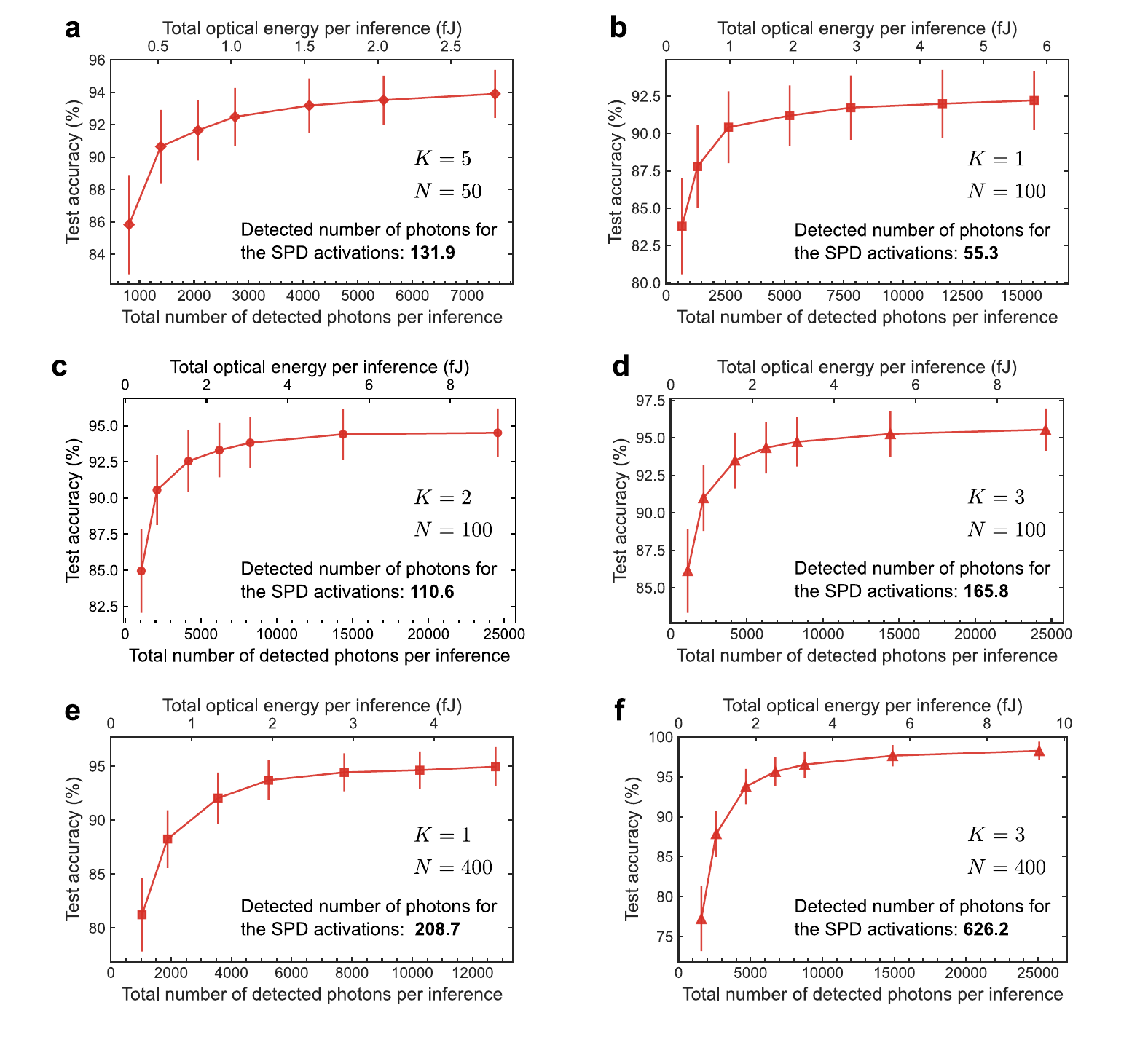}
\caption{\textbf{Experimental results of full-optical implementation with different SPDNN configurations.} 
In addition to Figure 3d in the main text, this figure shows the results obtained with different numbers of hidden neurons ($N$) and shots of SPD measurements per activation ($K$). Each model uses experimentally collected activation values as input for the optical implementation of the output layer. 
The number of detected photons in the first $784\rightarrow N$ layer to compute the SPD activations in each configuration is denoted in the corresponding plot. The noise-free test accuracies with full-precision output layer are shown in Supplementary Table \ref{tab:first_acc}.
The number of detected photons in the $N\rightarrow 10$ output layer is varied to control the noise in the optical implementation, which is reflected in the resultant test accuracy.
The total number of detected photons per inference is the sum of the photon budgets in the two layers.
}
\label{suppfig:l2_results}
\end{figure}

To visualize the impact of photon noise on accuracy in ONN inferences with a limited photon budget, the data collected from the last layer inference is depicted in Supplementary Figure \ref{suppfig:l2_cali}. In each panel, 6000 data points are plotted for either positive or negative output, considering the 100 input images, 6 repetitions in the first layer inference, and 10 output nodes. The ground truth dot product values are computed with high-precision operations on a computer. Both the raw camera pixel values and the corresponding photon count are shown on the vertical axes. As the number of detected photons per detection increases, the detected values become less noisy, resulting in a test accuracy that is closer to the ground truth of 99.2\% (Supplementary Table \ref{tab:first_acc}). Similar to conventional optical neural networks, the decrease in accuracy is primarily due to shot noise.

In addition, we performed the output layer optically for other configurations as well. The results are represented in  Supplementary Figure \ref{suppfig:l2_results}). The activation values collected in experiments of other choices of number of hidden neuron $N$ and shots of SPD readouts $K$ are used as the input for the output layer. If the output layer is implemented with full numerical precision, the test accuracies were shown in Supplementary Table \ref{tab:first_acc}. These accuracies are the upper bound for the full-optical implementation with the presence of noise in optical implementation. 
For these configurations of numbers of hidden neurons ($N$) and shots of SPD measurements per activation ($K$), one inference through the $784\rightarrow N$ hidden layer involves $N\times K$ SPD measurements to compute the activation vector in the hidden layer of size $N$. The detected number of photons for the SPD activation computation in the hidden layer of each configuration is denoted in the corresponding panel in Supplementary Figure \ref{suppfig:l2_results}. The total number of detected photons per inference is the summation of this number and the total number of photons detected in the $N\rightarrow 10$ output layer, similar to the procedure we discussed above for the configuration of $N=400$ and $K=5$.

Similar to the plot in Figure 3d in the main text, the test accuracies increase with the detected optical energy in a similar trend to that of the $N=400$, $K=5$ we discussed in detail above. Comparing these panels with different configurations, we can see that models with a smaller number of neurons $N$ exhibit greater resilience to noise when a similar number of photons are used in the output layer. For instance, comparing panels d and f, with approximately 2000 photons in the output layer (the second point from the left), the test accuracy declines more in the $N=400$ model (panel f) than in the $N=100$ model (panel d). Although the models with smaller $N$ and $K$ suffer from a lower noise-free accuracy due to a smaller network size and higher stochasticity, as shown in Supplementary Table \ref{tab:first_acc}.
The final test accuracy is a combination of these two factors.

\newpage
\part{Discussion}

\vspace{24pt}
\section{Robustness tests of SPDNNs}
\label{sec:rob}

The first thing to check is the errors induced by the single-photon detectors. 
The two key parameters to consider when choosing commercial SPDs are photon detection efficiency and dark count rate.
Photon detection efficiency refers to the amount of incident light that can be detected by the SPD. Although low photon detection efficiency is a common issue in many photon experiments, it does not add extra noise to our SPDNN models. This is because any attenuation to the light still follows a Poisson distribution and cannot be noisier than a single-photon detector. Hence, a low photon detection efficiency will only add to the overall transmission loss in the setup, and the input light power is usually redundant, so it will not affect the performance much.
On the other hand, dark count rates, or false clicks, could pose a greater challenge in experiments with SPDs. False clicks are hard to distinguish from real signals, and the output of the detection is binary. The dark count rate of a functional SPD is typically between $10^{-5}$ and $10^{-2}$ false clicks per signal, depending on the experimental configuration. In some extreme circumstances, such as when the exposure time is very long or when it is hard to remove ambient light, the dark count rate could be as high as one false click in tens of detections, ruining the results of the experiment.
However, our SPDNN models are resilient to high dark count rates. As shown in Supplementary Figure \ref{suppfig:robust_test}a, even with a false click in fewer than 10 measurements, we still obtain relatively good accuracy. The common range of $<10^{-2}$ barely affects the performance of the SPDNNs.

\vspace{20pt}
\begin{figure}[htp]
\includegraphics [width=0.86\textwidth] {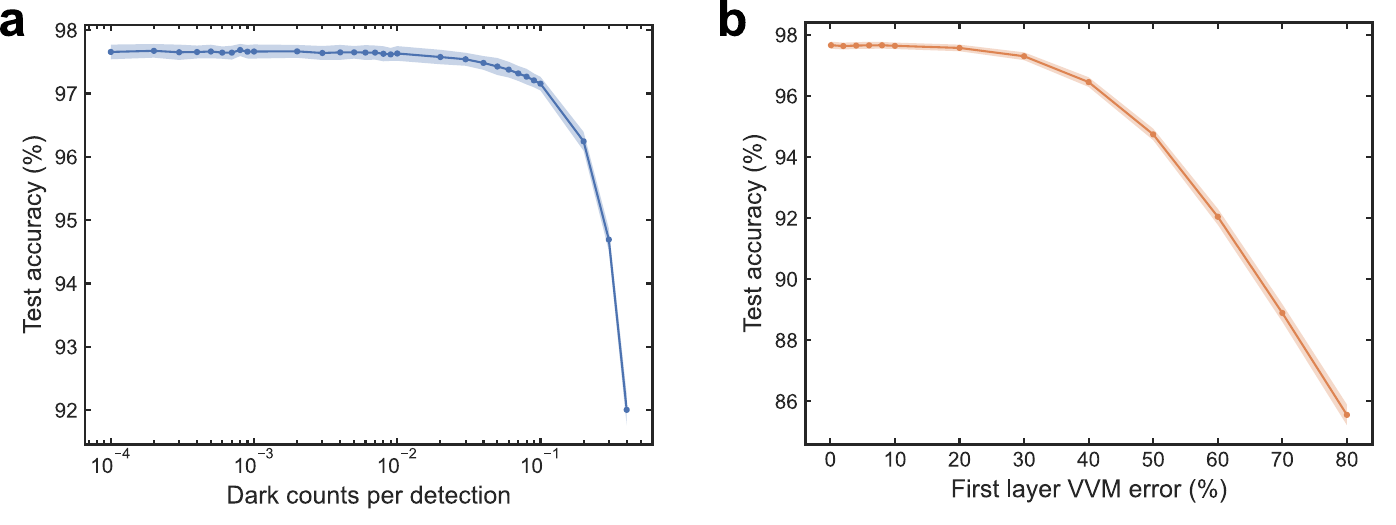}
\caption{\textbf{Robustness tests for the SPDNN model.} \textbf{a,} MNIST test accuracy as a function of dark count rate of each SPD activation measurement. The dark count rate is varied to test the robustness of the SPDNN to noise in the measurement process. 
\textbf{b,} MNIST test accuracy as a function of the relative errors in the vector-vector multiplications (VVMs, or dot products) in the first layer. The errors are introduced at a fixed ratio with respect to the dot product result to simulate systematic errors in the optical setup. Both panels present results obtained from the $784 \rightarrow 400 \rightarrow 10$ incoherent model with one shot per
inference (N = 400, K = 1), and the test accuracies are computed on the full test set of 10,000 images.}
\label{suppfig:robust_test}
\end{figure}

As introduced in \ref{sec:qcmos}, the dark count rate with our SPD setting is 0.01 per second per pixel. Given the exposure time of milliseconds, the effects due to dark counts are trivial in the experimental implementation.
In summary, the robustness of SPDNN models to noise obviates the need for selecting specialized SPDs for experimental realization. Cost-effective SPDs can be employed for implementing SPDNNs with high performance. Furthermore, considering the significant power consumption of cooling systems for state-of-the-art SPDs, relaxing the dark current requirement can greatly reduce the power consumption of the detection system.

The precision of linear operations is a crucial factor in neural network inferences. As discussed in \ref{sec:vvm}, the accuracy of vector-vector multiplication may not be optimal when using a single-pixel camera for single-photon detection. To assess the effect of errors in dot product calculations on the performance, we conducted a simulation test by adding different levels of random noise to the dot product results in the first layer, which serve as the pre-activations to the SPD activation function. The results, shown in Supplementary Figure \ref{suppfig:robust_test}b, indicate that SPDNNs are robust to errors in linear operations, even with up to $20\%$ relative noise. This robustness ensures the reliability of the experimental implementation.

\vspace{24pt}
\section{Noise resilience compared to conventional models}
In our SPD activation function, two key features set it apart from conventional neural networks: the quantization of activation values and the stochastic activation process. Both of these processes occur naturally through the detection of single photons.
The intrinsic quantization of energy and detection process results in a nonlinear response to the input light intensity, eliminating the need for additional nonlinear operations in the neural network. This nonlinearity is evident in the higher MNIST classification test accuracy of SPDNNs compared to linear models. Additionally, the intrinsic photon noise in the activation function makes the output values stochastic. 
With more averaging, the stochasticity is reduced, resulting in a more precise output as seen in the implementation of SPD activations in the fully-connected layers.
This may imply that the noise is unwanted in the neural network inferences.
However, the stochastic inference is inevitable in many real-world tasks with a physical device, our stochastic models demonstrated a high noise-resilience that can still yield reliable outputs regardless of this amount of stochasticity.

\begin{figure}[htp]
\includegraphics [width=0.96\textwidth] {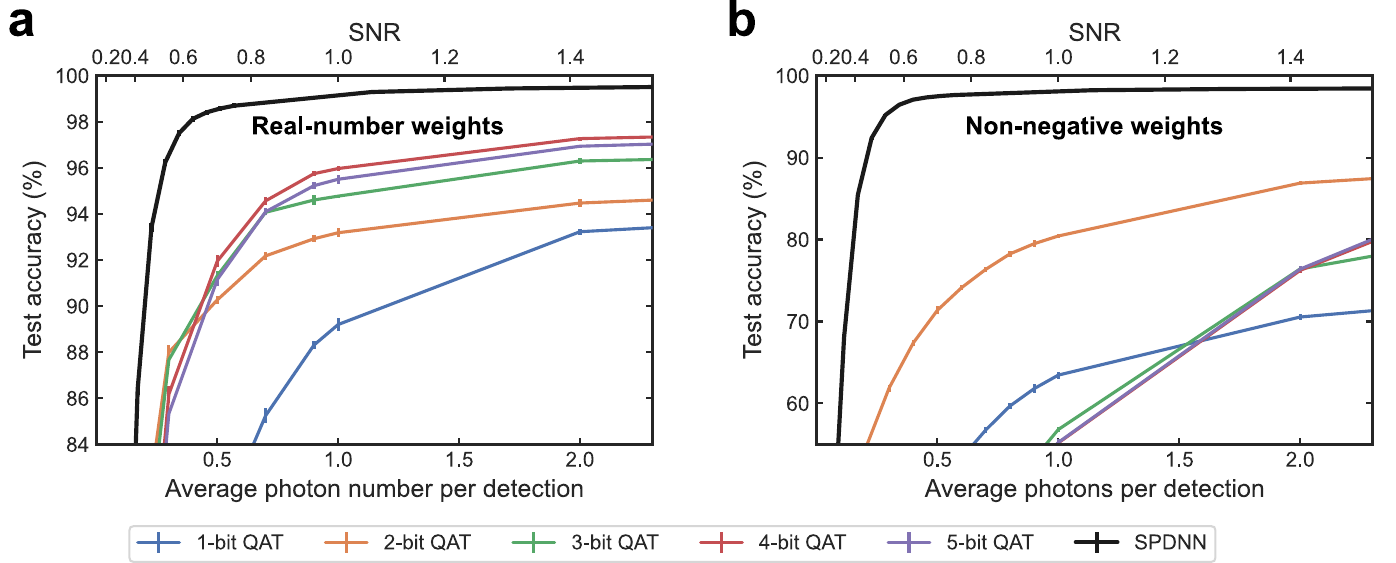}
\caption{\textbf{Comparison of SPDNNs and QAT models on MNIST classification.} \textbf{a,} The test accuracy of MNIST classification using models with real-number weights is shown as a function of photon budget in the hidden layer. The SPDNN model employs the coherent SPD activation function, while the QAT models use ReLU as the activation function. 
\textbf{b,} The test accuracy of MNIST classification using models with non-negative weights is shown as a function of photon budget in the hidden layer. The SPDNN model employs an incoherent setup with non-negative weights, while the QAT models use Sigmoid activation function. The test accuracies are calculated on the full test set of 10,000 images. 
Both panels present results obtained from the $784 \rightarrow 400 \rightarrow 10$ model with one shot per
inference (N = 400, K = 1), and the test accuracies are computed on the full test set of 10,000 images.}
\label{suppfig:qat}
\end{figure}

To evaluate the noise resilience of our SPDNNs against conventional continuous-variable models, we conducted experiments to compare the test accuracy of the models under varying levels of photon noise. We adopted quantization-aware training (QAT) as a popular noise-aware training method, which involves quantizing the weights during training to make the model more noise-resilient. We trained deterministic QAT models with the same multi-layer perceptron (MLP) structure of $784\rightarrow400\rightarrow10$ and quantized the weight precision to a specific number of bits. We then compared the MNIST test accuracy of these models to SPDNNs with the same level of photon noise added during the neural network inference of the hidden layer.

For the real-valued QAT models that are compared to the coherent SPDNNs, we chose to use the ReLU activation functions. The QAT models adopted a deterministic quantization function and quantized the weights to the corresponding precision. During inferences, we performed computations with full precision, with the photon noise added to the pre-activation values of the hidden neurons. Supplementary Figure \ref{suppfig:qat}a shows that the ReLU models exhibit high noise resilience, and harsh quantization does not significantly enhance the noise resilience but harms the overall precision. In fact, decreasing the quantization levels leads to decreased model performance at this photon noise level. The accuracy almost converges at a precision of 5 bits or higher.

For the non-negative QAT models that are compared to the incoherent SPDNNs, the non-negativity of the weights renders ReLU activation functions less effective. Hence, we use the Sigmoid activation function, more rigorously, the positive half of it, to train the QAT models. However, the models are not as noise resilient as with real-number operations, and stronger quantization is required to enhance the model robustness. 
As the simulation results show, the performance of models of precision 3 bits or more almost converges. It is worth noting that, despite having over $98\%$ test accuracy without photon noise, the performance of these models with 3-bit precision or more is worse under such noise levels. 
Decreasing the quantized precision is a tradeoff between noise resilience and overall accuracy. We observed that the 2-bit QAT model performs the best over other precisions.
These results showed that all the QAT models are inferior to SPDNNs in terms of accuracy under the same or lower photon budget. This finding indicates that SPDNNs are more effective in achieving high accuracy in photon-starved environments.

Our results suggest that natural quantization of optical energy enhances noise resilience in neural networks, and that stochasticity could aid in searching for more accurate and noise-resilient models. However, we do not claim that the SPD activation function is the best way to train a noisy neural network, and we are open to exploring other noise-aware training methods that could further improve resilience. Our findings demonstrate that with appropriate training that takes into account the stochastic and quantized nature of optical energy in the realistic physical computing system, ONNs can achieve high performance even at very high noise levels, which was not previously possible. What makes it more intriguing about our approach is that it exploits the natural single-photon detection process.

\vspace{24pt}
\section{Distribution of expectation values for SPD activations}

\begin{figure}[htp]
\includegraphics [width=0.96\textwidth] {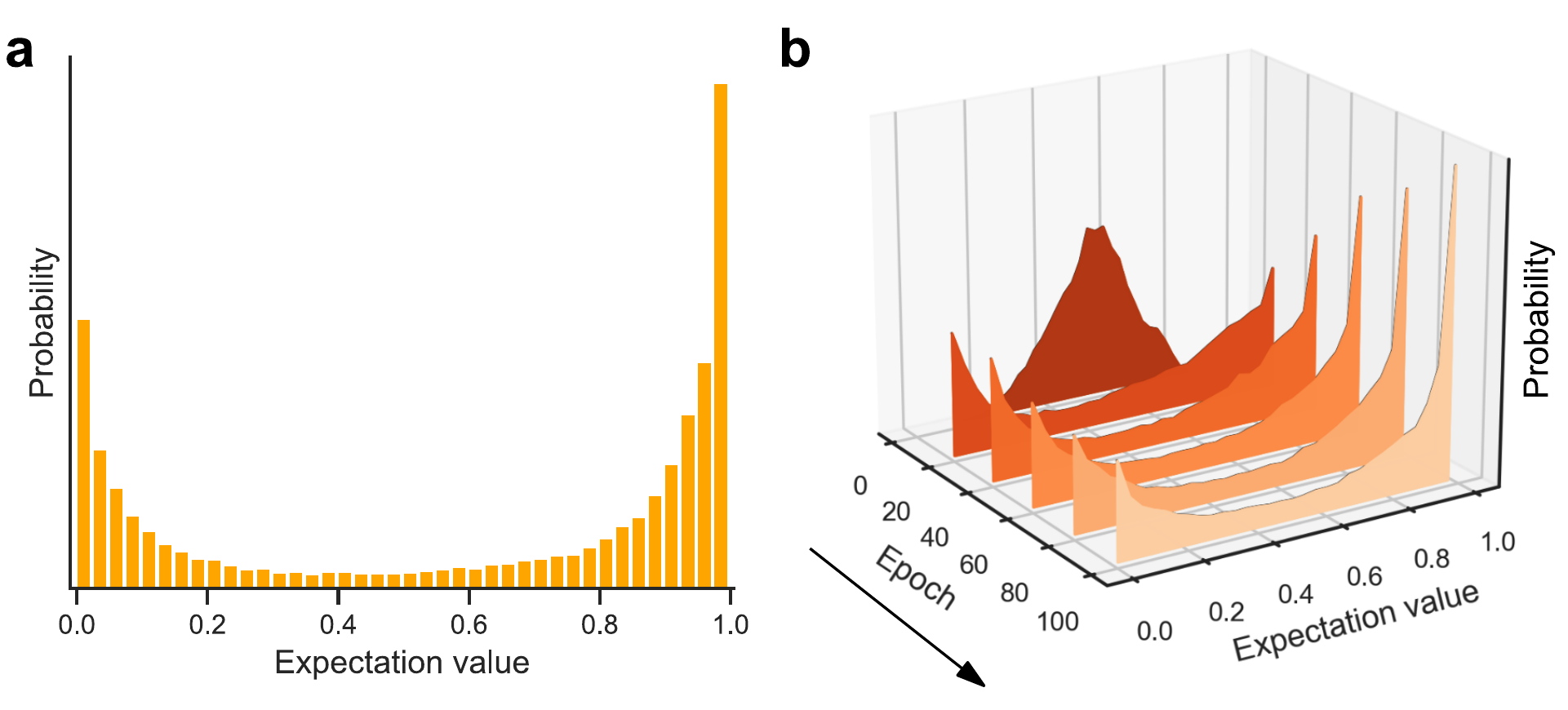}
\caption{\textbf{Distribution of expectation values for hidden neuron activations.} \textbf{a,} The distribution of expectation values for each hidden neuron during the inference of a trained model. The model follows a multi-layer perceptron (MLP) structure of $784\rightarrow400\rightarrow400\rightarrow10$ using a coherent SPD activation function and real-valued weights (Supplementary Note 2).
The expectation values of neuron activations in the first hidden layer are depicted. 
\textbf{b,} The evolution of the expectation value distribution during training. Different expectation value distributions of hidden neuron activations (as in \textbf{a}) are plotted at six training epochs to demonstrate changes throughout the training process.}
\label{suppfig:expct_distr}
\end{figure}

In this study, we explored the use of highly stochastic SPDNN models to achieve high performance in deterministic classification tasks. At first glance, this may seem counter-intuitive, as deterministic classification typically requires stable and reliable outputs, while stochastic models introduce inherent uncertainty. However, a closer examination of the characteristics of the activation values in SPDNN inferences provides a more intuitive understanding of how this approach can achieve such high accuracy.

In Supplementary Figure \ref{suppfig:expct_distr}a, we present the distribution of expectation values for hidden neuron activations. This distribution is obtained using a single shot of SPD readout ($K=1$). Since the activations are binary (either 0 or 1), the expectation value represents the probability of the activation being 1. We constructed this histogram by considering the inferences for all input images in the test set and all hidden neurons' activation values, so that the distribution is averaged over many different samples to show the overall picture of the general behavior of the network inference. For example, a layer with 400 hidden neurons and 10,000 test input images would yield $400\times10,000=4\times10^6$ expectation values included in the histogram. 
We utilized an optimized SPDNN model with an MLP structure of $784\rightarrow400\rightarrow400\rightarrow10$ to generate this histogram, and we also found that this distribution is consistent across models with varying numbers of hidden neurons or layers, as well as coherent or incoherent SPD detection schemes.

Interestingly, we observed that the majority of neuron activations exhibit more deterministic expectation values rather than pure randomness. While some models trained with experimental limitations cannot reach absolute zero values, the peak at zero value shifts to a less sharp bump close to zero, still distributing towards either end rather than the middle value of 0.5. In Bernoulli sampling, an expectation value of 0.5 signifies that the probability of being 0 or 1 is equivalent, indicating that there is no useful information in the process, and the entropy is at its maximum. Noisy channels with such characteristics cannot carry valuable information for neural network inference. Consequently, during the training process, the model should strive to learn from the training set and update the neural network weights accordingly to capture the essential features. This process involves storing information in the trained model, which can be reflected by decreasing the entropy of each stochastic binary neuron. 

In Supplementary Figure \ref{suppfig:expct_distr}b, we observe that as the model undergoes more training epochs, the expectation value distribution of activations becomes more concentrated towards 0 or 1. This indicates that the model retains more information and generates more reliable outputs.

However, it is important to note that while the entropy of each individual neuron decreases, at the network level, the average activation still tends to be around 0.5 photons when considering all the neurons, denoting maximum entropy. This suggests that the neural network is effectively utilizing its capacity to extract information using all its neurons by increasing the overall network entropy. In fact, a network with all neurons having the same expectation value (entropy of 0) would not be able to learn any meaningful features.

In summary, while SPDNNs are inherently stochastic, the distribution of expectation values for hidden neuron activations leans towards deterministic outcomes, allowing the model to effectively learn features and achieve high accuracy in deterministic classification tasks. The training process shapes the probabilistic distribution of the neurons and allocates different neurons close to either 0 or 1 to learn the patterns of input images and output reliable inferences. Remarkably, the implementation of this allocation is exceptionally efficient in optical energy, as each activation only involves a photon click.

\section{Background for stochastic computing}
The stochastic operation of neural networks has been extensively studied in computer science as part of the broader field of stochastic computing \cite{alaghi2013survey}. In the field of machine learning, binary stochastic neurons (BSNs) have been used to construct stochastic neural networks \cite{ackley1985learning,neal1990learning,neal1992connectionist,bengio2013estimating,tang2013learning,raiko2014techniques,hubara2016binarized}, with training being a major focus of study. Investigations of hardware implementations of stochastic computing neural networks, such as those in Refs.~\cite{ji2015hardware,lee2017energy} (with many more surveyed in Ref.~\cite{liu2020survey}), have typically been for deterministic complementary metal--oxide--semiconductor (CMOS) electronics, with the stochasticity introduced by random-number generators. While many studies of binary stochastic neural networks have been conducted with standard digital CMOS processors, there have also been proposals to construct them from beyond-CMOS hardware, motivated by the desire to minimize power consumption: direct implementation of binary stochastic neurons using bistable systems that are noisy by design---such as low-barrier magnetic tunnel junctions (MTJs)---has been explored \cite{vodenicarevic2017low,hassan2019low,chowdhury2023full}, and there have also been proposals to realize hardware stochastic elements for neural networks that could be constructed with noisy CMOS electronics or other physical substrates \cite{hylton2021vision,coles2023thermodynamic}. ONNs in which noise has been intentionally added \cite{wu2022harnessing,wu2022photonic,ma2023stochastic} have also been studied. Our work with low-photon-count optics is related but distinct from many of the studies cited here in its motivating assumption: instead of desiring noise and stochastic behavior---and purposefully designing devices to have them, we are concerned with situations in which physical devices have large and unavoidable noise but where we would like to nevertheless construct deterministic classifiers using these devices because of their potential for low-energy computing (Fig. 1 in main text).

\section{Prospects for future applications}
While we have demonstrated a fundamental point---that ONNs can be successfully operated in the few-photon-per-activation regime in which quantum shot noise causes very low SNR---an important practical consideration for the construction of ONNs is that the energy used by optical signals within the ONN is only part of the ONN's total energy consumption, and it is the total energy per inference that is generally what one wants to optimize for \cite{nahmias2019photonic,hamerly2019large,anderson2024optical}. A practical limitation of our experiments is that they were conducted with a relatively slow\footnote{19.8~kHz maximum frame rate.} single-photon-detector array, limiting the speed at which a single execution of a layer could be carried out, and the detector array was not optimized for energy efficiency. For our fundamental approach and methods to be applied to make ONNs that offer a practical advantage over state-of-the-art electronic processors as generic neural-network accelerators, there remains important work to be done in engineering an overall system that operates sufficiently fast while minimizing total energy cost. Recent progress in the development of large, fast arrays of single-photon detectors coupled with digital logic \cite{bruschini2023linospad2} suggest that there is a path towards this goal. Ref.~\cite{shainline2017superconducting} has also pointed out the possibility of using fast superconducting-nanowire single-photon detectors for realizing spiking neural networks. Furthermore, there is a complementary path toward utility in the nearer term: if instead of aiming to use ONNs to entirely replace electronic processors, one uses ONNs as a pre-processor for input data that is already optical \cite{wetzstein2020inference,wang2023image,huang2023photonic}, operating the ONN with single-photon detectors is a natural match with scenarios in which the optical input is very weak---for example, in low-light-imaging applications.

Our approach is not tied to a specific architecture of ONN---the free-space matrix-vector multiplier used in our experiments is just one of many possible choices of architecture. Other ONNs could be adapted to use our approach by replacing the photodetectors typically used for readout of neurons at the end of a layer with single-photon detectors. ONNs based on diffractive optics \cite{lin2018all,chang2018hybrid,zhou2021large}, Mach-Zehnder interferometer (MZI) meshes \cite{carolan2015universal,shen2017deep,bogaerts2020programmable}, and other on-chip approaches to matrix-vector multiplication \cite{tait2015demonstration,xu202111,feldmann2021parallel} all appear compatible.

In our optical experiments, we used single-photon detectors that output an electronic signal when a photon is detected. However, in multilayer ONNs, the input to each layer is optical. One can convert an electronic detector output to an optical input by modulating an optical source---which is what we did and what is often done in ONNs more generally \cite{wetzstein2020inference}---but an alternative is to construct a device that performs SPD with high efficiency and gives the measurement result as an \textit{optical} signal that can be directly used as an input to the next layer in the ONN. Designing and demonstrating such a device is an interesting potential avenue for future work in applied quantum nonlinear optics \cite{mazets2007multiatom,pinotsi2008single,sotier2009femtosecond,kiilerich2019input,li2023single,roques2023biasing}, and could lead to both lower electronic energy consumption and higher speed for single-photon-detection ONNs.

We trained our demonstration SPDNN \textit{in silico} using backpropagation, but if SPDNNs with high overall energy efficiency are built, it would be a boon use this efficient hardware not only for inference but also for training. To this end, it could be interesting to study how to adapt \textit{in situ} training \cite{zhou2020insitu,guo2021backpropagation,bandyopadhyay2022single,pai2023experimentally}, including backpropagation-free (e.g., Refs.~\cite{bengio2015towards,lillicrap2020backpropagation,hinton2023mortal,stern2023learning}), methods for SPDNNs. An open question related to training is whether it is possible to make SPDNNs that do not involve a final high-SNR layer while preserving task accuracy; this could help to reduce the overall energy per inference. Other future work could explore the extension of our research to neural networks with larger sizes (wider and more layers, which could both improve the capability of the neural network and further amortize the energy cost of the final, high-SNR layer, if used), more sophisticated classification tasks (beyond MNIST and CIFAR-10 image classification---such as has been shown with conventional binary neural networks \cite{rastegari2016xnor, bulat2019xnor, bulat2019matrix}), and generative or other probabilistic tasks---for which the stochasticity can be harnessed rather than merely tolerated. Beyond machine-learning tasks, an SPDNN layer could be used as the core of a single-photon-regime photonic Ising machine \cite{mohseni2022ising} for heuristically solving combinatorial-optimization problems, realizing an optical version of p-bit computing \cite{chowdhury2023full}.

The phenomena observed in our work seemingly relies on two key physical ingredients. First, the system's available states are effectively quantized, as in the photonic quantization of energy in our ONN demonstration, or the binarization that occurs in low-barrier, stochastic magnetic tunnel junctions \cite{grollier2020neuromorphic}. Second, the noise in the system results in the quantized outputs of the system being stochastic. This suggests that ultra-low-SNR physical neural networks should be possible in many physical hardware platforms beyond photonics. Systems in which shot noise dominates are natural matches with our approach and methods. Our approach could also be relevant to systems in which thermal (Johnson) noise dominates---as is typically the case in room-temperature electronics---but this will depend on not just the noise but also the system's dynamics. Which hardware platforms and system architectures can yield an overall energy benefit by being operated in a stochastic regime while maintaining computational accuracy is an important open question.

\bibliographystyle{mcmahonlab}
\bibliography{references}